\documentclass[11pt]{article}

\usepackage{vasilis}
\usepackage{enumitem}
\usepackage{tikz}
\usepackage{microtype}
\usepackage[normalem]{ulem}
\usepackage{array,booktabs}
\usepackage{longtable,tabularray}

\newcommand{\Hh}{\mathbb H}
\newcommand{\Leb}{\operatorname{Leb}}
\newcommand{\dist}{\operatorname{dist}}

\newcommand{\na}{\mathrm{na}}
\newcommand{\atom}{\mathrm{Atom}}

\newcommand{\planarinstance}[1][1]{%
\begin{tikzpicture}[
  x=2cm, y=3.3cm, font=\small,
  axis/.style={draw=black!55, line width=0.4pt},
  guide/.style={draw=black!20, densely dashed, line width=0.4pt},
  point/.style={circle, inner sep=0pt, minimum size=5pt},
  red point/.style={point, fill=red!80!black},
  blue point/.style={point, fill=blue!80!black},
  red mass/.style={text=red!80!black, inner sep=2pt},
  blue mass/.style={text=blue!80!black, inner sep=2pt},
  measure/.style={draw=black!75, line width=0.45pt, Latex-Latex},
  proj/.style={draw=black!75, line width=0.4pt}
]
  \def\epsdraw{0.18}
  \pgfmathsetmacro{\sqrteps}{sqrt(\epsdraw)}
  \def\farx{2.65} 
  \def\breakx{1.80}

  \def\centerdimy{0.09}
  \def\topdimraise{0.075}
  \def\bottomdimraise{-0.075}
  \def\fardimy{0.11}

  \draw[axis]
    ({-\farx-0.35},0) -- ({-\breakx-0.08},0);
  \draw[axis]
    ({-\breakx+0.08},0) -- ({\breakx-0.08},0);
  \draw[axis,-{Latex[length=4pt]}]
    ({\breakx+0.08},0) -- ({\farx+0.35},0)
    node[right] {$x$};

  \foreach \s in {-1,1} {
    \draw[axis]
      ({\s*\breakx-0.12},-0.035) --
      ({\s*\breakx-0.04}, 0.035)
      ({\s*\breakx+0.04},-0.035) --
      ({\s*\breakx+0.12}, 0.035);
  }

  \draw[axis,-{Latex[length=4pt]}]
    (0,{-\epsdraw-0.20}) -- (0,{\sqrteps+0.30})
    node[above] {$y$};

  \node[below=10pt] at (-0.75*\farx,0) {$-\frac{4}{3\varepsilon}$};
  \node[below=10pt] at ( 0.75*\farx,0) {$\frac{4}{3\varepsilon}$};

  \node[red point]  (RI) at (-\farx,0) {};
  \node[blue point] (BI) at ( \farx,0) {};

  \node[red mass, above=6pt] at (RI) {$\varepsilon$};
  \node[blue mass,above=6pt] at (BI) {$\varepsilon$};

  \node[red point]  (RII) at ( 1,-\epsdraw) {};
  \node[blue point] (BII) at (-1,-\epsdraw) {};

  \node[red mass, below=6pt] at (RII) {$1/2 - \eps/3$};
  \node[blue mass,below=6pt] at (BII) {$1/2 - \eps/3$};

  \node[red point]  (RIII) at ( \epsdraw,0) {};
  \node[blue point] (BIII) at (-\epsdraw,0) {};

  \node[red mass, above right=6pt] at (RIII) {$1/3 - \eps/3$};
  \node[blue mass, above left=6pt] at (BIII) {$1/3 - \eps/3$};

  \node[red point]  (RIV) at (-1,\sqrteps) {};
  \node[blue point] (BIV) at ( 1,\sqrteps) {};

  \node[red mass, above=6pt] at (RIV) {$1/6 - \eps/3$};
  \node[blue mass,above=6pt] at (BIV) {$1/6 - \eps/3$};

  \draw[measure] (-\epsdraw,0) -- (\epsdraw,0);
  \node[above right=1pt] at (0,0) {$2\varepsilon$};

  \draw[measure] (-1,\sqrteps) -- (1,\sqrteps);
  \node[above=2pt] at (-\epsdraw,\sqrteps) {$2$};

  \draw[measure] (-1,-\epsdraw) -- (1,-\epsdraw);
  \node[below left=2pt] at (0,-\epsdraw) {$2$};

  \draw[measure] (0,-\fardimy) -- (\farx,-\fardimy);
  \draw[measure] (-\farx,-\fardimy) -- (0,-\fardimy);

  \draw[measure] (1+\epsdraw,0) -- (1+\epsdraw,-\epsdraw);
  \node[right=2pt] at (1+\epsdraw,-0.3*\epsdraw) {$\varepsilon$};
  
  \draw[measure] (1+\epsdraw,0) -- (1+\epsdraw,\sqrteps);
  \node[right=2pt] at (1+\epsdraw,\sqrteps/2) {$\sqrt{\eps}$};

\end{tikzpicture}%
}

\usepackage[textsize=small,textwidth=2cm]{todonotes}

\title{Separating Measures by Hyperplanes}
\author{
Jos\'{e} Correa
\thanks{Department of Industrial Engineering, Universidad de Chile {\tt (correa@uchile.cl)}.
Partially supported by ANID Chile grant FONDECYT 1260036.}
\and
Svenja M.~Griesbach
\thanks{Faculty of Computer Science, RWTH Aachen University {\tt (griesbach@algo.rwth-aachen.de)}.}
\and
Elfarouk Harb
\thanks{Siebel School of Computer Science, University of Illinois Urbana-Champaign {\tt (eyfmharb@gmail.com)}.}
\and
Felix Hommelsheim\thanks{Institute for Computer Science, University of Cologne, Germany (\texttt{hommelsheim@cs.uni-koeln.de}). Funded by the Deutsche Forschungsgemeinschaft (DFG, German Research Foundation) -- Project Number 573939419.}
\and
Reda Jlibene
\thanks{Moroccan Center for Game Theory, Universit\'{e} Mohammed VI Polytechnique {\tt (reda.jlibene@um6p.ma)}.}
\and
Vasilis Livanos
\thanks{Department of Computer Science, University of Southern California {\tt (vas.livanos@gmail.com)}. Supported by ANID Chile grant FONDECYT 1260036 and by the US Air Force Office of Scientific Research under award number FA9550-24-1-0261.
}
\and
Eilon Solan
\thanks{School of Mathematical Sciences, Tel Aviv University {\tt (eilonsolan@gmail.com)}.\\
Part of this work was done while Correa, Griesbach, Hommelsheim, and Livanos were at the Center for Mathematical Modeling CNRS-IRL 2807 (Supported by ANID grant FB210005), Universidad de Chile.}
}

\date{}

\begin{document}
\maketitle

\begin{abstract}
Given red and blue points in $\mathbb{R}^d$, connect every red point to every blue point and weigh each segment by its length under a fixed norm. What fraction of the total weight can always be captured by a single affine hyperplane separating the segments' endpoints? We study this question for finite point configurations in affine general position and, more generally, for pairs of probability measures.

For dimension one, we show that choosing the best of three thresholds determined by the means of the two distributions captures at least one third of the total red-blue distance. We also give a matching upper bound. Embedding the line construction to higher dimensions shows that the guarantee of $1/3$ cannot be improved for arbitrary measures. Thus, we pivot to the family of measures whose atoms are in affine general position and whose non-atomic part places zero mass on all hyperplanes. Crucially, we establish that the worst-case guarantee is identical for finite configurations in affine general position and several natural classes of probability measures, thereby unifying the discrete and continuous formulations of the problem. 

For this family of measures and for higher dimensions, we extend our guarantee of $1/3$ for all norms. Our proofs combine the ham-sandwich theorem with a novel coupling argument and, in the planar case, an integral representation of norms. We also show that the optimal guarantee improves as the dimension grows, giving a lower bound of $1/2 - \operatorname{O}\left(1/\sqrt{d}\right)$ as well as an asymptotically matching upper bound of $1/2$ for every $d$. For the Euclidean distance, we refine the lower bounds to $8/(8 + 5\pi) \approx 0.337$ and $1/2 - \operatorname{O}\left(1/d\right)$ and, for $d = 2$, the upper bound to $5/11 \approx 0.455$. We also derive dimension-dependent guarantees for finite point configurations, including perfect separation of $n$ points of each color when $d \geq 2n - 1$.
\end{abstract}

\clearpage
\tableofcontents

\clearpage

\section{Introduction}
\label{sec:intro}

\begin{figure}[htbp]
\centering
\begin{tikzpicture}[x=1.15cm,y=0.85cm,font=\small]
  \draw[->,black!55] (-0.35,0) -- (5.45,0);
  \draw[thick,red!65!black,bend left=22] (0,0.10) to (3,0.10);
  \draw[thick,red!65!black,bend left=25] (0,0.10) to (4,0.10);
  \draw[thick,red!65!black,bend left=52] (2,0.10) to (3,0.10);
  \draw[thick,red!65!black,bend left=34] (2,0.10) to (4,0.10);
  \draw[thick,blue!65!black,bend right=27] (1,-0.10) to (5,-0.10);
  \draw[very thick,dashed] (2.5,-0.95) -- (2.5,1.25);
  \node[above] at (2.5,1.25) {$\tau = 5/2$};
  \foreach \x in {0,2,5}{
    \fill[red!80!black] (\x,0) circle (3.5pt);
    \node[below=5pt,fill=white,inner sep=1pt] at (\x,0) {$\x$};
  }
  \foreach \x in {1,3,4}{
    \node[rectangle,fill=blue!80!black,inner sep=0pt,minimum size=5.5pt] at (\x,0) {};
    \node[below=5pt,fill=white,inner sep=1pt] at (\x,0) {$\x$};
  }
\end{tikzpicture}
\caption{$R=\{0,2,5\}$ and $B=\{1,3,4\}$. 
The total red--blue length is $L = 19$.
The five highlighted pairs are separated by $\tau = 5/2$ and have total length $C(\tau) = 14$. This threshold is optimal.}
\label{ch6:fig:threshold-cut}
\end{figure}
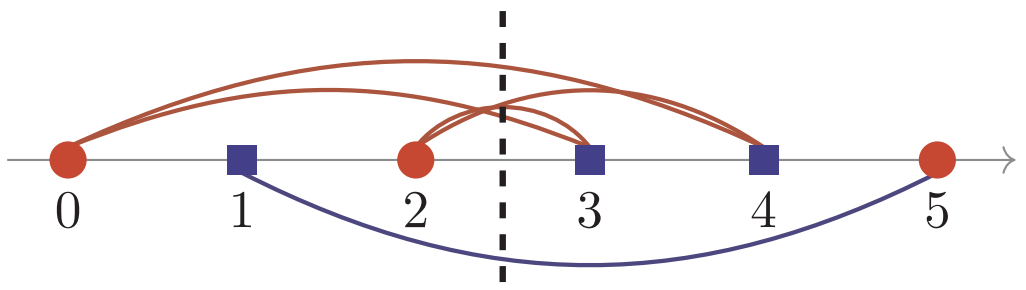

In this paper, we study a simple yet fundamental question in geometry: given red and blue points in $\R^d$, join every red point to every blue point by a segment. What fraction of their total length can a single affine hyperplane cut? We measure length by a fixed norm $\|\cdot\|$ and count a segment when its endpoints lie on opposite sides of the hyperplane. For finite non-empty sets $R,B \subset \R^d$, we define the total length and the cut value of a hyperplane $H$ as
\[
    L = \sum_{r \in R}\sum_{b \in B} \|r - b\|, \qquad C(H) = \sum_{\stack{r \in R,\ b \in B}{H \text{ strictly separates }r, b}} \|r - b\|,
\]
and we assume that $L > 0$. We seek a lower bound on $\sup_H C(H)/L$ that holds independently of the number and locations of the points. At first glance, it might be unclear why we cannot ensure that all red-blue intervals are cut, even on the line; Figure~\ref{ch6:fig:threshold-cut} gives such an example.

The question has a natural probabilistic formulation and generalization. Let $X \sim \mu$ and $Y \sim \nu$ be independent random vectors with finite first moments, representing a red point and a blue point.\footnote{We refer to $\mu$ as the red mass, or red points, and to $\nu$ as the blue mass, or blue points.} Define
\[
    L_{\mu,\nu} = \E[\|X - Y\|], \qquad C_{\mu,\nu}(H) = \E\brk{\|X - Y\| \, \1_{\set{H \text{ separates }X,Y}}}.
\]
We write $L$ and $C(H)$ when the measures are clear from context. Taking $\mu$ and $\nu$ uniform on $R$ and $B$ divides both sums by $|R||B|$ and leaves the ratio unchanged. Allowing general probability measures also gives meaning to the problem for weighted points and continuous distributions. In its most general form, for $d \geq 1$ and a family $\cF$ of (not necessarily atomic) probability measure pairs in $\R^d$, we investigate the ratio
\[
    \alpha(\cF) \coloneqq \inf_{\substack{(\mu,\nu)\in\cF\\L_{\mu,\nu}>0}} \sup_{H\in\Hh^d}\frac{C_{\mu,\nu}(H)}{L_{\mu,\nu}}.
\]
where $\Hh^d$ is the set of hyperplanes in $\R^d$. We call this the \emph{cut ratio} of $\cF$.

Correa, Cristi, D\"{u}tting, Hajiaghayi, Olkowski and Schewior~\cite{correa-trading-prophets} isolate the one-dimensional version of this question in their analysis of the trading prophets problem, aiming for a threshold that separates buying prices from selling prices. They prove a lower bound of $1/4$ by choosing the better of the two medians, an upper bound of $1/3$, and an exact ratio of $1/2$ when the two distributions are identical.

Due to its fundamental nature, this question has several applications in different areas, beyond the trading prophets problem. In bilateral trade with random initial ownership, a posted price captures the gain from a pair of valuations when it lies between them. In machine learning, the same objective measures the separation of differently labeled observations in a decision tree. For more on these applications, see Section~\ref{sec:related-work}.

\subsection{Our contributions}
\label{sec:contritions}

\begin{figure}[htbp]
\centering
\begin{tikzpicture}[x=1.15cm,y=0.85cm,font=\small]
  \def\gammadraw{0.18}
  \pgfmathsetmacro{\bluepos}{2.5*(1-\gammadraw)}
  \pgfmathsetmacro{\redpos}{2.5*(1+\gammadraw)}

  \draw[->,black!55] (-0.35,0) -- (5.45,0);

  \draw[thick,red!65!black,bend left=32]
    (0,0.10) to (5,0.10);
  \draw[thick,blue!65!black,bend left=45]
    (\bluepos,-0.10) to (\redpos,-0.10);

  \draw[very thick,dashed] (2.5,-1.05) -- (2.5,1.45);
  \node[above] at (2.5,1.45) {$\tau = 1$};

  \fill[red!80!black] (\bluepos,0) circle (3.5pt);
  \fill[red!80!black] (5,0) circle (3.5pt);

  \node[rectangle,fill=blue!80!black,inner sep=0pt,minimum size=5.5pt]
    at (0,0) {};
  \node[rectangle,fill=blue!80!black,inner sep=0pt,minimum size=5.5pt]
    at (\redpos,0) {};

  \node[below=5pt,fill=white,inner sep=1pt] at (0,0) {$0$};
  \node[below=5pt,fill=white,inner sep=1pt] at (\bluepos,0) {$1-\gamma$};
  \node[below=5pt,fill=white,inner sep=1pt] at (\redpos,0) {$1+\gamma$};
  \node[below=5pt,fill=white,inner sep=1pt] at (5,0) {$2$};

  \node[above=7pt,text=blue!80!black] at (0,0) {$\frac{1}{k+1}$};
  \node[above=7pt,text=red!80!black] at (\bluepos,0) {$\frac{k}{k+1}$};
  \node[above=7pt,text=blue!80!black] at (\redpos,0) {$\frac{k}{k+1}$};
  \node[above=7pt,text=red!80!black] at (5,0) {$\frac{1}{k+1}$};
\end{tikzpicture}
\caption{The one-dimensional upper-bound construction, with $\gamma=1/(2k)$.
Red circles and blue squares represent atoms, with their probability masses shown above. The optimal threshold $\tau=1$ separates the two highlighted pairs and achieves cut ratio $(k+2)/(3k+1)$, which tends to $1/3$ as $k\to\infty$.
The displayed spacing is illustrative.}
\label{fig:line-construction}
\end{figure}
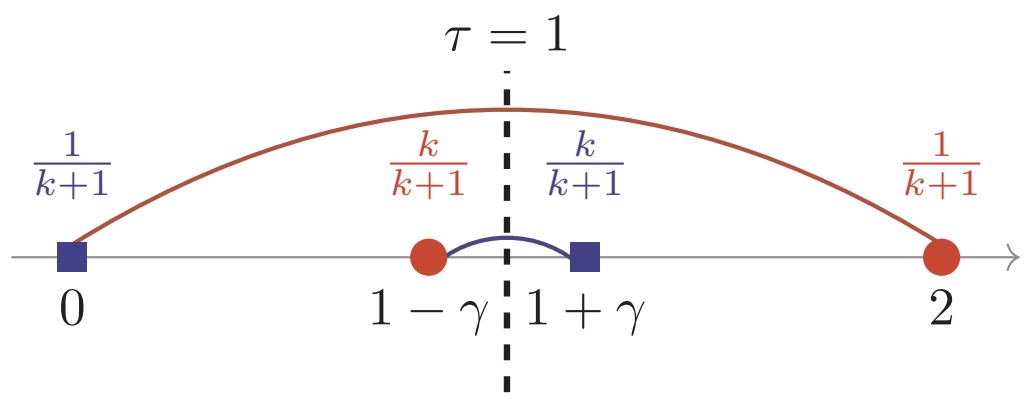

\paragraph{Dimension one.}
We start with the simplest case: two arbitrary, finite-expectation measures on the line. 
Let $\cP^1_f$ be the family of pairs of probability measures in $\R$ with finite first moments. We first characterize the exact optimal cut ratio for the Euclidean distance. 
For the lower bound, we consider three candidate thresholds: $\E_\mu[X]$, $\E_\nu[Y]$, and their midpoint $(\E_\mu[X]+\E_\nu[Y])/2$. When mass at a threshold is assigned to its right side, the best of these three cuts gives a cut ratio of $1/3$.
Approaching the selected threshold from the left then yields the same guarantee for the supremum over strictly separating cuts.
For the matching upper bound, for a fixed $k > 0$, we place a red and a blue cluster of mass $k/(k+1)$ at distance $1/k$ from each other, centered at point $1$, together with a blue outlier far to the left and a red outlier far to the right, both of mass $1/(k+1)$ and at distance $1$ from the center. Each of the three intervals captures asymptotically one-third of $L$, and taking $k \to \infty$ yields the result. The construction can be seen in Figure~\ref{fig:line-construction}.

Since every norm on $\R$ is a positive scalar of the Euclidean distance
\footnote{This follows by the absolute homogeneity axiom for norms: for every $\lambda, v \in \R$, we have $\|\lambda \: v\| = |\lambda| \: \|v\|$. Thus, for any $x \in \R$ we have $\|x\| = \|x \cdot 1\| = |x| \: \|1\|$. Setting $c = \|1\|$ yields that, for every $x \in \R$, $\|x\| = c \: |x|$.},
we obtain the following.

\begin{theorem}\label{thm:dim-1:exact}
For every norm on $\R$, 
\[
    \alpha(\cP^1_f) = \frac{1}{3}.
\]
\end{theorem}

Several natural threshold strategies fail to achieve this optimal guarantee, even on the line.
Choosing the better of the two color medians guarantees only $1/4$~\cite{correa-trading-prophets}, and adding the overall median as a candidate does not improve this worst-case bound.
Other natural quantile-based choices also fall short.
Nor does sampling a threshold with density proportional to the number or total length of red--blue intervals crossing it guarantee $1/3$.
For finite point sets, an optimal threshold can be found efficiently by evaluating the cut value between consecutive points, but proving a sharp guarantee directly for this optimum is challenging.

\paragraph{Measure equivalence.}
Next, we turn to higher dimensions. Embedding the previous line construction in $\R^d$ shows that no guarantee above $1/3$ is possible for arbitrary (finite-expectation) measures in higher dimensions. The issue here is that such measures may be supported entirely on lower-dimensional sets, such as a line, making the ambient dimension irrelevant. This motivates restricting attention to the family $\cP^d$ of pairs of probability measures with finite first moments, whose atoms are in affine general position and whose non-atomic parts assign zero mass to every hyperplane; this is the main family of pairs of probability measures that we study in this paper.

Even under these geometric assumptions, it is not clear a priori whether the optimal cut ratio depends on the particular subfamily from which our measures come. For example, there is no immediate reason why finite point configurations in affine general position\footnote{A set of points in $\R^d$ is in \emph{affine general position} if every subset of at most $d+1$ points is affinely independent. In other words, no three points are collinear for $d \geq 2$, no four points are coplanar for $d \geq 3$, etc.} (denoted by $\cP^d_\N$) and compactly supported, absolutely continuous distributions (denoted by $\cP^d_{bc}$) should have the same optimal ratio.

We show that, for every dimension $d$, the cut ratio is the same for $\cP^d$, $\cP^d_\N, \cP^d_{bc}$ as well as the family $\cP^d_b \subseteq \cP^d$ of probability measure pairs with bounded support. This equivalence lets us study the problem from several different angles and allows us to choose the most convenient family for each proof and transfer the resulting bounds to the others. Its proof can be found in Section~\ref{sec:measure-equivalence}.

\begin{theorem}\label{thm:measure-equivalence}
For every $d \geq 1$ and every norm on $\R^d$,
\[
    \alpha(\cP^d) = \alpha(\cP_b^d) = \alpha(\cP_{bc}^d) = \alpha(\cP_\N^d).
\]
\end{theorem}

\paragraph{Higher dimensions.}
Given this equivalence, we turn to studying $\alpha(\cP^d)$ for $d \geq 2$. Perhaps surprisingly, we show that the guarantee of $1/3$ for the cut ratio continues to hold for higher dimensions, independently of the chosen norm.

\begin{theorem}\label{thm:intro:lower-bound}
For every $d \geq 1$ and every norm on $\R^d$,
\[
    \alpha(\cP^d) \geq \frac{1}{3}.
\]
\end{theorem}

As in our proof for the line, we show this lower bound by finding three hyperplanes whose cut values sum to at least $L$. By Theorem~\ref{thm:measure-equivalence}, we may work with compactly supported, absolutely continuous measures. To choose the three hyperplanes, we use a signed\footnote{A signed measure permits positive and negative mass, and a hyperplane bisects it if the two sides have equal signed mass.} version of the ham-sandwich theorem~\cite{Matousek2003,akopyan-karasev}. For compactly supported, absolutely continuous measures, one hyperplane can bisect up to $d$ measures. Thus, for $d \geq 3$, we choose three hyperplanes and construct a coupling of two red samples $X, X' \sim \mu$ on opposite sides of the first two hyperplanes and the same side of the third. The idea here is to ensure that, whenever none of the three chosen hyperplanes separates the pair $(X, Y)$, at least two of them separate $(X', Y)$, thus compensating, in expectation, for the loss. Since this coupling argument works by counting the number of hyperplanes separating $(X,Y)$ and $(X',Y)$, it is independent of the chosen norm.

The case of $d = 2$ requires a separate argument because a line cannot satisfy three bisection conditions. Here, we choose three lines to bisect two measures each, and prove an inequality for separation probabilities, without distance weights. To pass from probabilities to distances, we use a property of planar norms: each can be approximated by positive sums of absolute values of linear functionals~\cite{pinelis-norm-approx}. Integrating the probability inequality in these projected directions yields $C(H_1) + C(H_2) + C(H_3) \geq L$ for the three chosen lines $H_1, H_2, H_3$.

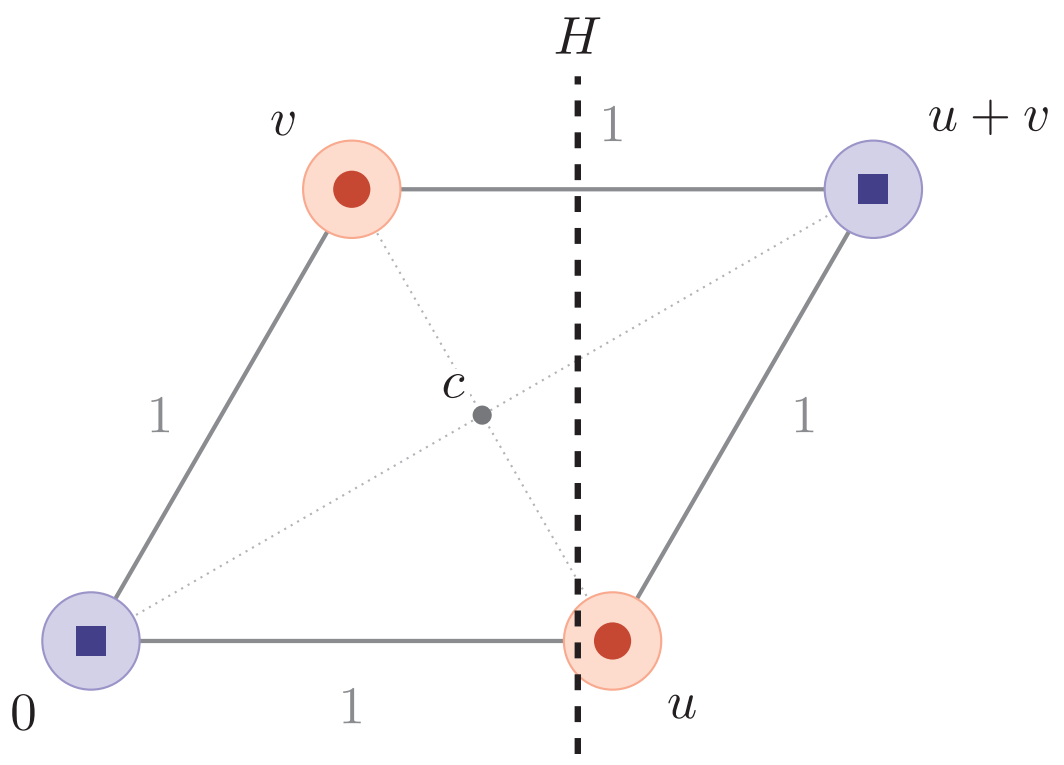
\begin{figure}[tb]
\centering
\begin{tikzpicture}[x=1.15cm,y=1.15cm,font=\small]
  \pgfmathsetmacro{\height}{1.5*sqrt(3)}

  \coordinate (O) at (0,0);
  \coordinate (U) at (3,0);
  \coordinate (V) at (1.5,\height);
  \coordinate (W) at (4.5,\height);
  \coordinate (C) at (2.25,{0.5*\height});

  \draw[thick,black!55]
    (O) -- node[midway,below=5pt] {$1$} (U)
        -- node[midway,right=5pt] {$1$} (W)
        -- node[midway,above=5pt] {$1$} (V)
        -- node[midway,left=5pt] {$1$} cycle;

  \draw[densely dotted,black!35] (O) -- (W);
  \draw[densely dotted,black!35] (U) -- (V);

  \foreach \point in {U,V}{
    \filldraw[fill=red!15,draw=red!40]
      (\point) circle (0.28);
  }
  \foreach \point in {O,W}{
    \filldraw[fill=blue!15,draw=blue!40]
      (\point) circle (0.28);
  }

  \draw[very thick,dashed]
    (2.8,-0.65) -- (2.8,{\height+0.65})
    node[above] {$H$};

  \fill[red!80!black] (U) circle (3.5pt);
  \fill[red!80!black] (V) circle (3.5pt);

  \node[rectangle,fill=blue!80!black,inner sep=0pt,minimum size=5.5pt]
    at (O) {};
  \node[rectangle,fill=blue!80!black,inner sep=0pt,minimum size=5.5pt]
    at (W) {};

  \node[below left=9pt] at (O) {$0$};
  \node[below right=9pt] at (U) {$u$};
  \node[above left=9pt] at (V) {$v$};
  \node[above right=9pt] at (W) {$u+v$};

  \fill[black!65] (C) circle (1.8pt);
  \node[above left=3pt,fill=white,inner sep=1pt] at (C) {$c$};
\end{tikzpicture}
\caption{The higher-dimensional upper-bound construction.
Each colored cloud carries mass $1/2$ of its color, distributed uniformly over a small norm ball.
The red and blue measures share the center of symmetry $c$, and all four red--blue distances between cloud centers equal $1$.
Every hyperplane separates a random red--blue pair with probability at most $1/2$; shrinking the clouds makes their distances uniformly close to $1$, yielding an optimal cut ratio tending to $1/2$.
}
\label{fig:upper-bound-parallelogram}
\end{figure}

To complement our lower bound, we also obtain an upper bound in higher dimensions. The idea behind it is the following: for $ d = 2$, consider the four corners of a parallelogram\footnote{A square would also work, but we present a more generalized counterexample.} with equal norm-length sides, and assign equal red mass to the two opposite corners and equal blue mass to the other two. It is easy to see that, for this configuration, no line can guarantee more than $1/2$ of $L$. 
Extend this construction to $d > 2$ by turning the corners of the parallelogram into clouds\footnote{By clouds, we mean solid norm balls that are absolutely continuous with respect to $d$-dimensional Lebesgue measure.} in $\R^d$ and assigning the mass uniformly over the cloud. See Figure~\ref{fig:upper-bound-parallelogram} for an illustration.

\begin{theorem}\label{thm:intro:upper-bound}
For every $d \geq 2$ and every norm on $\R^d$,
\[
    \alpha(\cP^d) \leq \frac{1}{2}.
\]
\end{theorem}

\paragraph{Improved guarantees in large dimensions.}
For our next set of results, we ask the following question: do the guarantees for the cut ratio of measures in $\cP^d$ improve as we increase the dimension? We answer this question affirmatively, in two different ways. First, we use the signed ham-sandwich theorem to select $d-1$ appropriate hyperplanes that bisect both measures. Control of the overlap of their separation events and an application of the Cauchy-Schwarz inequality yield the following result.

\begin{proposition}\label{prp:ratio-converges-to-half-intro}
For every $d \geq 2$ and every norm $\| \cdot \|$ on $\R^d$,
\[
    \alpha(\cP^d) \geq \frac{1}{2} - \sqrt{\frac{2}{d-1}}.
\]
\end{proposition}

This result separates $\alpha(\cP^d_f)$ from $\alpha(\cP^d)$ for large enough $d$; while the former cannot be greater than $1/3$ for any $d$, the geometric assumptions on $\cP^d$ yield a lower bound that increases with the dimension. Together with Theorem~\ref{thm:intro:upper-bound}, this result shows that $\lim_{d \to \infty} \alpha(\cP^d) = 1/2$, making our guarantee on the cut ratio asymptotically optimal. Next, we turn to atomic measures: fix the number $n$ of points, and let $\cP^d_n$ be the family of probability distribution pairs with exactly $n$ equally weighted points of each color in affine general position. We show that, for this family of measures, the optimal cut ratio goes from $1/3$ to $1$ as $d$ increases and is, in fact, equal to $1$ for $d \geq 2n-1$. 
Let $(x)^+ = \max\set{x,0}$.

\begin{proposition}\label{prp:exact-high-dimension-2}
For $n \geq 3$, $d \in \N$ and every norm on $\R^d$, we have
\[
    \alpha(\cP_n^d) \geq \begin{cases}
        \dfrac{(d-3)^+ + 2\floor{n - \frac{d}{2}}}{(d-3)^+ + 6\floor{n - \frac{d}{2}}}, & \text{if } d < 2n-1, \\
        1, & \text{if } d \geq 2n-1.
    \end{cases}
\]
\end{proposition}

Interestingly, this result is related to the fact that the Vapnik--Chervonenkis (VC) dimension of affine separators in $\R^d$, i.e., the largest size of a point set on which every assignment of two labels can be realized by an affine separator, is $d+1$.\footnote{A point set is said to be \emph{shattered} by a class $\cF$ if, for every assignment of two labels to its points, there exists a function $f \in \cF$ realizing that assignment. The \emph{VC dimension} of a class $\cF$ of separating functions is the largest cardinality of a point set shattered by $\cF$. Since every affinely independent set of at most $d+1$ points is shattered by affine separators and $d \geq 2n-1 \iff 2n \leq d+1$, we get that every red-blue labeling of $2n$ affinely independent points admits a hyperplane that strictly separates the two colors.}

\paragraph{Euclidean distance.}
Finally, we ask whether any of our guarantees can be improved when we specialize to the Euclidean distance. Indeed, neither the lower bound of $1/3$ nor the upper bound of $1/2$ are tight, even for $d = 2$. Proposition~\ref{prp:lower-bound:improved-planar} proves, for $d \geq 2$
\[
    \alpha(\cP^d) \geq \frac{8}{8+5\pi} \approx 0.337,
\]
by combining a hyperplane bisecting both measures with random hyperplanes through their means and the midpoint of those means. A planar construction in Proposition~\ref{prp:upper-bound:improved-planar} gives a complementary upper bound of $5/11 \approx 0.455$, and thus
\[
    \frac{8}{8+5\pi} \leq \alpha(\cP^2) \leq \frac{5}{11}.
\]
For the Euclidean distance, we are also able to improve the rate at which the lower bound on the optimal cut ratio goes to $1/2$ as $d \to \infty$. Proposition~\ref{prp:ratio-converges-to-half} shows that for every $d \geq 2$,
\[
    \alpha(\cP^d) \geq \frac{1}{2} - \bigO{\frac{1}{d}}.
\]

We summarize our results in Table~\ref{tab:results}.

\paragraph{Algorithmic considerations.}
On the line, sorting the points and scanning the linear number of intervals finds an optimal threshold in $\bigO{n \log{n}}$ time.
In higher dimensions, our bisection arguments use the ham-sandwich theorem; the realizable hyperplane partitions of $n$ points can be enumerated in $n^{\bigO{d}}$ time, given access to an oracle that evaluates the chosen norm. However, the running time can be improved for finite configurations in affine general position: we can approximately optimize the cut using a small sample of red--blue pairs by sampling pairs independently with probability proportional to their distance.
A hyperplane $H$ then separates a sampled pair with probability $C(H)/L$, so the fraction of sampled pairs it separates estimates its original cut ratio. Since the separation events have VC dimension $\bigO{d}$, for $0 < \eps,\delta < 1$, a sample of $s = \bigO{(d + \log(1/\delta))/\eps^2}$ pairs estimates these ratios within additive error $\eps/6$, simultaneously for all hyperplanes, with probability at least $1-\delta$~\cite[Theorem~2.12(ii)]{har-peled-sharir-approximations}. Therefore, the hyperplane separating the most sampled pairs loses at most $\eps L/3$ compared with the optimum. Because $\max_H C(H) \geq L/3$, this gives a hyperplane achieving a $(1 - \varepsilon)$-approximation. Crucially, the number of samples is independent of the number of points $n$. Thus, we can find an optimal hyperplane for the sample in $\bigO{s^{\bigO{d}} + n^2}$ time by enumerating the hyperplane partitions of its endpoints, where the $\bigO{n^2}$ term accounts for computing the pair distances $\|r - b\|$, yielding a polynomial randomized approximation scheme for the problem for fixed $d$.

\begin{table}[!t]
\centering
\begingroup
\small
\SetTblrTemplate{head}{empty}
\DefTblrTemplate{lastfoot}{default}{%
  \UseTblrTemplate{caption}{default}%
}
\begin{talltblr}[
  caption = {Our results.},
  label = {tab:results},
]{
  width = \textwidth,
  colspec = {
    X[0.19,c,m]
    X[0.19,c,m]
    X[0.17,c,m]
    X[0.45,c,m]
  },
  colsep = 4pt,
  rowsep = 5pt,
  rowhead = 1,
  row{1} = {font=\bfseries,c},
  hlines = {0.4pt},
  vline{2-4} = {0.4pt},
}
Norm & Measures & Dimension & Guarantee \\

\SetCell[r=5]{c,m} Any norm
& $\cP^1_f$
& $d = 1$
& {$\alpha(\cP^1_f) = \dfrac13$ \\ \vspace{0.5em}
   (Theorem~\ref{thm:dim-1:exact})} \\

& \SetCell[r=2]{c,m} $\cP^d$
& \SetCell[r=2]{c,m} $d\geq2$
& {$\dfrac13 \leq \alpha(\cP^d) \leq \dfrac12$ \\ \vspace{0.5em}
   (Theorems~\ref{thm:intro:lower-bound} \&~\ref{thm:intro:upper-bound})} \\

&
&
& {$\alpha(\cP^d)\geq\dfrac12 - \bigO{\dfrac1{\sqrt{d}}}$ \\ \vspace{0.5em}
   (Proposition~\ref{prp:ratio-converges-to-half-intro})} \\

& \SetCell[r=2]{c,m} {$\cP_n^d$}
& $d < 2n - 1$
& {$\alpha(\cP_n^d)\geq \dfrac{(d-3)^+ + 2 \floor{n - \frac{d}{2}}}{(d-3)^+ + 6 \floor{n - \frac{d}{2}}}$ \\ \vspace{0.5em}
   (Proposition~\ref{prp:exact-high-dimension-2})} \\

&
& $d\geq2n-1$
& {$\alpha(\cP_n^d)=1$ \\ \vspace{0.5em}
   (Proposition~\ref{prp:exact-high-dimension-2})} \\

\SetCell[r=3]{c,m} Euclidean distance
& \SetCell[r=3]{c,m} {$\cP^d$}
& $d=2$
& {$\dfrac8{8+5\pi} \approx 0.337 \leq \alpha(\cP^2) \leq \dfrac5{11} \approx 0.455$ \\ \vspace{0.5em}
   (Propositions~\ref{prp:lower-bound:improved-planar} and~\ref{prp:upper-bound:improved-planar})} \\

&
& \SetCell[r=2]{c,m} $d\geq2$
& {$\dfrac8{8+5\pi} \approx 0.337 \leq \alpha(\cP^d) \leq \dfrac12$ \\ \vspace{0.5em}
   (Proposition~\ref{prp:lower-bound:improved-planar} \& Theorem~\ref{thm:intro:upper-bound})} \\

&
&
& {$\alpha(\cP^d) \geq \dfrac12 - \bigO{\dfrac1d}$ \\ \vspace{0.5em}
   (Proposition~\ref{prp:ratio-converges-to-half})} \\
\end{talltblr}
\endgroup
\end{table}

\subsection{Applications and related work}
\label{sec:related-work}

We next discuss applications of our separation guarantees and their connections to existing work.

\paragraph{Trading prophets.}
In Trading Prophets~\cite{correa-trading-prophets}, a trader starts without an item, observes prices sequentially, and may hold at most one item at a time. The objective is to maximize expected profit, with performance measured relative to a prophet who knows the entire price sequence in advance. 
Prices are non-negative with finite means, and there is no budget constraint. A threshold policy buys below $\tau$ when empty, sells at or above $\tau$ when holding the item, and always liquidates any remaining item at the end. For two independent prices $X,Y$ in uniformly random order, the expected prophet profit is $(1/2) \: \E[|X - Y|]$. If neither probability measure has an atom at $\tau$, the expected threshold profit is $C(\tau)/2$. Our theorem therefore gives a $1/3$ guarantee, matching the two-period upper bound of \cite{correa-trading-prophets}. Assigning values at the threshold consistently to one side extends the guarantee to atomic prices.

In fact, we can use our improved two-period guarantee to also improve the uniform guarantee for Trading Prophets with zero initial inventory and unit capacity. Let $X_1, \dots, X_n$ be independent non-negative prices with finite means, revealed in uniformly random order, $W = \sum_{i < j} \E[|X_i - X_j|]$, and $Q(\tau)$ be the same sum, restricted to pairs that lie on opposite sides of $\tau$, assigning values equal to $\tau$ consistently to one side. The profit identities of \cite{correa-trading-prophets} give $\E[\mathrm{OPT}] = W / n$ and $\E[\mathrm{ALG}_\tau] = Q(\tau) / n$ for the policy that buys below $\tau$, sells at or above $\tau$, and liquidates any remaining item at the final date. A uniformly random balanced partition of the $X_i$'s separates each pair $X_i, X_j$ with probability at least $1/2$. Therefore, by linearity of expectation, there exists a partition into groups $A$ and $B$, whose sizes differ by at most one, for which the sum that keeps only $X_i$'s from one group and $X_j$'s from the other is at least $W/2$. Next, draw $X_i$ and $X_j$ uniformly at random from $A$ and $B$, respectively, and independently of each other. Applying our $1/3$ result to $X_i, X_j$ gives a common threshold $\tau$ that captures at least $1/3$ of the cross-group sum $W/2$. Since $Q(\tau)$ includes the non-negative contributions from within-group pairs, we obtain $Q(\tau) \geq W/6$ and thus a $1/6$-approximation, improving the uniform $1/16$ bound of \cite{correa-trading-prophets}. This uniform guarantee complements the horizon-dependent guarantee of Rajput, Chiplunkar, and Vaish~\cite{rajput-trading-prophets-25} in the same model and is orthogonal to the $1/3$ guarantee of Azar, Buchbinder, Levin, and Vardi~\cite{azar-trading-prophets} which assumes that both the trader and the prophet start with one initial item.

\paragraph{Gains from trade.}
In bilateral trade, a seller and a buyer trade a single indivisible item \cite{ms-embt-83}. Given distributions for the seller's and buyer's valuations, the goal is to design a truthful mechanism that maximizes gains from trade, on expectation. An important special mechanism is the fixed price mechanism in which we offer a single price and trade occurs if this price is below the buyer's value and above the seller's \cite{KangPV22,BlumrosenD21}. While in general no constant factor approximation can be obtained, in the symmetric case (i.e, when the valuations are i.i.d.) the optimal guarantee is $1/2$ \cite{mcafee2008gains}.

Our result applies to bilateral trade with independent valuations when the initial owner is chosen uniformly at random.
It also applies to partnership dissolution~\cite{cramton-gibbons-klemperer-87,FS12}.
Suppose two partners each own half of an asset and have independent private valuations $X$ and $Y$ for the entire asset.
At a posted price $\tau$ for the entire asset, the lower-valuing partner sells their half for $\tau/2$ whenever the valuations lie strictly on opposite sides of $\tau$.
The expected gain from trade is therefore $C(\tau)/2$, whereas the first-best expected gain is $\mathbb{E}[|X-Y|]/2$.
Our one-dimensional result thus gives a sharp worst-case guarantee of $1/3$ for a single posted price in this equal-share setting.

\paragraph{Decision trees and recursive separation.}

Bodine and Hochbaum~\cite{bodine-hochbaum-maxcut-tree-22} study the construction of decision trees for supervised classification, with the aim of predicting the labels of unseen observations. Consider observations $x_1, \dots, x_n \in \R^d$ with binary labels $y_1, \dots, y_n \in \set{0, 1}$, and a decision tree defined as follows. Each internal node $u$ in a decision tree has a pair $(v_u, t_u) \in \R^d \times \R$ where $v_u$ specifies the projection direction used at $u$ and $t_u$ specifies a threshold, and each leaf is assigned a label in $\set{0, 1}$. The tree routes an observation $x$ by testing $v_u^{\mathsf T} x \leq t_u$ at each visited internal node $u$, sending it to the left child if the test holds and to the right child otherwise, and outputs the label at the resulting leaf as a prediction. At each node $u$, let $I = I_u \subseteq [n]$ denote the indices of the $x_i$'s that reach $u$. Given a finite set $\cV_u \subset \R^d$ of non-zero candidate projection directions, the Max-Cut rule selects $v_u \in \cV_u$ and $t_u$ by maximizing
\[
    Q_I(v, t) = \sum_{\substack{i \in I_-(v, t), \: j \in I_+(v,t) \\ y_i \neq y_j}} \|v^{\mathsf T} (x_i - x_j) \|,
\]
where $I_-(v, t) = \set{i \in I \midd v^{\mathsf T} x_i \leq t}$ and $I_+(v, t) = I \setminus I_-(v, t)$. Each pair $x_i, x_j$ with $y_i \neq y_j$ contributes its projected distance $\| v^{\mathsf T} x_i - v^{\mathsf T} x_j \|$ to the sum exactly when $x_i, x_j$ are sent to different children. For each fixed projection direction $v$, this is precisely our one-dimensional objective, and thus our result gives
\[
    \max_t Q_I(v, t) \geq \frac{1}{3} L_I(v), \qquad \text{where} \qquad L_I(v) = \sum_{\substack{i,j \in I, \: i < j \\ y_i \neq y_j}} \|v^{\mathsf T} (x_i - x_j)\|,
\]
giving a worst-case guarantee on the Max-Cut splitting criterion studied by \cite{bodine-hochbaum-maxcut-tree-22}.

Our higher-dimensional results give recursive guarantees for a variant that uses the fixed distances $\|x_i - x_j\|$ as pair weights and permits arbitrary affine hyperplanes. Define the residual weight in a node by $\Phi(I) = \sum_{i, j \in I, \: i < j, \: y_i \neq y_j} \|x_i - x_j\|$. For distinct observations in affine general position, our theorem provides a split separating at least $1/3$ of $\Phi(I)$. The two children therefore satisfy $\Phi(I_-) + \Phi(I_+) \leq (2/3) \Phi(I)$. Splitting every node containing observations of both labels (i.e., with $\Phi(I) > 0$) for $k$ levels, and letting $\cL_k$ denote the resulting leaf index sets, has each observation undergoing at most $k$ affine tests and yields
\[
    \sum_{I \in \cL_k} \Phi(I) \leq \prn{\frac{2}{3}}^k \Phi(\set{1, \dots, n}).
\]

\paragraph{Organization.}
Section~\ref{sec:preliminaries} contains the formal setup of the problem as well as useful definitions and lemmas. In Section~\ref{sec:lower-bounds} we show our lower bound results and in Section~\ref{sec:upper-bounds} we show our upper bound results. The improved bounds for the Euclidean distance are proved in Section~\ref{sec:improved-bounds}. Section~\ref{sec:measure-equivalence} establishes the equivalence of the finite and continuous formulations.
\section{Preliminaries}
\label{sec:preliminaries}

We formulate the problem for pairs of probability measures, treating finite point configurations and continuous distributions within the same framework. Unless stated otherwise, $\|\cdot\|$ denotes an arbitrary fixed norm on $\R^d$.

\subsection{Measures and hyperplane cuts}\label{subsec:measures-cuts}

Fix $d\geq 1$.
For a probability measure $\mu$ on $\R^d$, let $\atom(\mu)$ denote its set of atoms, $\delta_x$ be the Dirac probability measure at $x$, and
\[
    \mu_{\na}\coloneqq\mu-\sum_{x\in\atom(\mu)}\mu(\{x\})\delta_x
\]
for its non-atomic part.
Let $\cP_f^d$ be the family of pairs $(\mu,\nu)$ of probability measures on $\R^d$ with finite first moments:
\[
    \int\|x\|\,\dif\mu(x)<\infty,
    \qquad
    \int\|y\|\,\dif\nu(y)<\infty.
\]
For a finite dimension, this condition is independent of the choice of norm.

A set of points in $\R^d$ is in \emph{affine general position} if every subset of at most $d+1$ points is affinely independent.
Equivalently, for every $k=0,\dots,d-1$, no $k$-dimensional affine subspace contains $k+2$ distinct points of the set.
In dimension one, this imposes no restriction on a set of distinct points; in the plane, it excludes three collinear points.

Let $\cP^d\subseteq\cP_f^d$ consist of the pairs $(\mu,\nu)$ satisfying the following geometric conditions:
\begin{enumerate}[label=(\roman*)]
    \item The atom sets are disjoint, and $\atom(\mu)\cup\atom(\nu)$ is in affine general position.
    \item For every affine hyperplane $H\subset\R^d$, we have $\mu_{\na}(H)=\nu_{\na}(H)=0$.
\end{enumerate}
These conditions rule out geometric degeneracies while allowing both atomic and continuous measures.
In particular, every pair of absolutely continuous probability measures with finite first moments belongs to $\cP^d$.

Let $\Hh^d$ be the set of affine hyperplanes in $\R^d$.
For an oriented hyperplane
\[
    H=\set{x\in\R^d\midd u\cdot x=c},
    \qquad u\neq 0,
\]
define its two open half-spaces by
\[
    H^+\coloneqq\set{x\in\R^d\midd u\cdot x>c},
    \qquad \text{and} \qquad
    H^-\coloneqq\set{x\in\R^d\midd u\cdot x<c}.
\]
The set of ordered pairs strictly separated by $H$ is
\[
    \text{Sep}(H)\coloneqq(H^+\times H^-)\cup(H^-\times H^+).
\]
This set is unchanged when the orientation of $H$ is reversed. For $(\mu,\nu)\in\cP_f^d$, define the expected total red--blue distance and the cut value of $H$ by
\begin{align*}
    L_{\mu,\nu}
    &\coloneqq\iint\|x-y\|\,\dif\mu(x)\,\dif\nu(y), \qquad \text{and} \\
    C_{\mu,\nu}(H)
    &\coloneqq\iint_{\text{Sep}(H)}\|x-y\|\,\dif\mu(x)\,\dif\nu(y),
\end{align*}
respectively. Equivalently, if $X\sim\mu$ and $Y\sim\nu$ are independent, then
\[
    L_{\mu,\nu}=\E[\|X-Y\|],
    \qquad \text{and} \qquad
    C_{\mu,\nu}(H)=\E\brk{\|X-Y\|\cdot\1_{\set{(X,Y) \in \text{Sep}(H)}}}.
\]
We refer to $\mu$ as the red measure and $\nu$ as the blue measure, and write $L$ and $C(H)$ when the measures are clear from context.
Finite first moments ensure that $0\leq C(H)\leq L<\infty$.
Moreover, $L_{\mu,\nu}=0$ if and only if $\mu=\nu=\delta_x$ for some $x\in\R^d$.
Such a pair does not belong to $\cP^d$, so every pair in $\cP^d$ has $L_{\mu,\nu}>0$.

For finite non-empty red and blue point sets $R,B$, the uniform measures
\[
    \mu=\frac{1}{|R|}\sum_{r\in R}\delta_r,
    \qquad
    \nu=\frac{1}{|B|}\sum_{b\in B}\delta_b
\]
divide both the total distance and each cut value of the point configuration by $|R||B|$.
The cut ratio is therefore unchanged.
When working directly with finite point configurations, we use the unnormalized sums over red--blue pairs.
In dimension one, a hyperplane is a threshold $\tau$, and we write $C(\tau)$ for $C(\{\tau\})$.

For the approximation arguments in Section~\ref{sec:measure-equivalence}, we also use Euclidean balls and strips.
Write $\Leb$ for $d$-dimensional Lebesgue measure and $\dist(x,H)$ for the Euclidean distance from $x$ to $H$.
For $r>0$, let $B(x,r)$ be the closed Euclidean ball of radius $r$ centered at $x$.
For $\eta>0$, define the strip
\[
    B(H,\eta)\coloneqq\set{x\in\R^d\midd\dist(x,H)\leq\eta}.
\]
The strip $B(H,\eta)$ contains the points close enough to $H$ that a displacement of size at most $\eta$ could change their side.

\subsection{The optimal ratio and families of measures}\label{subsec:ratio-families}

The ratio $C(H)/L$ is the fraction of the total red--blue distance captured by a hyperplane $H$. We optimize over hyperplanes for each pair of measures and then take the worst case over a prescribed family, omitting pairs with $L_{\mu,\nu} = 0$.

\begin{definition}\label{def:ratio-alpha}
For a family $\cF\subseteq\cP_f^d$ containing at least one pair with $L_{\mu,\nu}>0$, define its \emph{cut ratio} by
\[
    \alpha(\cF)\coloneqq
    \inf_{\substack{(\mu,\nu)\in\cF\\L_{\mu,\nu}>0}}
    \sup_{H\in\Hh^d}\frac{C_{\mu,\nu}(H)}{L_{\mu,\nu}}.
\]
\end{definition}

We use the following subclasses of $\cP^d$.
\begin{definition}
Let $\cP_b^d$ consist of the pairs in $\cP^d$ for which both measures have bounded support, and let $\cP_{bc}^d\subseteq\cP_b^d$ consist of those pairs for which both measures are absolutely continuous with respect to Lebesgue measure.
For $n\geq 1$, let $\cP_n^d$ consist of the pairs in $\cP^d$ for which each measure is uniform on a set of $n$ points, and let $\cP_\N^d = \bigcup_{n = 1}^{\infty} \cP_n^d$.
\end{definition}
The family $\cP_\N^d$ allows the common number of atoms to vary; all atom sets inherit the disjointness and general-position requirements of $\cP^d$.
Theorem~\ref{thm:measure-equivalence}, proved in Section~\ref{sec:measure-equivalence}, shows that
\[
    \alpha(\cP^d)=\alpha(\cP_b^d)=\alpha(\cP_{bc}^d)=\alpha(\cP_\N^d).
\]
Thus, we may work with whichever of these families is most convenient for a proof without changing the optimal guarantee.

\subsection{Signed measures and bisection}\label{subsec:signed-bisection}

For an oriented hyperplane $H$, let $\sigma_H:\R^d\to\{-1,0,+1\}$ be $+1$ on $H^+$, $-1$ on $H^-$, and $0$ on $H$.
For a finite signed measure $\lambda$ and a $\lambda$-integrable function $f$, write $\E_\lambda[f]\coloneqq\int f\,\dif\lambda$.
We say that $H$ \emph{bisects} $\lambda$ if
\[
    \lambda(H^+)=\lambda(H^-),
    \qquad\text{or, equivalently,}\qquad
    \E_\lambda[\sigma_H]=0.
\]
If $\lambda(H)=0$, each half-space then has signed mass $\lambda(\R^d)/2$.
For a probability measure assigning no mass to $H$, this is the usual requirement that each side has probability $1/2$.
For a bounded Borel function $S:\R^d\to\R$, let $S\lambda$ be the signed measure defined by $\dif(S\lambda)=S\,\dif\lambda$.
In particular, $\sigma_H\lambda$ retains the signed mass on $H^+$, reverses its sign on $H^-$, and assigns zero mass to $H$.

Bisection of signed measures allows us to coordinate different hyperplanes.
For example, a hyperplane $K$ bisects $\sigma_H\mu$ exactly when $\E_\mu[\sigma_H\sigma_K]=0$.
If $\mu(H\cup K)=0$, this balances the total $\mu$-mass of the regions where the two signs agree against that of the regions where they differ.
The following signed version of the ham-sandwich theorem lets us impose such conditions together with ordinary bisection requirements.

\begin{lemma}[Signed ham-sandwich theorem in $\R^d$; {\cite[Chapter 3]{Matousek2003}, \cite[Theorem 2.2]{akopyan-karasev}}]\label{lem:unified-signed-ham}
Let $\lambda_1,\dots,\lambda_n$ be compactly supported, absolutely continuous finite signed measures on $\R^d$, where $n\leq d$.
If $\lambda_1$ is a non-zero positive measure, then some affine hyperplane simultaneously bisects $\lambda_1,\dots,\lambda_n$.
\end{lemma}

Once the hyperplanes have been chosen, we express their separation probabilities in terms of sign expectations.
A hyperplane $K$ strictly separates $x,y\in\R^d$ exactly when $\sigma_K(x)\sigma_K(y)=-1$.
Away from $K$, this product is $+1$ when the points lie on the same side and $-1$ when they lie on opposite sides.

\begin{lemma}\label{lem:aux-separated-by-one-hyperplane}
Let $X\sim\mu$ and $Y\sim\nu$ be independent, $K \in \Hh^d$ such that $\mu(K) = \nu(K) = 0$, $p_K \coloneqq \E_\mu[\sigma_K]$, and $q_K \coloneqq \E_\nu[\sigma_K]$. Then,
\[
    \Pr[K\text{ strictly separates }X,Y]=\frac{1-p_Kq_K}{2}.
\]
\end{lemma}
\begin{proof}
Since $X$ and $Y$ avoid $K$ almost surely,
\[
    \1_{\set{K\text{ strictly separates }X,Y}}
    =\frac{1-\sigma_K(X)\sigma_K(Y)}{2}
\]
almost surely.
Taking expectations and using independence gives the result.
\end{proof}

The quantities $p_K$ and $q_K$ measure the imbalance of the red and blue mass across $K$.
In particular, if $K$ bisects either measure, the separation probability is $1/2$.
For two hyperplanes, the joint separation probability also depends on whether their signs agree within each measure, as captured by the product $\sigma_K\sigma_{K'}$.

\begin{lemma}\label{lem:aux-separated-by-two-hyperplanes}
Let $X\sim\mu$ and $Y\sim\nu$ be independent, $K,K'\in\Hh^d$ such that $\mu(K\cup K')=\nu(K\cup K')=0$, $p_K\coloneqq\E_\mu[\sigma_K]$, $p_{K'}\coloneqq\E_\mu[\sigma_{K'}]$, $q_K\coloneqq\E_\nu[\sigma_K]$, $q_{K'}\coloneqq\E_\nu[\sigma_{K'}]$, $p_{KK'}\coloneqq\E_\mu[\sigma_K\sigma_{K'}]$, and $q_{KK'}\coloneqq\E_\nu[\sigma_K\sigma_{K'}]$. Then,
\[
    \Pr[\text{both }K\text{ and }K'\text{ strictly separate }X,Y]
    =\frac{1-p_Kq_K-p_{K'}q_{K'}+p_{KK'}q_{KK'}}{4}.
\]
\end{lemma}
\begin{proof}
The indicator of joint separation is the product of the two separation indicators, so, almost surely,
\[
    \1_{\set{\text{both }K\text{ and }K'\text{ strictly separate }X,Y}}
    =\frac{1-\sigma_K(X)\sigma_K(Y)}{2}
    \cdot\frac{1-\sigma_{K'}(X)\sigma_{K'}(Y)}{2}.
\]
Expanding the product, taking expectations, and using the independence of $X$ and $Y$ gives
\begin{align*}
    &4\Pr[\text{both }K\text{ and }K'\text{ strictly separate }X,Y]\\
    &\qquad =1-\E[\sigma_K(X)\sigma_K(Y)]-\E[\sigma_{K'}(X)\sigma_{K'}(Y)] +\E[\sigma_K(X)\sigma_{K'}(X)\sigma_K(Y)\sigma_{K'}(Y)] \\
    &\qquad=1-p_Kq_K-p_{K'}q_{K'}+p_{KK'}q_{KK'}.
    \qedhere
\end{align*}
\end{proof}

\subsection{Couplings and Wasserstein distance}\label{subsec:measure-equivalence-preliminaries}

In Section~\ref{sec:measure-equivalence}, we replace probability measures by measures with bounded support, by absolutely continuous measures, or by measures with finite support.
The Wasserstein distance measures the cost of moving mass from the original measure to its replacement.

For probability measures $\mu,\mu'$ on $\R^d$, let $\Pi(\mu,\mu')$ denote the set of probability measures on $\R^d\times\R^d$ with marginals $\mu$ and $\mu'$.
Each such measure is a \emph{coupling}, specifying how to match the mass of $\mu$ with that of $\mu'$.
For measures with finite first moments, their $1$-Wasserstein distance is
\[
    W_1(\mu,\mu')\coloneqq\inf_{\pi\in\Pi(\mu,\mu')}\iint\|x-x'\|\,\dif\pi(x,x').
\]
Thus, $W_1(\mu,\mu')$ is the smallest average distance that mass must move to transform $\mu$ into $\mu'$.
The following bound relates this distance to the change in the expected total red--blue distance.

\begin{lemma}\label{lem:L-cont}
For every $d \geq 1$, every norm $\|\cdot\|$ on $\R^d$ and probability measures $\mu,\nu,\mu',\nu'$ on $\R^d$ with finite first moments,
\[
    |L_{\mu,\nu}-L_{\mu',\nu'}| \leq W_1(\mu,\mu')+W_1(\nu,\nu').
\]
\end{lemma}
\begin{proof}
Choose a coupling $(X,X')$ of $\mu,\mu'$ and, independently, a coupling $(Y,Y')$ of $\nu,\nu'$.
Then $X,Y$ are independent, as are $X',Y'$, and the triangle inequality gives
\[
    \bigl|\|X-Y\|-\|X'-Y'\|\bigr|\leq\|X-X'\|+\|Y-Y'\|.
\]
Taking expectations and then the infimum over the two couplings proves the claim.
\end{proof}
\section{Lower Bounds}\label{sec:lower-bounds}

In this section, we present our lower bounds on $\alpha(\cP^d_f)$ and $\alpha(\cP^d)$. We first discuss a natural heuristic: fix a small set of directions, project the points onto each direction, solve the one-dimensional threshold problem in each projection, and lift the best projected threshold back to a hyperplane. As we show in Appendix~\ref{app:projections}, using only the coordinate directions can lose a factor $\bigO{d}$ for any norm on $\R^d$.

We then prove that $\alpha(\cP^d) \geq 1/3$ for every $d \geq 1$ and every norm on $\R^d$ (Theorem~\ref{thm:intro:lower-bound}).
For $d = 1$, the proof is particularly simple: the guarantee is achieved by the best of three thresholds, namely, the mean of the red measure, the mean of the blue measure, and the midpoint of these two means.

For $d \geq 3$, the proof uses a simple and elegant coupling argument.
We apply the signed ham-sandwich theorem to choose three hyperplanes, depending on $\mu$ and $\nu$, and show that $X \sim \mu$ can be coupled with another $X' \sim \mu$ so that, whenever $(X,Y)$ is not separated by any of the three hyperplanes, $(X',Y)$ is separated by at least two of them.
This compensates for the loss and ensures that at least one of the three hyperplanes achieves the $1/3$ guarantee.
The condition $d \geq 3$ is needed to ensure that one of the hyperplanes simultaneously bisects three carefully chosen measures.

The case $d = 2$ requires an additional idea.
The proof resembles the coupling argument for $d \geq 3$, but the signed ham-sandwich theorem only guarantees simultaneous bisection of two measures.
To overcome this obstacle, we first establish an inequality involving separation probabilities.
We then use a special property of the plane: every norm on $\R^2$ admits an integral representation by absolute values of linear functionals.
This representation allows us to express distances as integrals of separation indicators and integrate the probability inequality to obtain the desired guarantee.

Finally, we investigate the behavior of $\alpha(\cP^d)$ as the dimension grows, considering both absolutely continuous and atomic measures.
Specifically, we show that $\alpha(\cP^d_{bc})$ tends to $1/2$ as $d \to \infty$, whereas $\alpha(\cP^d_n)$ tends to $1$ and equals $1$ for $d \geq 2n-1$.

\subsection{The \texorpdfstring{$d = 1$}{d = 1} case}
\label{sec:dimension-one-simple}

In this section, we prove Theorem~\ref{thm:intro:lower-bound} for $d = 1$ and measures in $\cP^1_f$.

\begin{theorem}\label{thm:lower-bound:dimension-one}
For every norm in $\R$, we have
\[
    \alpha(\cP^1_f) \geq \frac{1}{3}.
\]
\end{theorem}
\begin{proof}
Every norm on $\R$ is a positive multiple of the absolute value and thus, to prove the theorem, it suffices to show it for the Euclidean distance. Let $(\mu, \nu) \in \cP^1_f$ be a pair of probability measures and $X \sim \mu$ and $Y \sim \nu$ be independent. Recall that $L = \E[|X - Y|]$, and in dimension one, the cut value specializes to
 \begin{align}
    C(\tau)= \E[|X-Y|\cdot\1_{\{X<\tau<Y\} \cup \{Y<\tau<X\}}].
    \label{eq:dim1:C}
\end{align}

When either measure has an atom at $\tau$, assigning that atom to a side can give a larger value than the strictly separated value $C(\tau)$.
We therefore distinguish the cut value with the tie convention of assigning such atoms to the right side by defining
\[
    C'(\tau)\coloneqq\E\brk{|X-Y|\cdot\1_{\{X<\tau\leq Y\}\cup\{Y<\tau\leq X\}}}.
\]

First, we will show that under this definition of the cut, the mean of the red mass, the mean of the blue mass, or the average of those two means gives the desired guarantee, i.e.,
\[
    \max\left\{C'(\E[X]),\ C'\left(\frac{\E[X]+\E[Y]}{2}\right),\ C'(\E[Y])\right\}\geq\frac{L}{3}.
\]
Afterward, we will recover the same guarantee for strict cuts by approaching the selected threshold from the left.

In the analysis, we distinguish whether the two means coincide.
In both cases, we write $F(\tau) = \Pr[X < \tau]$ and $G(\tau)=\Pr[Y<\tau]$. 
Thus, $1-F(\tau)$ and $1-G(\tau)$ include any mass at $\tau$.

We first consider the case $\E[X]\neq\E[Y]$.
By exchanging the colors if necessary, we may assume that $\E[X]<\E[Y]$.
An increasing affine transformation then allows us to normalize the instance so that
\begin{align*}
    \E[X]=-1 \quad\text{ and }\quad \E[Y]=1.
\end{align*}
Exchanging the colors leaves the cut values unchanged, while the increasing affine transformation preserves the tie convention and scales all distances by the same positive factor.
Hence, it suffices to prove that
\begin{align*}
     \max\left\{C'(-1),C'(0),C'(1)\right\}\geq \frac{1}{3}L.
 \end{align*}
 In fact, we prove the stronger inequality
 \begin{align}
    \label{eq:dim-1}
    C'(-1)+C'(0)+C'(1)-(\E[|X|]+\E[|Y|])\geq 0.
\end{align}
Together with the triangle inequality $\E[|X|]+\E[|Y|]\geq L$, this gives the desired bound.

 To this end, we define 
 \begin{align*}
     L_X(\tau) &\coloneqq \E[(\tau-X)^+],
     & R_X(\tau) &\coloneqq \E[(X-\tau)^+] \\
     L_Y(\tau) &\coloneqq \E[(\tau-Y)^+],
     & R_Y(\tau) &\coloneqq \E[(Y-\tau)^+].
 \end{align*}
 Since $z=(z)^+-(-z)^+$ for any $z\in\R$, we have
 \begin{align}
 \begin{split}
     R_X(\tau)-L_X(\tau)&=\E[X]-\tau=-1-\tau\\
     R_Y(\tau)-L_Y(\tau)&=\E[Y]-\tau=1-\tau.        
 \end{split}
 \label{eq:dim1:R-L}
 \end{align}
Together with $|z|=(z)^++(-z)^+$ for any $z\in\R$, this gives
\begin{align}
    \E[|X|]+\E[|Y|]
    &=L_X(0)+R_X(0)+L_Y(0)+R_Y(0)
    =2L_X(0)+2L_Y(0).
    \label{eq:dim1:EX+EY}
\end{align}

Independence of $X$ and $Y$ together with \eqref{eq:dim1:R-L} and the convention that mass on a threshold belongs to the right side yields
 \begin{align*}
     C'(\tau)&=(1-G(\tau))L_X(\tau)+F(\tau)R_Y(\tau)+(1-F(\tau))L_Y(\tau)+G(\tau)R_X(\tau)\\
     &=L_X(\tau)+L_Y(\tau)+F(\tau)(1-\tau)+G(\tau)(-1-\tau).
 \end{align*}
 Evaluating this at $\tau\in\{-1,0,1\}$ and using \eqref{eq:dim1:EX+EY}, we obtain
 \begin{align*}
    C'(-1)+C'(0)+C'(1)-(\E[|X|]+\E[|Y|]) &=
    \big(L_X(-1)-L_X(0)+L_X(1)+F(0)+2F(-1)\big) \\
    &\qquad + \big(L_Y(-1)-L_Y(0)+L_Y(1)-G(0)-2G(1)\big).
 \end{align*}
We consider both terms separately.
For the first one, define
\[
    \phi(x)\coloneqq (-1-x)^+-(-x)^++(1-x)^++2\cdot \1_{x<-1}+\1_{x<0}
\]
such that
\[
    L_X(-1)-L_X(0)+L_X(1)+F(0)+2F(-1)=\E[\phi(X)].
\]
Further, observe that 
\begin{align*}
    \phi(x)=
    \begin{cases}
        3-x&\text{if }x<-1\\
        2&\text{if }-1\leq x<0\\
        1-x&\text{if }0\leq x<1\\
        0&\text{if }1\leq x
    \end{cases}
\end{align*}
and in particular, $\phi(x)\geq 1-x$ for all $x$.
Similarly, for the second one, define
\[
    \psi(y)\coloneqq (-1-y)^+-(-y)^++(1-y)^+-\1_{y<0}-2\cdot \1_{y<1}
\]
such that 
\[
    L_Y(-1)-L_Y(0)+L_Y(1)-G(0)-2G(1)=\E[\psi(Y)].
\]
Moreover, 
\begin{align*}
    \psi(y)=
    \begin{cases}
        -y-3&\text{if }y<-1\\
        -2&\text{if }-1\leq y<0\\
        -y-1&\text{if }0\leq y<1\\
        0&\text{if }1\leq y
    \end{cases}
\end{align*}
and in particular, $\psi(y)\geq -2$ for all $y$.
Combining these bounds and using $\E[X]=-1$ yields
\begin{align*}
    C'(-1)+C'(0)+C'(1)-(\E[|X|]+\E[|Y|])
    =\E[\phi(X)]+\E[\psi(Y)]
    \geq 1-\E[X]-2=0
\end{align*}
which proves \eqref{eq:dim-1}.

It remains to consider the case $\E[X]=\E[Y]$.
Here, all three thresholds coincide, and we denote their common value by 
\begin{align*}
    \tau^*\coloneqq \E[X]=\E[Y].
\end{align*}
We show the even stronger guarantee $C'(\tau^*)\geq \frac{1}{2}L.$

Since $\E[X-\tau^*]=0$, the positive and negative parts have equal expectations, i.e.,
\begin{align*}
    \E[(X-\tau^*)^+]=\E[(\tau^*-X)^+]=\frac{1}{2}\E[|X-\tau^*|].
\end{align*}
Analogously,
\begin{align*}
    \E[(Y-\tau^*)^+]=\E[(\tau^*-Y)^+]=\frac{1}{2}\E[|Y-\tau^*|].
\end{align*}
The independence of $X$ and $Y$ now yields
\begin{align*}
    C'(\tau^*)
    &=\E[|X-Y|\cdot\1_{\{X<\tau^*\leq Y\} \cup \{Y<\tau^*\leq X\}}]\\
    &=\E[(\tau^*-X)^+](1-G(\tau^*))
    +\E[(X-\tau^*)^+]G(\tau^*) \\
    &\qquad +\E[(\tau^*-Y)^+](1-F(\tau^*))
    +\E[(Y-\tau^*)^+]F(\tau^*)\\
    &=\frac{1}{2}\E[|X-\tau^*|]+\frac{1}{2}\E[|Y-\tau^*|]\\
    &\geq \frac{1}{2}L
\end{align*}
where the last inequality follows from the triangle inequality.

Finally, we pass from $C'$ to the strict cut value $C$.
For every fixed $\tau$, the indicators defining $C(\tau-h)$ converge point-wise to the indicator defining $C'(\tau)$ as $h\downarrow 0$.
Since the integrands are bounded by the integrable random variable $|X-Y|$, dominated convergence gives
\[
    C'(\tau)=\lim_{h\downarrow 0}C(\tau-h).
\]
Consequently,
\[
    \sup_{s\in\R}C(s)
    \geq\max\left\{C'(\E[X]),\ C'\left(\frac{\E[X]+\E[Y]}{2}\right),\ C'(\E[Y])\right\}
    \geq\frac{L}{3}.
\]
Taking the infimum over the pairs in $\cP_f^1$ at non-zero distance proves the theorem.
\end{proof}

\subsection{The \texorpdfstring{$d \geq 3$}{d >= 3} case}
\label{sec:dimension-three-or-above}

In this section, we prove Theorem~\ref{thm:intro:lower-bound} for $d \geq 3$ and absolutely continuous measures on $\R^d$ using a coupling argument.
The result for measures in $\cP^d$ then follows from the measure equivalence established in Section~\ref{sec:measure-equivalence}.

\begin{theorem}\label{thm:lower-bound:dimension-three-or-above}
For every $d \geq 3$ and every norm in $\R^d$, we have
\[
    \alpha\prn{\cP^d} \geq \frac{1}{3}.
\]
\end{theorem}
\begin{proof}
Fix a norm $\|\cdot\|$ on $\R^d$.
By the measure equivalence established in Section~\ref{sec:measure-equivalence}, it suffices to consider $(\mu,\nu) \in \cP^d_{bc}$.
We show that there exists a hyperplane $H \in \Hh^d$ such that
\[
    C(H) \geq \frac{L}{3}.
\]

Using Lemma~\ref{lem:unified-signed-ham}, we choose three hyperplanes $H_1,H_2,H_3$ as follows:
\begin{enumerate}
    \item $H_1$ bisects $\mu$.
    \item $H_2$ bisects $\mu$ and the signed measure
    $\sigma_{H_1}\nu$.
    \item $H_3$ bisects $\nu$ and the signed measures
    $\sigma_{H_1}\mu$ and $\sigma_{H_2}\mu$.
\end{enumerate}
The last step uses $d \geq 3$.
Since $\mu$ and $\nu$ are absolutely continuous, they assign zero mass to these hyperplanes.
The bisection conditions give
\begin{align}
    \label{eq:dim-3:red}
    \E_{X \sim \mu}[\sigma_{H_1}(X)]
    = \E_{X \sim \mu}[\sigma_{H_2}(X)]
    = \E_{X \sim \mu}[\sigma_{H_1}(X)\sigma_{H_3}(X)]
    = \E_{X \sim \mu}[\sigma_{H_2}(X)\sigma_{H_3}(X)]
    &= 0, \\
    \label{eq:dim-3:blue}
    \text{and } \qquad \E_{Y \sim \nu}[\sigma_{H_3}(Y)]
    = \E_{Y \sim \nu}[\sigma_{H_1}(Y)\sigma_{H_2}(Y)]
    &= 0.
\end{align}

For independent $X \sim \mu$ and $Y \sim \nu$, let
\[
    C_i \coloneqq C(H_i)
    = \E\brk{
        \|X-Y\|
        \1_{\set{\sigma_{H_i}(X) \neq \sigma_{H_i}(Y)}}
    }.
\]
We will prove that $C_1+C_2+C_3 \geq L$, which implies that at least one of the three hyperplanes achieves the desired guarantee.

Let $N(x,y)$ denote the number of hyperplanes among $H_1,H_2,H_3$ that strictly separate $x$ and $y$, and define
\[
    F(x,y) \coloneqq (N(x,y)-1)\|x-y\|.
\]
Thus,
\[
    \E[F(X,Y)] = C_1+C_2+C_3-L.
\]
The only negative contribution occurs when none of the three hyperplanes separates $X$ and $Y$.
Our strategy is to couple $X \sim \mu$ with another $X' \sim \mu$ so that, whenever this happens, $X'$ and $Y$ are separated by exactly two of the hyperplanes.
We then use the triangle inequality and the blue bisection conditions to show that the excess is non-negative after averaging over $Y$.

We first construct the coupling.
Fix $c \in \set{-1,+1}$.
Since $\sigma_{H_3}(X) \in \set{-1,+1}$ almost surely,
\[
    \1_{\set{\sigma_{H_3}(X)=c}}
    = \frac{1+c\,\sigma_{H_3}(X)}{2}
\]
almost surely.
For $i=1,2$, it follows from \eqref{eq:dim-3:red} that
\[
    \E_{X \sim \mu}\brk{
        \sigma_{H_i}(X)\1_{\set{\sigma_{H_3}(X)=c}}
    }
    =
    \frac{
        \E_{X \sim \mu}[\sigma_{H_i}(X)]
        +c\,\E_{X \sim \mu}[
            \sigma_{H_i}(X)\sigma_{H_3}(X)
        ]
    }{2}
    =0.
\]
In other words, within each side of $H_3$, both $H_1$ and $H_2$ split the red mass equally between their two signs.

Denote the four red masses within $\set{\sigma_{H_3}(X) = c}$ by
\[
 \begin{array}{c|cc}
 & \sigma_{H_2}(X) = +1 & \sigma_{H_2}(X) = -1\\ \hline
 \sigma_{H_1}(X) = +1 & \mu^{++} & \mu^{+-} \\
 \sigma_{H_1}(X) = -1 & \mu^{-+} & \mu^{--}
 \end{array}
\]
In particular, for this fixed $c$
\[
    \mu^{ab}
    \coloneqq
    \mu\prn{\set{
        x :
        \sigma_{H_1}(x)=a,\;
        \sigma_{H_2}(x)=b,\;
        \sigma_{H_3}(x)=c
    }},
    \qquad a,b \in \set{-1,+1}.
\]
The two equal-mass conditions become
\[
    \mu^{++}+\mu^{+-}=\mu^{-+}+\mu^{--},
    \qquad \text{and} \qquad
    \mu^{++}+\mu^{-+}=\mu^{+-}+\mu^{--}.
\]
Adding and subtracting these equalities yields
\[
    \mu^{++}=\mu^{--},
    \qquad \text{and} \qquad
    \mu^{+-}=\mu^{-+}.
\]

We now sample $X \sim \mu$ and, conditional on its signs $(a,b,c)$, sample $X'$ from $\mu$ conditioned on the cell with signs $(-a,-b,c)$.
Cells of zero mass can be ignored. Since each cell has the same mass as its paired cell, $X'$ also has distribution $\mu$.
By construction,
\[
    \sigma_{H_1}(X')=-\sigma_{H_1}(X),
    \qquad
    \sigma_{H_2}(X')=-\sigma_{H_2}(X),
    \qquad
    \sigma_{H_3}(X')=\sigma_{H_3}(X)
\]
almost surely.

Draw $Y \sim \nu$ independently of $(X,X')$.
Since $X$ and $X'$ both have distribution $\mu$,
\[
    \E[F(X,Y)+F(X',Y)]
    =2\,\E[F(X,Y)].
\]
It therefore suffices to show that
\[
    \E_{Y \sim \nu}[F(x,Y)+F(x',Y)] \geq 0
\]
for almost every realization $(x,x')$ of the coupling.

Fix such a pair $(x,x')$. For this calculation, reverse the orientations of the hyperplanes as needed so that
\[
    (\sigma_{H_1}(x),\sigma_{H_2}(x),\sigma_{H_3}(x))
    =(+1,+1,+1),
\]
and hence
\[
    (\sigma_{H_1}(x'),\sigma_{H_2}(x'),\sigma_{H_3}(x'))
    =(-1,-1,+1).
\]
These reversals preserve the zero expectations in \eqref{eq:dim-3:red} and \eqref{eq:dim-3:blue}.
The triangle inequality gives
\[
    \bigl|\|x-y\|-\|x'-y\|\bigr|
    \leq \|x-x'\|
    \leq \|x-y\|+\|x'-y\|.
\]
We distinguish four cases according to the signs at $y$.

\begin{enumerate}
    \item[(A)]
    If $\sigma_{H_3}(y)=+1$ and $\sigma_{H_1}(y)=\sigma_{H_2}(y)$, then one of the pairs $(x,y)$ and $(x',y)$ is separated by no hyperplane and the other by exactly two. Therefore,
    \[
        F(x,y)+F(x',y)
        \geq -\bigl|\|x-y\|-\|x'-y\|\bigr|
        \geq -\|x-x'\|.
    \]

    \item[(B)]
    If $\sigma_{H_3}(y)=+1$ and $\sigma_{H_1}(y)\neq\sigma_{H_2}(y)$, then each pair is separated by exactly one hyperplane. Therefore,
    \[
        F(x,y)+F(x',y)=0.
    \]

    \item[(C)]
    If $\sigma_{H_3}(y)=-1$ and $\sigma_{H_1}(y)=\sigma_{H_2}(y)$, then one pair is separated by exactly one hyperplane and the other by all three. Therefore,
    \[
        F(x,y)+F(x',y)
        \geq 2\min\set{\|x-y\|,\|x'-y\|}
        \geq 0.
    \]

    \item[(D)]
    If $\sigma_{H_3}(y)=-1$ and $\sigma_{H_1}(y)\neq\sigma_{H_2}(y)$, then each pair is separated by exactly two hyperplanes. Therefore,
    \[
        F(x,y)+F(x',y)
        =\|x-y\|+\|x'-y\|
        \geq \|x-x'\|.
    \]
\end{enumerate}

By \eqref{eq:dim-3:blue}, cases (A) and (B) together have probability $1/2$, as do cases (A) and (C).
Consequently, if $p$ is the probability of case (A), then the probabilities of cases (A), (B), (C), and (D) are, respectively,
\[
    \prn{p,\;\frac{1}{2}-p,\;\frac{1}{2}-p,\;p},
    \qquad p \in [0,1/2].
\]
Thus, the possible negative contribution from case (A) is offset by the contribution from case (D), while the remaining cases contribute non-negatively:
\[
    \E_{Y \sim \nu}[F(x,Y)+F(x',Y)]
    \geq -p\|x-x'\|+p\|x-x'\|
    =0.
\]
Averaging over $(X,X')$ gives $\E[F(X,Y)] \geq 0$.
Hence, $C_1+C_2+C_3 \geq L$, completing the proof.
\end{proof}

\subsection{The \texorpdfstring{$d = 2$}{d = 2} case}
\label{sec:dimension-two}

In this section, we prove Theorem~\ref{thm:intro:lower-bound} for $d = 2$ and absolutely continuous measures on $\R^2$.
The argument combines an inequality for separation probabilities with an integral representation of planar norms.
The result for measures in $\cP^2$ then follows from the measure equivalence established in Section~\ref{sec:measure-equivalence}.

\begin{theorem}\label{thm:lower-bound:dimension-two}
For every norm on $\R^2$, we have
\[
    \alpha\prn{\cP^2} \geq \frac{1}{3}.
\]
\end{theorem}
\begin{proof}
Fix a norm $\|\cdot\|$ on $\R^2$.
By the measure equivalence established in Section~\ref{sec:measure-equivalence}, it suffices to consider $(\mu,\nu) \in \cP^2_{bc}$.
We show that there exists a line $H \in \Hh^2$ such that
\[
    C(H) \geq \frac{L}{3}.
\]

Using Lemma~\ref{lem:unified-signed-ham}, we choose three lines $H_1,H_2,H_3$ as follows:
\begin{enumerate}
    \item $H_1$ bisects both $\mu$ and $\nu$.
    \item $H_2$ bisects $\mu$ and the signed measure $\sigma_{H_1}\nu$.
    \item $H_3$ bisects $\nu$ and the signed measure $\sigma_{H_1}\mu$.
\end{enumerate}
Since $\mu$ and $\nu$ are absolutely continuous, they assign zero mass to every line.
The bisection conditions give
\begin{align*}
    \E_\mu[\sigma_{H_1}]
    =\E_\mu[\sigma_{H_2}]
    =\E_\mu[\sigma_{H_1}\sigma_{H_3}]
    &=0, \\
    \E_\nu[\sigma_{H_1}]
    =\E_\nu[\sigma_{H_3}]
    =\E_\nu[\sigma_{H_1}\sigma_{H_2}]
    &=0.
\end{align*}
We will prove that
\[
    C(H_1)+C(H_2)+C(H_3) \geq L,
\]
which implies that at least one of the three lines achieves the $1/3$ guarantee.

The bisection conditions have a useful interpretation.
The line $H_1$ places half of each color on each side.
The line $H_2$ also bisects the red mass, but need not bisect the blue mass.
Instead, within each side of $H_2$, the blue mass is balanced across $H_1$: this follows from $\E_\nu[\sigma_{H_1}]=\E_\nu[\sigma_{H_1}\sigma_{H_2}]=0$.
Similarly, $H_3$ bisects the blue mass and, within each side of $H_3$, the red mass is balanced across $H_1$.
Figure~\ref{fig:unified-three-hyperplanes} illustrates these conditions.

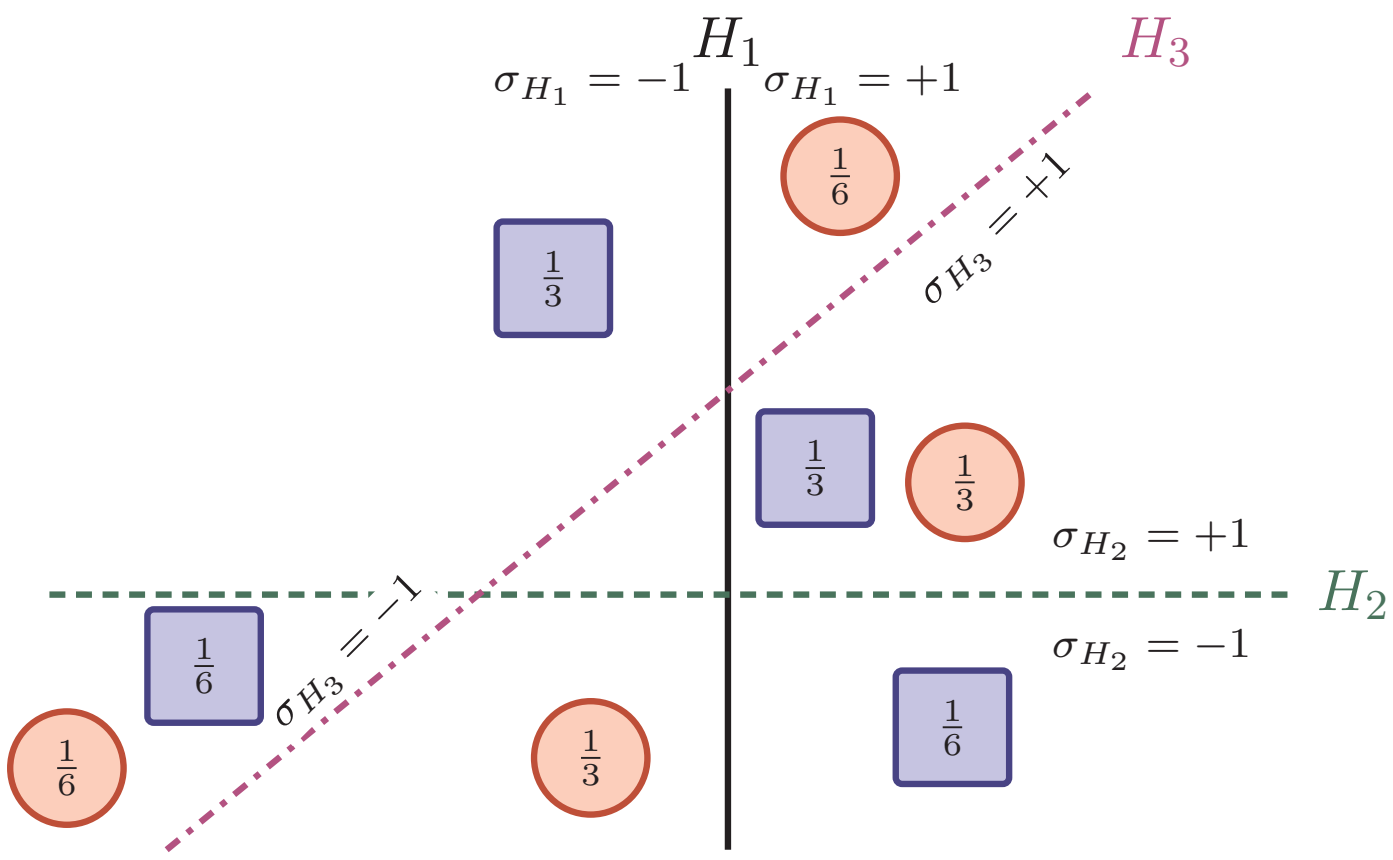
\begin{figure}[tbp]
\centering
\begin{tikzpicture}[
  x=1.65cm,
  y=1.35cm,
  redmass/.style={
    circle,
    draw=red!75!black,
    fill=red!22,
    very thick,
    minimum size=7.5mm,
    inner sep=0pt,
    font=\scriptsize
  },
  bluemass/.style={
    rectangle,
    rounded corners=1.2pt,
    draw=blue!75!black,
    fill=blue!20,
    very thick,
    minimum size=7.5mm,
    inner sep=0pt,
    font=\scriptsize
  }
]
  \draw[very thick]
    (0,-1.25) -- (0,2.48)
    node[above] {$H_1$};
  \node[font=\scriptsize,anchor=south east] at (-0.06,2.30) {$\sigma_{H_1} = -1$};
  \node[font=\scriptsize,anchor=south west] at (0.06,2.30) {$\sigma_{H_1} = +1$};

  \draw[very thick,densely dashed,green!45!black]
    (-2.72,0) -- (2.27,0)
    node[right] {$H_2$};
  \node[font=\scriptsize,anchor=south east] at (2.18,0.07) {$\sigma_{H_2} = +1$};
  \node[font=\scriptsize,anchor=north east] at (2.18,-0.07) {$\sigma_{H_2} = -1$};

  \draw[very thick,dash dot,magenta!70!black]
    (-2.25,-1.25) -- (1.48,2.48)
    node[above right] {$H_3$};
  \node[
    font=\scriptsize,
    rotate=45,
    fill=white,
    inner sep=1pt
  ] at (-1.52,-0.29) {$\sigma_{H_3} = -1$};
  \node[
    font=\scriptsize,
    rotate=45,
    fill=white,
    inner sep=1pt
  ] at (1.08,1.78) {$\sigma_{H_3} = +1$};

  \node[redmass] at (0.95,0.55) {$\frac13$};
  \node[redmass] at (-0.55,-0.80) {$\frac13$};
  \node[redmass] at (0.45,2.05) {$\frac16$};
  \node[redmass] at (-2.65,-0.85) {$\frac16$};

  \node[bluemass] at (0.35,0.62) {$\frac13$};
  \node[bluemass] at (-0.70,1.55) {$\frac13$};
  \node[bluemass] at (0.90,-0.65) {$\frac16$};
  \node[bluemass] at (-2.10,-0.35) {$\frac16$};
\end{tikzpicture}
\caption{A planar example of the three bisections. Each red circle represents
a smooth $\mu$-cloud and each blue square a smooth $\nu$-cloud, supported
in a sufficiently small disk; the number inside is its mass.  Here
$H_1=\{x=0\}$, $H_2=\{y=0\}$, and $H_3=\{y=x+1\}$, with the displayed
orientations.  The line $H_2$ splits $\mu$ as $1/2$--$1/2$ but splits
$\nu$ as $2/3$--$1/3$; nevertheless, on each side of $H_2$, the blue
mass on the two sides of $H_1$ agrees. Similarly, $H_3$ splits $\nu$ as
$1/2$--$1/2$ but splits $\mu$ as $2/3$--$1/3$, with the red mass
balanced across $H_1$ inside each side of $H_3$.}
\label{fig:unified-three-hyperplanes}
\end{figure}

Unlike the argument for $d \geq 3$, we first work with separation probabilities rather than distances.
We show that, for every line $K$, the total probability that $K$ and one of the $H_i$ both separate a red--blue pair is at least the probability that $K$ separates the pair.
At the end of the proof, we integrate this inequality to recover the distance-weighted guarantee.

Let $D(K)$ be the probability that a fixed line $K$ strictly separates independent random variables $X \sim \mu$ and $Y \sim \nu$, and let $I_i(K)$ be the probability that both $K$ and $H_i$ strictly separate them.
Write
\[
    p_0=p_0(K)\coloneqq \E_\mu[\sigma_K],
    \qquad
    q_0=q_0(K)\coloneqq \E_\nu[\sigma_K].
\]
For a measurable function $S:\R^2\to\{-1,0,+1\}$, let
\[
    p_S=p_S(K)\coloneqq \E_\mu[\sigma_K S],
    \qquad
    q_S=q_S(K)\coloneqq \E_\nu[\sigma_K S].
\]
For brevity, write $p_i\coloneqq p_{\sigma_{H_i}}$ and $q_i\coloneqq q_{\sigma_{H_i}}$, and define
\[
    \Phi_K \coloneqq p_0q_0-\sum_{i=1}^3 p_iq_i.
\]

By Lemma~\ref{lem:aux-separated-by-one-hyperplane},
\begin{equation}\label{eq:D}
    D(K)=\frac{1-p_0q_0}{2}.
\end{equation}
Similarly, Lemma~\ref{lem:aux-separated-by-two-hyperplanes} gives
\begin{align*}
    4I_i(K)
    &=1-p_0q_0-\E_\mu[\sigma_{H_i}]\E_\nu[\sigma_{H_i}]+p_iq_i \\
    &=1-p_0q_0+p_iq_i,
\end{align*}
where the last equality follows because each $H_i$ bisects at least one of $\mu$ and $\nu$.
Summing over $i$ and using \eqref{eq:D}, we obtain
\[
    4\prn{\sum_{i=1}^3 I_i(K)-D(K)}
    =1-p_0q_0+\sum_{i=1}^3 p_iq_i
    =1-\Phi_K.
\]
Thus, our goal is to prove that, for every line $K$,
\begin{equation}\label{eq:dim-2:phi-at-most-one}
    \Phi_K\leq 1
    \iff
    \sum_{i=1}^3 I_i(K)-D(K)\geq 0.
\end{equation}

We first observe a simple inequality.
It expresses a tradeoff: a bounded random variable cannot have both a large mean and a large correlation with a balanced sign.

\begin{claim}\label{clm:shared-budget-sign}
Let $U$ be a random variable taking each value in $\{-1,+1\}$ with probability $1/2$, and let $W$ be a random variable supported on $[-1,1]$.
Then
\[
    |\E[W]|+|\E[UW]|\leq 1.
\]
\end{claim}

\begin{proof}
Let
\[
    w_+=\E\brk{W\midd U=+1},
    \qquad
    w_-=\E\brk{W\midd U=-1}.
\]
Since $U$ is equally likely to have either sign,
\begin{align*}
    |\E[W]|+|\E[UW]|
    &=\frac{|w_++w_-|+|w_+-w_-|}{2} 
    =\max\set{|w_+|,|w_-|}
    \leq 1.
    \qedhere
\end{align*}
\end{proof}

We now prove that $\Phi_K\leq 1$.
By reversing the orientations of $H_2$ and $H_3$ if necessary, we may assume that $p_2,p_3\geq 0$.
These reversals preserve both the bisection conditions and $\Phi_K$.

Applying Claim~\ref{clm:shared-budget-sign} to the bisection conditions gives the following six inequalities:
\begin{itemize}
    \item[(A)] Under $\mu$, take $U=\sigma_{H_1}$ and $W=\sigma_K$, then $|p_0|+|p_1|\leq 1.$
    \item[(B)] Under $\nu$, take $U=\sigma_{H_1}$ and $W=\sigma_K$, then $|q_0|+|q_1|\leq 1.$
    \item[(C)] Under $\mu$, take $U=\sigma_{H_2}$ and $W=\sigma_K$, then $|p_0|+p_2\leq 1.$
    \item[(D)] Under $\nu$, take $U=\sigma_{H_3}$ and $W=\sigma_K$, then $|q_0|+|q_3|\leq 1.$
    \item[(E)] Under $\mu$, take $U=\sigma_{H_1}\sigma_{H_3}$ and $W=\sigma_K\sigma_{H_1}$.
    Since $(\sigma_{H_1})^2=1$ almost surely, we obtain $|p_1|+p_3\leq 1.$
    \item[(F)] Under $\nu$, take $U=\sigma_{H_1}\sigma_{H_2}$ and $W=\sigma_K\sigma_{H_1}$.
    Again using $(\sigma_{H_1})^2=1$ almost surely, we obtain $|q_1|+|q_2|\leq 1.$
\end{itemize}

We use these inequalities to obtain two complementary bounds on $\Phi_K$.
The first shows that $\Phi_K>1$ is possible only if $p_2+p_3>1$.
By (D),
\begin{align*}
    p_0q_0-p_3q_3
    &\leq |p_0||q_0|+p_3|q_3| 
    \leq \max\set{|p_0|,p_3}(|q_0|+|q_3|) 
    \leq \max\set{|p_0|,p_3}.
    \intertext{Similarly, by (F),}
    -p_1q_1-p_2q_2
    &\leq |p_1||q_1|+p_2|q_2| 
    \leq \max\set{|p_1|,p_2}(|q_1|+|q_2|) 
    \leq \max\set{|p_1|,p_2}.
\end{align*}
Combining these bounds and applying (A), (C), and (E), we obtain
\begin{align}
    \Phi_K
    &\leq \max\set{|p_0|,p_3}+\max\set{|p_1|,p_2}
    \nonumber \\
    &=\max\set{|p_0|+|p_1|,\;|p_0|+p_2,\;p_3+|p_1|,\;p_3+p_2}
    \nonumber \\
    &\leq \max\set{1,p_2+p_3}.
    \label{eq:phi-k-1}
\end{align}

For the second bound, (C) and (E) give
\begin{align}
    \Phi_K
    &\leq |p_0||q_0|+|p_1||q_1|-p_2q_2-p_3q_3
    \nonumber \\
    &\leq (1-p_2)|q_0|+(1-p_3)|q_1|-p_2q_2-p_3q_3.
    \label{eq:phi-k-2}
\end{align}
To apply (D) and (F), we group the terms involving $q_0,q_3$ and those involving $q_1,q_2$.
Using
\[
    p_2=(1-p_3)+(p_2+p_3-1),
    \qquad
    p_3=(1-p_2)+(p_2+p_3-1),
\]
we can rewrite the last two terms as
\begin{align*}
    -p_2q_2-p_3q_3
    &=-(1-p_3)q_2-(1-p_2)q_3-(p_2+p_3-1)(q_2+q_3).
\end{align*}
Substituting into \eqref{eq:phi-k-2}, we obtain
\begin{align}
    \Phi_K
    &\leq (1-p_2)(|q_0|-q_3)+(1-p_3)(|q_1|-q_2)-(p_2+p_3-1)(q_2+q_3)
    \nonumber \\
    &\leq (1-p_2)(|q_0|+|q_3|)+(1-p_3)(|q_1|+|q_2|)-(p_2+p_3-1)(q_2+q_3)
    \nonumber \\
    &\leq 2-p_2-p_3-(p_2+p_3-1)(q_2+q_3)
    \nonumber \\
    &=1+(p_2+p_3-1)(-q_2-q_3-1).
    \label{equ:mid}
\end{align}
Here the second inequality uses $1-p_2,1-p_3\geq 0$, and the third uses (D) and (F).

Suppose, towards a contradiction, that $\Phi_K>1$.
By \eqref{eq:phi-k-1}, we must have $p_2+p_3>1$.
It then follows from \eqref{equ:mid} that $-q_2-q_3>1$ as well.

These two inequalities have a geometric consequence.
The condition $p_2+p_3>1$ forces red mass on specified sides of $K$, while $-q_2-q_3>1$ forces blue mass on the opposite sides.
We show that this happens in every sign region determined by $H_1$ and $H_2$, which is impossible for a line $K$ in the plane.

\begin{claim}\label{clm:sides-mass-lower-bound}
For every $a,b\in\{-1,+1\}$, we have
\begin{align*}
    \mu(\sigma_{H_1}=a,\ \sigma_{H_2}=\sigma_K=b)
    &>0, \\
    \nu(\sigma_{H_1}=a,\ \sigma_{H_2}=-\sigma_K=b)
    &>0.
\end{align*}
\end{claim}

\begin{proof}
We prove the stronger inequalities
\begin{align*}
    \mu(\sigma_{H_1}=a,\ \sigma_{H_2}=\sigma_{H_3}=\sigma_K=b)
    &\geq \frac{p_2+p_3-1}{4}>0, \\
    \nu(\sigma_{H_1}=a,\ \sigma_{H_2}=\sigma_{H_3}=-\sigma_K=b)
    &\geq \frac{-q_2-q_3-1}{4}>0.
\end{align*}

Fix $a,b\in\{-1,+1\}$ and let
\[
    E_{a,b}
    =\set{\sigma_{H_1}=a,\ \sigma_{H_2}=b,\ \sigma_{H_3}=b,\ \sigma_K=b}.
\]
Define
\[
    Z=\sigma_K(\sigma_{H_2}+\sigma_{H_3})+b\sigma_{H_2}+ab\sigma_{H_1}\sigma_{H_3}.
\]
The last two terms have zero expectation under $\mu$, so
\begin{equation}\label{equ:mid1}
    \E_\mu[Z]
    =\E_\mu[\sigma_K\sigma_{H_2}]+\E_\mu[\sigma_K\sigma_{H_3}]
    =p_2+p_3.
\end{equation}
These zero-mean terms allow us to bound $Z$ by $2$ on one side of $H_1$ and by $0$ on the other, except on the desired event $E_{a,b}$.

Indeed, on $\set{\sigma_{H_1}=a}$,
\[
    Z=(\sigma_K+b)(\sigma_{H_2}+\sigma_{H_3}).
\]
If $\sigma_K=-b$, this is zero.
If $\sigma_K=b$, it is positive only when $\sigma_{H_2}=\sigma_{H_3}=b$, in which case it equals $4$.
Thus, on this side of $H_1$,
\[
    Z\leq 4\cdot\1_{E_{a,b}}.
\]
On $\set{\sigma_{H_1}=-a}$,
\[
    Z=(\sigma_K+b)\sigma_{H_2}+(\sigma_K-b)\sigma_{H_3}.
\]
Exactly one of the two coefficients is zero, and the other has absolute value $2$, so $Z\leq 2$.

Combining the two bounds gives, almost surely,
\[
    Z\leq 4\cdot\1_{E_{a,b}}+2\cdot\1_{\set{\sigma_{H_1}=-a}}.
\]
Taking expectations and using $\mu(\sigma_{H_1}=-a)=1/2$, we obtain
\[
    p_2+p_3=\E_\mu[Z]\leq 4\mu(E_{a,b})+1.
\]
Therefore,
\[
    \mu(E_{a,b})\geq \frac{p_2+p_3-1}{4}>0.
\]

For the blue measure, let
\[
    E'_{a,b}
    =\set{\sigma_{H_1}=a,\ \sigma_{H_2}=b,\ \sigma_{H_3}=b,\ \sigma_K=-b}
\]
and
\[
    Z'=-\sigma_K(\sigma_{H_2}+\sigma_{H_3})+b\sigma_{H_3}+ab\sigma_{H_1}\sigma_{H_2}.
\]
Since $\E_\nu[\sigma_{H_3}]=\E_\nu[\sigma_{H_1}\sigma_{H_2}]=0$,
\[
    \E_\nu[Z']
    =-\E_\nu[\sigma_K(\sigma_{H_2}+\sigma_{H_3})]
    =-q_2-q_3.
\]
On $\set{\sigma_{H_1}=a}$,
\[
    Z'=(b-\sigma_K)(\sigma_{H_2}+\sigma_{H_3}),
\]
which is positive only on $E'_{a,b}$, where it equals $4$.
On $\set{\sigma_{H_1}=-a}$,
\[
    Z'=(b-\sigma_K)\sigma_{H_3}-(\sigma_K+b)\sigma_{H_2}\leq 2,
\]
again because exactly one coefficient is zero and the other has absolute value $2$.
Hence, almost surely,
\[
    Z'\leq 4\cdot\1_{E'_{a,b}}+2\cdot\1_{\set{\sigma_{H_1}=-a}}.
\]
Taking expectations and using $\nu(\sigma_{H_1}=-a)=1/2$ gives
\[
    -q_2-q_3\leq 4\nu(E'_{a,b})+1,
\]
and therefore
\[
    \nu(E'_{a,b})\geq \frac{-q_2-q_3-1}{4}>0.
    \qedhere
\]
\end{proof}

Claim~\ref{clm:sides-mass-lower-bound} implies that each of the four open sets
\[
    H_1^+\cap H_2^+,\qquad
    H_1^+\cap H_2^-,\qquad
    H_1^-\cap H_2^+,\qquad
    H_1^-\cap H_2^-
\]
contains points on both sides of $K$.
In particular, all four sets are non-empty, so $H_1$ and $H_2$ are distinct and non-parallel.
Moreover, each set is convex, so a segment joining points on opposite sides of $K$ lies within the set and intersects $K$.
Thus, $K$ must intersect all four open regions.
This is impossible: the two lines $H_1$ and $H_2$ can cut $K$ at at most two points, allowing $K$ to pass through at most three of their open regions.
We have reached a contradiction, proving \eqref{eq:dim-2:phi-at-most-one}.

It remains to turn this probability inequality into a bound on distances.
The key planar property is that every norm on $\R^2$ admits an integral representation by absolute values of linear functionals.
Specifically, by \cite[Theorem 1.1]{pinelis-norm-approx}, there exists a finite non-negative Borel measure $\rho$ on $A=\R^2\setminus\{0\}$ such that
\[
    \|z\|=\int_A |a\cdot z|\,\rho(\dif a)
    \qquad (z\in\R^2).
\]
This representation allows us to express distances as integrals of separation indicators.

For $a\in A$ and $t\in\R$, let
\[
    K_{a,t}=\set{x\midd a\cdot x=t}.
\]
The thresholds strictly separating $a\cdot x$ and $a\cdot y$ form an interval of length $|a\cdot(x-y)|$.
Hence,
\[
    |a\cdot(x-y)|
    =\int_\R \1_{\set{K_{a,t}\text{ strictly separates }x,y}}\,\dif t.
\]
Integrating the inequality $\sum_{i=1}^3 I_i(K)\geq D(K)$ over these lines, and using Tonelli's theorem to interchange the non-negative integrals and expectations, we obtain
\begin{align*}
    \sum_{i=1}^3 C(H_i)
    &=\int_A\int_\R \sum_{i=1}^3 \Pr\brk{\text{both }H_i\text{ and }K_{a,t}\text{ strictly separate }X,Y}\,\dif t\,\rho(\dif a) \\
    &\geq \int_A\int_\R \Pr\brk{K_{a,t}\text{ strictly separates }X,Y}\,\dif t\,\rho(\dif a) \\
    &=\E[\|X-Y\|]
    =L.
\end{align*}
Thus, at least one of $H_1,H_2,H_3$ satisfies $C(H_i)\geq L/3$.

\end{proof}

\subsection{Large dimension}
\label{subsec:large-dimension}
In this section, we investigate how the optimal guarantee changes as the dimension grows. We first prove Proposition~\ref{prp:ratio-converges-to-half-intro}, showing that $\alpha(\cP^d)$ tends to $1/2$ as $d\to\infty$, for every norm, making the upper bound of Theorem~\ref{thm:intro:upper-bound} asymptotically tight. We then consider pairs of uniform measures on $n$ points each and give a dimension-dependent lower bound on $\alpha(\cP^d_n)$, which reaches $1$ when $d\geq 2n-1$.

For the first result, we use the signed ham-sandwich theorem (Lemma~\ref{lem:unified-signed-ham}) to select $d-1$ hyperplanes that bisect both measures and whose sign functions are orthonormal under their average. This orthogonality allows us to control the combined separation probabilities and obtain a guarantee approaching $1/2$.

\begin{proposition}\label{prp:ratio-converges-to-half}
For every $d \geq 2$ and every norm on $\R^d$,
\[
    \alpha(\cP^d) \geq \frac{1}{2} - \sqrt{\frac{2}{d-1}}.
\]
For the special case of the Euclidean distance,
\[
    \alpha(\cP^d) \geq \frac{1}{2} - \frac{1}{2(d-1)}.
\]
\end{proposition}
\begin{proof}
By Theorem~\ref{thm:measure-equivalence}, it suffices to prove the claimed bounds for an arbitrary pair $(\mu,\nu)\in\cP^d_{bc}$. We start with the special case of the Euclidean distance. Fix $(\mu,\nu)\in\cP^d_{bc}$ and let $\eta=(\mu+\nu)/2$ be the average of the red and blue measures.

Using Lemma~\ref{lem:unified-signed-ham}, we choose $d-1$ hyperplanes $H_1,\dots,H_{d-1}$ successively.
Each $H_i$ bisects the $i+1$ measures
\[
    \mu,\ \nu,\ \sigma_{H_1}\eta,\ \dots,\ \sigma_{H_{i-1}}\eta.
\]
This is possible because $i+1\leq d$.
Thus, every $H_i$ satisfies
\[
    \E_\mu[\sigma_{H_i}]
    =\E_\nu[\sigma_{H_i}]
    =0,
\]
and, for $j<i$,
\[
    \E_\eta[\sigma_{H_j}\sigma_{H_i}]=0.
\]
Since $\mu$ and $\nu$ are absolutely continuous, all hyperplanes have zero mass.
Consequently, the sign functions $\sigma_{H_1},\dots,\sigma_{H_{d-1}}$ are orthonormal under $\eta$.
This orthogonality will allow us to control their combined contribution.

We claim that
\begin{equation}\label{eq:d-tends-to-half-1}
    \sum_{i=1}^{d-1} C(H_i)\geq \prn{\frac{d}{2}-1}L.
\end{equation}
This implies the desired bound, since
\[
    \sup_{H\in\Hh^d}\frac{C(H)}{L}
    \geq \frac{1}{d-1}\sum_{i=1}^{d-1}\frac{C(H_i)}{L}
    \geq \frac{d-2}{2(d-1)}
    =\frac{1}{2}-\frac{1}{2(d-1)}.
\]
Taking the infimum over $(\mu,\nu)\in\cP^d_{bc}$ and applying Theorem~\ref{thm:measure-equivalence} then proves the proposition.

As in Section~\ref{sec:dimension-two}, we first establish an inequality for separation probabilities and then integrate it to obtain \eqref{eq:d-tends-to-half-1}.
Fix a hyperplane $K$, and let $D(K)$ be the probability that $K$ strictly separates independent random variables $X\sim\mu$ and $Y\sim\nu$.
Let $I_i(K)$ be the probability that both $K$ and $H_i$ strictly separate them.
Write
\[
    p_0=p_0(K)\coloneqq \E_\mu[\sigma_K],
    \qquad
    q_0=q_0(K)\coloneqq \E_\nu[\sigma_K].
\]
For a measurable function $S:\R^d\to\{-1,0,+1\}$, let
\[
    p_S=p_S(K)\coloneqq \E_\mu[\sigma_KS],
    \qquad
    q_S=q_S(K)\coloneqq \E_\nu[\sigma_KS].
\]
For brevity, write $p_i\coloneqq p_{\sigma_{H_i}}$ and $q_i\coloneqq q_{\sigma_{H_i}}$.

Since every $H_i$ bisects both $\mu$ and $\nu$, Lemmas~\ref{lem:aux-separated-by-one-hyperplane} and~\ref{lem:aux-separated-by-two-hyperplanes} give
\begin{equation}\label{eq:d-tends-to-half-2}
    D(K)=\frac{1-p_0q_0}{2},
    \qquad \text{and} \qquad
    I_i(K)=\frac{1-p_0q_0+p_iq_i}{4}.
\end{equation}
The key step is to show that
\begin{equation}\label{eq:d-tends-to-half-3}
    \sum_{i=1}^{d-1}p_iq_i\geq -(1-p_0q_0).
\end{equation}
Thus, although some products $p_iq_i$ may be negative, their total negative contribution is bounded independently of the number of hyperplanes.

To prove this, we use the orthogonality of the sign functions under $\eta$.
First, since every $H_i$ bisects both measures,
\begin{align}
    \E_{X\sim\mu}\brk{\sum_{i=1}^{d-1}\frac{p_i-q_i}{2}\sigma_{H_i}(X)}
    &=0,
    \label{eq:d-tends-to-half-4} \\
    \E_{Y\sim\nu}\brk{\sum_{i=1}^{d-1}\frac{p_i-q_i}{2}\sigma_{H_i}(Y)}
    &=0.
    \label{eq:d-tends-to-half-5}
\end{align}
We now compare the centered sign $\sigma_K$ with a linear combination of the $\sigma_{H_i}$, using opposite signs for the red and blue measures.
This choice produces the differences $p_i-q_i$ when the squares are expanded.
Let
\begin{align}
    \Delta
    &\coloneqq
    \frac{1}{2}\E_{X\sim\mu}\brk{
        \prn{\sigma_K(X)-p_0-\sum_{i=1}^{d-1}\frac{p_i-q_i}{2}\sigma_{H_i}(X)}^2
    }
    \nonumber \\
    &\quad+
    \frac{1}{2}\E_{Y\sim\nu}\brk{
        \prn{\sigma_K(Y)-q_0+\sum_{i=1}^{d-1}\frac{p_i-q_i}{2}\sigma_{H_i}(Y)}^2
    }
    \geq 0.
    \label{equ:Delta:positive}
\end{align}

Expanding the first square, using $\sigma_K(X)^2=1$ almost surely and \eqref{eq:d-tends-to-half-4}, gives
\begin{align}
    &\E_{X\sim\mu}\brk{
        \prn{\sigma_K(X)-p_0-\sum_{i=1}^{d-1}\frac{p_i-q_i}{2}\sigma_{H_i}(X)}^2
    }
    \nonumber \\
    &\qquad=
    1-p_0^2
    +\E_{X\sim\mu}\brk{
        \prn{\sum_{i=1}^{d-1}\frac{p_i-q_i}{2}\sigma_{H_i}(X)}^2
    }
    -2\sum_{i=1}^{d-1}\frac{p_i-q_i}{2}p_i.
    \label{eq:d-tends-to-half-8}
\end{align}
Similarly, using \eqref{eq:d-tends-to-half-5},
\begin{align}
    &\E_{Y\sim\nu}\brk{
        \prn{\sigma_K(Y)-q_0+\sum_{i=1}^{d-1}\frac{p_i-q_i}{2}\sigma_{H_i}(Y)}^2
    }
    \nonumber \\
    &\qquad=
    1-q_0^2
    +\E_{Y\sim\nu}\brk{
        \prn{\sum_{i=1}^{d-1}\frac{p_i-q_i}{2}\sigma_{H_i}(Y)}^2
    }
    +2\sum_{i=1}^{d-1}\frac{p_i-q_i}{2}q_i.
    \label{eq:d-tends-to-half-9}
\end{align}

The cross terms inside the squared sums need not vanish under $\mu$ or $\nu$ separately.
They do vanish after averaging, because $\eta=(\mu+\nu)/2$ and the sign functions are orthonormal under $\eta$.
Specifically,
\begin{align}
    &\frac{1}{2}\E_{X\sim\mu}\brk{
        \prn{\sum_{i=1}^{d-1}\frac{p_i-q_i}{2}\sigma_{H_i}(X)}^2
    }
    +\frac{1}{2}\E_{Y\sim\nu}\brk{
        \prn{\sum_{i=1}^{d-1}\frac{p_i-q_i}{2}\sigma_{H_i}(Y)}^2
    }
    \nonumber \\
    &\qquad=
    \E_{Z\sim\eta}\brk{
        \prn{\sum_{i=1}^{d-1}\frac{p_i-q_i}{2}\sigma_{H_i}(Z)}^2
    }
    \nonumber \\
    &\qquad=
    \sum_{i=1}^{d-1}\sum_{j=1}^{d-1}
    \frac{p_i-q_i}{2}\frac{p_j-q_j}{2}
    \E_\eta[\sigma_{H_i}\sigma_{H_j}]
    \nonumber \\
    &\qquad=
    \sum_{i=1}^{d-1}\prn{\frac{p_i-q_i}{2}}^2,
    \label{eq:d-tends-to-half-6}
\end{align}
where the last equality uses
\[
    \E_\eta[\sigma_{H_i}\sigma_{H_j}]
    =
    \begin{cases}
        1, & i=j, \\
        0, & i\neq j.
    \end{cases}
\]
The remaining terms combine as
\begin{equation}\label{eq:d-tends-to-half-7}
    \sum_{i=1}^{d-1}\frac{p_i-q_i}{2}(p_i-q_i)
    =
    2\sum_{i=1}^{d-1}\prn{\frac{p_i-q_i}{2}}^2.
\end{equation}
Averaging \eqref{eq:d-tends-to-half-8} and \eqref{eq:d-tends-to-half-9}, and using \eqref{eq:d-tends-to-half-6} and \eqref{eq:d-tends-to-half-7}, therefore yields
\[
    \Delta
    =1-\frac{p_0^2+q_0^2}{2}
    +\sum_{i=1}^{d-1}\prn{\frac{p_i-q_i}{2}}^2
    -2\sum_{i=1}^{d-1}\prn{\frac{p_i-q_i}{2}}^2 = 1-\frac{p_0^2+q_0^2}{2}
    -\sum_{i=1}^{d-1}\prn{\frac{p_i-q_i}{2}}^2.
\]

Since $\Delta\geq 0$, the squared differences are bounded by $1-(p_0^2+q_0^2)/2$.
Using this bound and $p_0^2+q_0^2\geq 2p_0q_0$, we obtain
\begin{align*}
    \sum_{i=1}^{d-1}p_iq_i
    &=\sum_{i=1}^{d-1}\prn{\frac{p_i+q_i}{2}}^2
    -\sum_{i=1}^{d-1}\prn{\frac{p_i-q_i}{2}}^2 = \sum_{i=1}^{d-1}\prn{\frac{p_i+q_i}{2}}^2
    +\Delta-\prn{1-\frac{p_0^2+q_0^2}{2}} \\
    &\geq -\prn{1-\frac{p_0^2+q_0^2}{2}} \geq -(1-p_0q_0).
\end{align*}
This proves \eqref{eq:d-tends-to-half-3}. Combining \eqref{eq:d-tends-to-half-2} and \eqref{eq:d-tends-to-half-3}, we conclude that
\begin{equation}\label{eq:d-tends-to-half-10}
    \sum_{i=1}^{d-1}I_i(K) = \frac{(d-1)(1-p_0q_0)+\sum_{i=1}^{d-1}p_iq_i}{4} \geq \frac{(d-2)(1-p_0q_0)}{4} = \frac{d-2}{2}D(K).
\end{equation}

Finally, for the Euclidean distance, Crofton's formula~\cite{Santalo2004} expresses distances as integrals of separation indicators. There is a translation- and rotation-invariant measure $\zeta_d$ on $\Hh^d$, normalized so that
\[
    \|x-y\|
    =\int_{\Hh^d}\1_{\set{K\text{ strictly separates }x,y}}\,\dif{\zeta_d}(K)
    \qquad (x,y\in\R^d).
\]
Thus, longer segments are strictly separated by a larger measure of hyperplanes. For independent $X\sim\mu$ and $Y\sim\nu$, Tonelli's theorem gives, for Euclidean distances,
\begin{align*}
    L_{\mu,\nu}
    &=\int_{\Hh^d}\Pr[K\text{ strictly separates }X,Y]\,\dif{\zeta_d}(K), \\
    C_{\mu,\nu}(H)
    &=\int_{\Hh^d}\Pr[\text{both }H\text{ and }K\text{ strictly separate }X,Y]\,\dif{\zeta_d}(K).
\end{align*}
Using the identities above, we can integrate \eqref{eq:d-tends-to-half-10} to obtain
\[
    \sum_{i=1}^{d-1}C(H_i)\geq \frac{d-2}{2} L,
\]
which gives \eqref{eq:d-tends-to-half-1}, upon rearrangement.

We now turn to an arbitrary norm $\|\cdot\|$ on $\R^d$.
Translate both measures so that $\E[X]=0$, which does not change distances or crossing values.
We repeat the preceding hyperplane construction, replacing $\eta$ by a measure weighted by the distance from the origin:
\[
    \dif {\eta_\rho}(x)
    \coloneqq
    \frac{\|x\|}{\E[\|X\|]+\E[\|Y\|]}
    \prn{\dif\mu(x)+\dif\nu(x)}.
\]
The denominator is positive, and $\eta_\rho$ is a compactly supported, absolutely continuous probability measure.
Let $H_1,\dots,H_{d-1}$ be the resulting hyperplanes.
Each $H_i$ bisects both $\mu$ and $\nu$, and their sign functions are orthonormal under $\eta_\rho$.
The weighting will allow us to control the distance-weighted correlations.

Since hyperplanes have zero mass under both measures,
\[
    C(H_i)
    =\E\brk{\|X - Y\| \cdot \1_{\set{H_i \text{ separates } X,Y}}} 
    =\E\brk{\|X-Y\|\cdot\frac{1-\sigma_{H_i}(X)\sigma_{H_i}(Y)}{2}}.
\]
Summing over $i$ gives
\begin{equation}\label{eq:lower-bound:large-dim-1}
    \sum_{i=1}^{d-1}C(H_i)
    =
    \frac{d-1}{2}L
    -\frac{1}{2}\E\brk{\|X-Y\|\sum_{i=1}^{d-1}\sigma_{H_i}(X)\sigma_{H_i}(Y)}.
\end{equation}
Our goal is to bound the correlation term on the right using the triangle inequality and the Cauchy--Schwarz inequality.

Because each $H_i$ bisects both measures and $X,Y$ are independent,
\[
    \E\brk{
        \prn{\|X\|+\|Y\|}
        \sum_{i=1}^{d-1}\sigma_{H_i}(X)\sigma_{H_i}(Y)
    }
    =0.
\]
Moreover, the triangle inequality gives
\[
    0
    \leq \|X\|+\|Y\|-\|X-Y\|
    \leq 2\min\set{\|X\|,\|Y\|}
    \leq 2\sqrt{\|X\|\|Y\|}.
\]
Subtracting the preceding zero-expectation term therefore yields
\begin{align}
    \E\brk{\|X-Y\|\sum_{i=1}^{d-1}\sigma_{H_i}(X)\sigma_{H_i}(Y)}
    &\leq
    \abs{\E\brk{
        \prn{\|X-Y\|-\|X\|-\|Y\|}
        \sum_{i=1}^{d-1}\sigma_{H_i}(X)\sigma_{H_i}(Y)
    }}
    \nonumber \\
    &\leq
    2\;\E\brk{
        \sqrt{\|X\|\|Y\|}
        \abs{\sum_{i=1}^{d-1}\sigma_{H_i}(X)\sigma_{H_i}(Y)}
    }
    \nonumber \\
    &\leq
    2\sqrt{
        \E\brk{
            \|X\|\|Y\|
            \prn{\sum_{i=1}^{d-1}\sigma_{H_i}(X)\sigma_{H_i}(Y)}^2
        }
    },
    \label{eq:lower-bound:large-dim-2}
\end{align}
where the last inequality follows from the Cauchy--Schwarz inequality.

We now use orthogonality under $\eta_\rho$ to bound this second moment.
For $i\neq j$, the definition of $\eta_\rho$ gives
\[
    0
    =
    \prn{\E_\mu[\|X\|]+\E_\nu[\|Y\|]}
    \E_{\eta_\rho}[\sigma_{H_i}\sigma_{H_j}] =
    \E_\mu\brk{\|X\|\sigma_{H_i}(X)\sigma_{H_j}(X)}
    +\E_\nu\brk{\|Y\|\sigma_{H_i}(Y)\sigma_{H_j}(Y)}.
\]
Thus,
\begin{equation}\label{eq:lower-bound:large-dim-3}
    \E_\mu\brk{\|X\|\sigma_{H_i}(X)\sigma_{H_j}(X)}
    =
    -\E_\nu\brk{\|Y\|\sigma_{H_i}(Y)\sigma_{H_j}(Y)}.
\end{equation}
In particular, the red and blue weighted correlations have opposite signs.
When we expand the square and use independence, every off-diagonal term is therefore non-positive:
\begin{align}
    &\E\brk{\|X\|\|Y\|\prn{\sum_{i=1}^{d-1}\sigma_{H_i}(X)\sigma_{H_i}(Y)}^2}
    \nonumber \\
    &\qquad=
    (d-1)\E_\mu[\|X\|]\E_\nu[\|Y\|]+
    \sum_{\substack{1\leq i,j\leq d-1\\i\neq j}}
    \E_\mu\brk{\|X\|\sigma_{H_i}(X)\sigma_{H_j}(X)}
    \E_\nu\brk{\|Y\|\sigma_{H_i}(Y)\sigma_{H_j}(Y)}
    \nonumber \\
    &\qquad=
    (d-1)\E_\mu[\|X\|]\E_\nu[\|Y\|]
    -
    \sum_{\substack{1\leq i,j\leq d-1\\i\neq j}}
    \prn{\E_\mu\brk{\|X\|\sigma_{H_i}(X)\sigma_{H_j}(X)}}^2
    \nonumber \\
    &\qquad\leq
    (d-1)\E_\mu[\|X\|]\E_\nu[\|Y\|].
    \label{eq:lower-bound:large-dim-4}
\end{align}

It remains to relate the expected distances from the origin to $L$.
Since $\E[X]=0$, Jensen's inequality, applied for each fixed $Y$, gives
\[
    \E_\nu[\|Y\|] = \E_\nu\brk{\|\E_\mu[X] - Y\|} \leq \E[\|X - Y\|] = L.
\]
The triangle inequality then yields
\[
    \E_\mu[\|X\|] \leq \E[\|X - Y\|] + \E_\nu[\|Y\|] \leq 2 L.
\]
Substituting these bounds into \eqref{eq:lower-bound:large-dim-4}, we obtain
\[
    \E\brk{
        \|X\|\|Y\|
        \prn{\sum_{i=1}^{d-1}\sigma_{H_i}(X)\sigma_{H_i}(Y)}^2
    }
    \leq 2(d-1)L^2.
\]
Combining this with \eqref{eq:lower-bound:large-dim-1} and \eqref{eq:lower-bound:large-dim-2} gives
\[
    \sum_{i=1}^{d-1}C(H_i)
    \geq \frac{d-1}{2}L-\sqrt{2(d-1)}\,L.
\]
Averaging over the $d-1$ hyperplanes, we conclude that
\[
    \max_{1\leq i\leq d-1}\frac{C(H_i)}{L}
    \geq
    \frac{1}{d-1}\sum_{i=1}^{d-1}\frac{C(H_i)}{L}
    \geq
    \frac{1}{2}-\sqrt{\frac{2}{d-1}}.
\]
Taking the infimum over $(\mu,\nu)\in\cP^d_{bc}$ and applying Theorem~\ref{thm:measure-equivalence} proves the bound for arbitrary norms.
\end{proof}

We next turn to pairs of uniform measures on $n$ points each.
When $d\geq 2n-1$, the $2n$ points are affinely independent by the general-position assumption, so a hyperplane can strictly separate the two colors.
The following proposition also gives a quantitative lower bound below this threshold.

\begin{repproposition}{prp:exact-high-dimension-2}
For $n \geq 3$, $d \in \N$ and every norm on $\R^d$, we have
\[
    \alpha(\cP_n^d) \geq \begin{cases}
        \dfrac{(d-3)^+ + 2\floor{n - \frac{d}{2}}}{(d-3)^+ + 6\floor{n - \frac{d}{2}}}, & \text{if } d < 2n-1 \\
        1, & \text{if } d \geq 2n-1
    \end{cases}.
\]
\end{repproposition}
\begin{proof}
Let $R$ and $B$ be sets of $n$ red and $n$ blue points, respectively, such that $R\cup B$ is in affine general position in $\R^d$.
It suffices to find a hyperplane $H\in\Hh^d$ satisfying
\begin{equation}\label{eq:prop:exact:high}
    \frac{C(H)}{L} \geq \begin{cases}
        \dfrac{(d-3)^+ + 2\floor{n - \frac{d}{2}}}{(d-3)^+ + 6\floor{n - \frac{d}{2}}}, & \text{if } d < 2n-1 \\
        1, & \text{if } d \geq 2n-1
    \end{cases}.
\end{equation}

First, suppose that $d\geq 2n-1$.
Assign label $+1$ to every point in $R$ and label $-1$ to every point in $B$.
Since $2n\leq d+1$ and $R\cup B$ is in affine general position, its points are affinely independent.
Consequently, the linear system
\[
    a\cdot p+b=
    \begin{cases}
        +1, & p\in R, \\
        -1, & p\in B
    \end{cases}
    \qquad (p\in R\cup B)
\]
has a solution $(a^*,b^*)$.
The hyperplane $H=\set{x\midd a^*\cdot x+b^*=0}$ strictly separates the two colors and therefore cuts every red--blue segment.
Hence, $C(H)=L$, proving \eqref{eq:prop:exact:high} and giving $\alpha(\cP_n^d)=1$ in this case.

For $d\leq 3$, we have
\[
    \dfrac{(d-3)^+ + 2\floor{n-\frac{d}{2}}}{(d-3)^+ + 6\floor{n-\frac{d}{2}}}
    =\frac{1}{3},
\]
so \eqref{eq:prop:exact:high} follows from Theorem~\ref{thm:intro:lower-bound}.

It remains to consider $3<d<2n-1$.
Let
\[
    C^*=\max_{H\in\Hh^d}C(H).
\]
This maximum is attained because hyperplanes induce only finitely many separation patterns on the finite set $R\cup B$.

For each point $p\in R\cup B$, define its weight to be the total length of its incident red--blue pairs:
\[
    w_p=\sum_{\substack{q\in R\cup B\\q\text{ has the opposite color of }p}}\|p-q\|.
\]
Let $S$ consist of the $d-3$ points with largest weights, breaking ties arbitrarily, and let
\[
    t=\max_{p\notin S}w_p.
\]
By the choice of $S$,
\begin{equation}
    \sum_{p\in S}w_p\geq (d-3)t.
\end{equation}
For $j=0,1,2$, let $L_j$ be the total length of the red--blue pairs with exactly $j$ endpoints in $S$.

The next claim combines the $1/3$ guarantee for pairs outside $S$ with stronger guarantees for pairs incident to $S$.
The idea is to choose a hyperplane through $S$ and then perturb it so that the red and blue points in $S$ lie on opposite sides.

\begin{claim}\label{clm:finite-pinned-cut}
We have
\[
    C^*\geq \frac{L_0}{3}+\frac{L_1}{2}+L_2.
\]
\end{claim}
\begin{proof}
We first construct a hyperplane containing $S$, together with side assignments for the remaining points on that hyperplane, such that the expected crossing value from pairs outside $S$ is at least $L_0/3$.

Suppose first that both colors have points outside $S$.
Replace each point outside $S$ by a smooth cloud of unit mass supported in a ball of radius $\rho>0$ in the chosen norm around that point, choosing $\rho$ small enough that the balls are disjoint.
Let $\mu'$ and $\nu'$ be the resulting red and blue measures.

We apply the construction from Theorem~\ref{thm:lower-bound:dimension-three-or-above}, requiring all three hyperplanes to contain $S$.
To enforce this requirement, add an auxiliary absolutely continuous, radially symmetric measure supported in a ball around each point of $S$, with positive density throughout the ball.
A hyperplane bisects such a measure if and only if it passes through its center.
Thus, requiring each hyperplane to bisect these $d-3$ auxiliary measures forces it to contain $S$.
Since each step of the construction already requires at most three bisections, the total number of measures to bisect is at most $d$, so Lemma~\ref{lem:unified-signed-ham} applies.

After normalizing $\mu'$ and $\nu'$ to probability measures and then rescaling, the construction gives three hyperplanes through $S$ whose total crossing value in the smoothed instance is at least
\[
    \int\int\|x-y\|\,\dif {\mu'}(x)\,\dif{\nu'}(y).
\]
Choose one of these hyperplanes, denoted by $H_\rho$, whose crossing value is at least one third of this quantity.

Let $\rho$ tend to zero.
Since every $H_\rho$ contains the non-empty set $S$, we may pass to a subsequence along which the oriented hyperplanes converge to a hyperplane $H$ containing $S$.
Passing to a further subsequence, we may also assume that, for each point outside $S$, the fraction of its cloud on the positive side of $H_\rho$ converges.

For each point outside $S$ that lies on $H$, independently assign it to the positive side with probability equal to this limiting fraction, and otherwise assign it to the negative side.
Points outside $H$ retain their actual sides.
The resulting expected crossing value from pairs outside $S$ is the limit of the crossing values of $H_\rho$ in the smoothed instances.
Indeed, the side probabilities converge, and the distance between points in two clouds differs from the distance between their centers by at most $2\rho$.
Since the total red--blue distance in the smoothed instance tends to $L_0$, this expected crossing value is at least $L_0/3$.

If one color has no point outside $S$, then $L_0=0$.
In this case, simply choose any hyperplane $H$ containing $S$ and assign arbitrary sides to the points outside $S$ that lie on $H$.

We now assign sides to the points in $S$.
Independently of the preceding assignments, use a fair coin to choose between placing all red points in $S$ on the positive side and all blue points in $S$ on the negative side, or the reverse.
Every pair counted in $L_2$ then has its endpoints assigned to opposite sides, while every pair counted in $L_1$ has its endpoints assigned to opposite sides with probability $1/2$.

It remains to justify that these side assignments can be realized by a hyperplane.
Since the original points are in affine general position, $H$ contains at most $d$ of them, including the $d-3$ points in $S$.
These points are affinely independent, so there exists an affine function taking any prescribed values $+1$ and $-1$ on them.
Adding a sufficiently small multiple of this function to an affine function defining $H$ realizes the prescribed sides of all points on $H$, while preserving the sides of all points outside $H$.

Thus, every outcome of the random assignments is realized by a perturbed hyperplane.
Its expected crossing value is at least
\[
    \frac{L_0}{3}+\frac{L_1}{2}+L_2.
\]
Since $C^*$ is at least the crossing value of every such hyperplane,
\[
    C^*\geq \frac{L_0}{3}+\frac{L_1}{2}+L_2.
    \qedhere
\]
\end{proof}

Since $L=L_0+L_1+L_2$ and each pair with both endpoints in $S$ contributes twice to the sum of the weights in $S$, we have
\[
    \sum_{p\in S}w_p=L_1+2L_2.
\]
Thus, Claim~\ref{clm:finite-pinned-cut} gives
\begin{equation}\label{eq:high-dimension-n-1}
    3C^*
    \geq L_0+\frac{3L_1}{2}+3L_2
    =L+\frac{1}{2}\sum_{p\in S}w_p+L_2
    \geq L+\frac{(d-3)t}{2}.
\end{equation}

We obtain a second bound by constructing a hyperplane that places all but a small number of points on their intended sides.
Let $S'\subseteq(R\cup B)\setminus S$ be any set of three points, and let $T=S\cup S'$.
Since $|T|=d$ and the points are in affine general position, $T$ determines a unique hyperplane $H_T$, which contains no other points of $R\cup B$.

Our intended assignment places every red point on the positive side and every blue point on the negative side.
Reversing the orientation of $H_T$ exchanges the correctly and incorrectly placed points outside $T$.
Since there are $2n-d$ such points, we can choose an orientation with at most
\[
    \floor{\frac{2n-d}{2}}=\floor{n-\frac{d}{2}}
\]
incorrectly placed points.

We can then perturb $H_T$ to place every point of $T$ on its intended side while preserving the sides of all points outside $T$.
Indeed, write
\[
    H_T=\set{x\in\R^d\midd f(x)=0}
\]
for an affine function $f$ with the chosen orientation.
Since the points of $T$ are affinely independent, there exists an affine function $g$ taking value $+1$ on the red points of $T$ and $-1$ on the blue points of $T$.
For sufficiently small $\eps>0$, the hyperplane
\[
    H_{T,\eps}\coloneqq\set{x\in\R^d\midd f(x)+\eps g(x)=0}
\]
therefore realizes the intended assignment on $T$ without changing the sides of any other points.

Every red--blue pair not separated by $H_{T,\eps}$ has exactly one incorrectly placed endpoint.
Charging each non-separated pair to its incorrectly placed endpoint therefore bounds the total length of non-separated pairs by $t\floor{n-\frac{d}{2}}$.
Consequently,
\begin{equation}\label{eq:high-dimension-n-2}
    C^*\geq C(H_{T,\eps})\geq L-t\floor{n-\frac{d}{2}}.
\end{equation}

The two bounds complement each other: \eqref{eq:high-dimension-n-1} improves as $t$ increases, whereas \eqref{eq:high-dimension-n-2} improves as $t$ decreases.
Multiplying \eqref{eq:high-dimension-n-1} by $2\floor{n-\frac{d}{2}}$ and \eqref{eq:high-dimension-n-2} by $d-3$, then adding, eliminates $t$ and gives
\[
    \prn{d-3+6\floor{n-\frac{d}{2}}}C^*
    \geq
    \prn{d-3+2\floor{n-\frac{d}{2}}}L.
\]
Dividing by $L$ and the positive coefficient of $C^*$ yields
\[
    \frac{C^*}{L}
    \geq
    \dfrac{d-3+2\floor{n-\frac{d}{2}}}{d-3+6\floor{n-\frac{d}{2}}},
\]
which proves \eqref{eq:prop:exact:high} in the remaining case $3<d<2n-1$.
\end{proof}

\section{Upper Bounds}
\label{sec:upper-bounds}

In this section, we establish upper bounds on $\alpha(\cP^d)$ for every norm on $\R^d$. We first give a simple construction on the line showing that $\alpha(\cP^1) \leq 1/3$, matching the lower bound of Theorem~\ref{thm:lower-bound:dimension-one}. Then, for $d \geq 2$, we construct pairs of measures in $\cP^d$ for which no hyperplane captures more than a fraction arbitrarily close to $1/2$ of the expected total red--blue distance.
This proves that $\alpha(\cP^d) \leq 1/2$.

\subsection{The \texorpdfstring{$d = 1$}{d = 1} case}
\label{subsec:upper-bounds:dim-one}

We begin with a construction in $\cP^1$, consisting of two atoms of each color on the line. By choosing their masses and locations appropriately, we obtain an optimal cut ratio tending to $1/3$. Since $\cP^1 \subseteq \cP^1_f$, it follows that $\alpha(\cP^1_f) \leq \alpha(\cP^1) \leq 1/3$.

\begin{proposition}\label{prp:upper-bound:dim-1}
For every norm on $\R$, we have
\[
    \alpha(\cP^1) \leq \frac{1}{3}.
\]
\end{proposition}
\begin{proof}
Every norm on $\R$ is a positive multiple of the absolute value, so changing the norm scales all distances by the same factor and leaves the cut ratio unchanged. It therefore suffices to consider the Euclidean distance. Fix $\eps > 0$. Choose an integer $k \geq 1$ such that
\[
    \frac{k+2}{3k+1} < \frac{1}{3} + \eps,
\]
and let $\gamma = 1/(2k)$ and
\[
    \mu = \frac{k}{k+1} \delta_{1 - \gamma} + \frac{1}{k+1} \delta_2, \qquad \nu = \frac{1}{k+1} \delta_0 + \frac{k}{k+1} \delta_{1 + \gamma}.
\]
In other words, most of the red and blue mass lies at two nearby points, located at $1 - \gamma$ and $1 + \gamma$, with a small blue mass at $0$ and a small red mass at $2$. The construction can be seen in Figure~\ref{fig:line-construction}. All four atoms are distinct, so $(\mu, \nu) \in \cP^1$.

The expected total red--blue distance is
\[
    L = \frac{k (1 - \gamma) + k^2 (2 \gamma) + k (1 - \gamma) + 2}{(k + 1)^2} = \frac{3k + 1}{(k + 1)^2}.
\]
There are three gaps between consecutive atoms in which to place a threshold. A threshold in the left-most gap separates the blue atom at $0$ from both red atoms, giving
\[
    C(\tau) = \frac{k(1-\gamma)+2}{(k+1)^2} = \frac{k+3/2}{(k+1)^2}, \qquad 0 < \tau < 1 - \gamma.
\]
A threshold in the middle gap separates the two central atoms and the two outer atoms, giving
\[
    C(\tau)
    =\frac{k^2(2\gamma)+2}{(k+1)^2}
    =\frac{k+2}{(k+1)^2},
    \qquad 1-\gamma<\tau<1+\gamma.
\]
By symmetry, the right-most gap gives the same cut value as the left-most gap. Thresholds outside $[0,2]$ have cut value zero. A threshold at an atom cannot improve on these values, since shifting it slightly preserves every pair it strictly separates. Therefore,
\[
    \sup_{\tau \in \R} \frac{C(\tau)}{L} = \frac{k+2}{3k+1}
    < \frac{1}{3} + \eps.
\]
Since $(\mu, \nu) \in \cP^1$ and $\eps > 0$ was arbitrary, we conclude that $\alpha(\cP^1) \leq 1/3$.
\end{proof}

\subsection{The \texorpdfstring{$d \geq 2$}{d >= 2} case}
\label{subsec:upper-bounds:dim-two-or-above}

We next construct compactly supported, absolutely continuous measures showing that $\alpha(\cP^d_{bc}) \leq 1/2$ (and thus also $\alpha(\cP^d) \leq 1/2$) for every $d \geq 2$ and every norm on $\R^d$.

\begin{reptheorem}{thm:intro:upper-bound}
For every $d\geq 2$ and every norm on $\R^d$, we have
\[
    \alpha(\cP^d)\leq \frac{1}{2}.
\]
\end{reptheorem}

\begin{proof}
Fix a norm $\| \cdot \|$ on $\R^d$. It suffices to construct, for every $\eps > 0$, a pair $(\mu, \nu) \in \cP^d_{bc}$ such that
\[
    \sup_{H \in \Hh^d} \frac{C(H)}{L} \leq \frac{1}{2}+\eps.
\]
The construction starts with the vertices of a parallelogram, with opposite vertices receiving the same color and all four red--blue distances equal to $1$.
We replace each vertex by a small uniform cloud, preserving the central symmetry of each color about the center of the parallelogram.
This common symmetry ensures that every hyperplane separates a random red--blue pair with probability at most $1/2$.
Since all red--blue distances remain close to $1$, the distance-weighted cut ratio is also close to $1/2$.
See Figure~\ref{fig:upper-bound-parallelogram} in the introduction for an illustration.

More formally, choose linearly independent vectors $u,v\in\R^d$ with $\|u\|=\|v\|=1$, and let
\[
    K=\set{x\in\R^d\midd \|x\|\leq 1}.
\]
For $0<\rho<1/2$ and $z\in\{0,u,v,u+v\}$, let $U_z$ denote the uniform probability measure on $z+\rho K$.
Also let
\[
    \mu=\frac{1}{2}(U_u+U_v),
    \qquad
    \nu=\frac{1}{2}(U_0+U_{u+v})
\]
and $X\sim\mu$ and $Y\sim\nu$ be independent.
These measures are compactly supported and absolutely continuous, so $(\mu,\nu)\in\cP^d_{bc}$.
Let $c = (u+v)/2$ be the center of the parallelogram.
Reflecting about $c$ exchanges the two red and the two blue clouds. Thus, both measures are centrally symmetric about $c$.

We first bound the separation probability for an arbitrary hyperplane $H \in \Hh^d$. If $H$ passes through $c$, central symmetry implies that it bisects both measures.
Otherwise, orient $H$ so that $c\in H^+$.
The half-space $H^+$ contains one of the open half-spaces bounded by the parallel hyperplane through $c$, which bisects both measures.
Consequently, in either case, we can choose the orientation so that
\[
    \mu(H^+)\geq \frac{1}{2},
    \qquad
    \nu(H^+)\geq \frac{1}{2}.
\]
Since both measures assign zero mass to $H$, their sign expectations satisfy
\[
    p_H\coloneqq \E_\mu[\sigma_H]=2\mu(H^+)-1\geq 0,
    \qquad
    q_H\coloneqq \E_\nu[\sigma_H]=2\nu(H^+)-1\geq 0.
\]
Lemma~\ref{lem:aux-separated-by-one-hyperplane} therefore gives
\[
    \Pr[H\text{ strictly separates }X,Y]
    =\frac{1-p_Hq_H}{2}
    \leq \frac{1}{2}.
\]

It remains to account for the distance weights.
Every red cloud center is at norm-distance $1$ from every blue cloud center, and each point lies within distance $\rho$ of its center.
The triangle inequality thus gives, almost surely,
\[
    1-2\rho\leq \|X-Y\|\leq 1+2\rho.
\]
Hence, for every hyperplane $H$,
\[
    C(H)
    \leq (1+2\rho)\Pr[H\text{ strictly separates }X,Y]
    \leq \frac{1+2\rho}{2},
    \qquad \text{and} \qquad
    L\geq 1-2\rho.
\]
Therefore,
\[
    \frac{C(H)}{L}
    \leq \frac{1+2\rho}{2(1-2\rho)}
    =\frac{1}{2}+\frac{2\rho}{1-2\rho}.
\]
Choosing
\[
    0<\rho<\frac{\eps}{2(1+\eps)}
\]
makes the last term smaller than $\eps$, uniformly over all hyperplanes $H$.
Since $(\mu,\nu)\in\cP^d_{bc}\subseteq\cP^d$ and $\eps>0$ was arbitrary, we conclude that $\alpha(\cP^d)\leq 1/2$.
\end{proof}

\section{Improved Guarantees for the Euclidean Distance}
\label{sec:improved-bounds}

In this section we focus on the Euclidean distance and measures in $\cP^d$. We show that neither the lower bound of $1/3$ nor upper bound of $1/2$ are tight, even for $d = 2$. Specifically, for every dimension $d \geq 2$, we improve the lower bound by combining a common bisecting hyperplane with random hyperplanes through the midpoint of the two means and through the individual means. This shows, for $d \geq 2$, that
\[
    \alpha(\cP^d) \geq \frac{8}{8+5\pi} \approx 0.337.
\]
To complement this, we also present a simple planar construction showing that no ratio larger than $5/11 \approx 0.4545$ can be guaranteed when $d = 2$.

\subsection{Lower bound}
\label{sec:improved-lower-bound}

We begin with the lower bound.
\begin{proposition}\label{prp:lower-bound:improved-planar}
For the Euclidean distance and every $d \geq 2$, we have
\[
    \alpha(\cP^d) \geq \frac{8}{8 + 5 \pi}.
\]
\end{proposition}

To prove the proposition above, we need the following elementary lemma.

\begin{lemma}\label{lem:pointwise-mean}
Let $d\geq2$, and let $x, y$ denote two non-zero vectors in $\R^d$ that form an angle $\varphi \in [0, \pi]$. Then
\begin{equation}
    \|x-y\| \cdot \varphi \geq \frac{4}{3} \|x\|(1 - \cos\varphi).
    \label{equ:pointwise-mean}
\end{equation}
\end{lemma}

\begin{proof}
The two vectors lie in a subspace of dimension at most two, so the argument is independent of the ambient dimension. When $\varphi = 0$,
both sides of \eqref{equ:pointwise-mean} vanish,
and hence \eqref{equ:pointwise-mean} holds with equality. Fix $\|x\|$ and $\varphi$ and consider the problem of minimizing $\|x-y\|$. For an acute angle, the closest point on the ray containing $y$ is the orthogonal projection of $x$ onto this ray, giving
\[
    \|x-y\| \geq \|x\| \sin\varphi.
\]
Hence it is enough to show
\[
    \frac{\varphi \sin\varphi}{1 - \cos\varphi}
    \geq \frac{4}{3}.
\]
Using the half-angle identity for the tangent function, we have
\[
    \frac{\varphi \sin\varphi}{1 - \cos\varphi} = \frac{\varphi}{\tan(\varphi/2)}.
\]
Since $\tan t$ is convex on $[0, \pi/4]$, it lies below the chord joining $(0,0)$ and $(\pi/4, 1)$. Therefore
$\tan(\varphi/2) \leq \frac{2\varphi}{\pi}$,
and thus
\[
    \frac{\varphi}{\tan(\varphi/2)} \geq \frac{\pi}{2} > \frac43.
\]

For an obtuse angle, the closest point on the ray is the origin, giving
$\|x-y\| \geq \|x\|$.
It therefore suffices to prove
\begin{equation}\label{warmup:angle}
    1 - \cos\varphi \leq \frac{3}{4} \varphi.
\end{equation}
On $[\pi/2,\pi]$, the function $1 - \cos\varphi$ is concave, so its tangent at $3 \pi/4$ is an upper bound:
\[
    1 - \cos\varphi \leq 1 + \frac{\sqrt2}{2} + \frac{\sqrt2}{2} \left(\varphi - \frac{3\pi}{4}\right).
\]
After subtracting $3\varphi/4$, the right-hand side is decreasing in $\varphi$, and at $\varphi = \pi/2$ it equals
\[
    \frac{8+4\sqrt2-\pi(3+\sqrt2)}{8} < 0.
\]
Thus \eqref{warmup:angle} holds.
\end{proof}

We are now ready to prove Proposition~\ref{prp:lower-bound:improved-planar}.

\begin{proof}[Proof of Proposition~\ref{prp:lower-bound:improved-planar}]
By Theorem~\ref{thm:measure-equivalence}, it suffices to prove the claimed bounds for an arbitrary pair $(\mu,\nu)\in\cP^d_{bc}$. To prove the proposition, it suffices to show that there exists an affine hyperplane $H \in \Hh^d$ satisfying
\[
    C(H) \geq \frac{8}{8 + 5 \pi} L.
\]
Let $X \sim \mu$ and $Y \sim \nu$ be independent.
Translate $\R^d$ so that $p := \E[X] = -\E[Y]$.
Denote
\[
    M\coloneq \max_H C(H).
\]
The maximum exists: if both supports lie in a ball of radius $R$ centered at the origin, it suffices to consider hyperplanes $H=\{x:n\cdot x=t\}$ with $(n,t)\in S^{d-1}\times[-R,R]$. Absolute continuity and dominated convergence imply that $C(H)$ is continuous in these parameters, and the parameter space is compact.

The proof combines three types of hyperplanes. A hyperplane bisecting both measures will give an upper bound on $L$ in terms of the average distance of $X$ and $Y$ from a suitable center. We then bound this average distance by considering random hyperplanes through the midpoint of the two means and random hyperplanes through the two individual means. The first estimate is strongest when $X$ and $Y$ tend to point in different directions from the center, while the second is strongest when they tend to point in similar directions.

We begin with the bisection argument. By the signed ham-sandwich theorem (Lemma~\ref{lem:unified-signed-ham}), there exists a hyperplane $H_0$ that bisects both $\mu$ and $\nu$. For any $z\in\R^d$, let
\[
    S_z \coloneqq \E[\|X-z\|] + \E[\|Y-z\|].
\]
Let $A$ be the event that $X$ and $Y$ lie on the same side of $H_0$. The pairs in $A$ are precisely those not cut by $H_0$, and hence by the triangle inequality,
\[
\begin{aligned}
    L-C(H_0)
    &=\E\left[\|X-Y\|\mathbf 1_A\right] 
    \leq
    \E\left[\|X-z\|\mathbf 1_A\right]
    +
    \E\left[\|Y-z\|\mathbf 1_A\right].
\end{aligned}
\]
Since $H_0$ bisects both measures and $X$ and $Y$ are independent, conditioning on $X$ gives
\[
    \E\left[\|X-z\| \cdot \mathbf 1_A\right] = \E\left[ \|X-z\| \cdot \E\brk{\mathbf 1_A\midd X} \right] = \frac{1}{2} \E[\|X-z\|].
\]
Similarly, we have $\E\left[\|Y-z\| \cdot \mathbf 1_A\right] = \E[\|Y-z\|] / 2$. Thus $L - C(H_0) \leq S_z / 2$. Since $C(H_0) \leq M$, we obtain
\begin{equation}\label{warmup:bisector}
    L \leq M + \frac{1}{2} S_z.
\end{equation}
It therefore remains to control $S_z$ in terms of $M$.
Choose $z=0$.
Let $X = au$ and $Y = bv$, where $a = \|X\|$, $b = \|Y\|$, and $u, v$ are unit vectors. Absolute continuity ensures that $X$ and $Y$ avoid the origin and both means almost surely, so all directions used below are well-defined almost surely. Thus
\begin{equation}
\label{equ:S0}
    S_0 = \E[a + b].
\end{equation}
To keep track of whether the two directions tend to agree, let
\begin{equation}\label{eq:direction-agreement}
    T \coloneqq \E[(a+b) (u \cdot v)].
\end{equation}

We first consider a random hyperplane through the origin,
\[
    H_N=\{x\in\R^d:N\cdot x=0\},
\]
where $N$ is uniform on $S^{d-1}$ and independent of $X,Y$. For fixed $X$ and $Y$, let $\theta\in[0,\pi]$ denote the angle between $u$ and $v$. The probability that $H_N$ separates them is $\theta/\pi$, independently of $d$. To see this, write $N=G/\|G\|$ for a standard Gaussian vector $G\in\R^d$. When $0<\theta<\pi$, the projection of $G$ onto $\operatorname{span}\{u,v\}$ is a rotationally symmetric two-dimensional Gaussian. Its direction is uniform, and the directions giving opposite signs to $G\cdot u$ and $G\cdot v$ form two sectors with total angle $2\theta$. The cases $\theta=0$ and $\theta=\pi$ follow directly. Averaging over these hyperplanes gives
\begin{equation}\label{eq:random-angle}
    M \geq \frac{1}{\pi} \E[\|X-Y\| \cdot \theta].
\end{equation}
We next lower-bound the contribution of a fixed pair in terms of its radial distances and the angle between its directions. By the cosine rule, we have
\[
    \|X-Y\|^2 = a^2 + b^2 - 2ab \cos\theta.
\]
Hence,
\begin{align*}
    \|X-Y\|^2-(a+b)^2\sin^2(\theta/2) &= (a^2+b^2)\cos^2(\theta/2) - 2ab\cos(\theta) - 2ab\sin^2(\theta/2) \\
    &= (a-b)^2\cos^2(\theta/2) \geq 0.
\end{align*}
Therefore,
\[
    \|X-Y\| \geq (a+b)\sin(\theta/2).
\]
Since $\theta \geq 2\sin(\theta/2)$,
we obtain
\[
    \|X-Y\| \cdot \theta \geq 2(a + b)\sin^2(\theta/2) = (a + b)(1 - \cos\theta).
\]
Since $\cos\theta = u \cdot v$, and by \eqref{eq:random-angle} and \eqref{eq:direction-agreement}, it follows that
\begin{equation}\label{warmup:midpoint}
    S_0 - T \leq \pi M.
\end{equation}

This inequality is loose when $u$ and $v$ frequently point in similar directions. To handle this case, we instead average over hyperplanes through the individual means of the two measures. 
Consider first a random hyperplane through the red mean $p$, with normal uniform on $S^{d-1}$, and let $\varphi$ be the angle between $X - p$ and $Y - p$. Averaging over its orientation and applying
Lemma~\ref{lem:pointwise-mean} gives
\[
    \pi M \geq \E[\|X-Y\| \cdot \varphi] \geq \frac{4}{3} \E\left[\|X-p\|(1-\cos\varphi) \right].
\]
We argue that the cosine term vanishes. Indeed,
\[
    \|X-p\|\cos\varphi = (X-p)\cdot\frac{Y-p}{\|Y-p\|},
\]
and independence together with $\E[X-p] = 0$ yields
\[
    \E\left[(X-p)\cdot\frac{Y-p}{\|Y-p\|}\right] = \E[X-p] \cdot \E\left[\frac{Y-p}{\|Y-p\|}\right] = 0.
\]
Hence,
$M \geq \frac{4}{3 \pi}\E[\|X-p\|]$.
Exchanging the roles of $X$ and $Y$ and averaging over hyperplanes through the blue mean $-p$ yields
$M \geq \frac{4}{3\pi}\E[\|Y+p\|]$.
Adding these two inequalities, we obtain
\begin{equation}\label{warmup:means-pre}
    \frac{3\pi}{2} M \geq \E[\|X-p\|] + \E[\|Y+p\|].
\end{equation}
We next relate the right-hand side to $S_0$ and $T$. By the supporting-hyperplane inequality for the Euclidean norm, we have
\[
    \|X - p\| \geq \|X\| + \frac{\left\langle X, (X - p) - X \right\rangle}{\|X\|}, \qquad \|Y + p\| \geq \|Y\| + \frac{\left\langle Y, (Y + p) - Y \right\rangle}{\|Y\|},
\]
and thus
$\|X - p\| \geq a - p \cdot u$.
Similarly,
$\|Y + p\| \geq b + p \cdot v$.
Therefore,
\begin{equation}
\label{equ:15.2}
    \E[\|X - p\|] + \E[\|Y + p\|] \geq \E[a + b + p \cdot (v - u)].
\end{equation}
By \eqref{eq:direction-agreement}, since $X = a u$ and $Y = b v$, 
and since $X$ and $Y$ are independent,
\begin{equation}
\label{equ:15.1}
    T = \E[(a+b) u \cdot v] = \E[X \cdot v] + \E[Y \cdot u] = p \cdot (\E[v] - \E[u]).
\end{equation}
By \eqref{warmup:means-pre}, \eqref{equ:15.2}, \eqref{equ:S0}, and \eqref{equ:15.1},
\begin{equation}\label{warmup:means}
    \frac{3\pi}{2}M \geq \E[\|X - p\|] + \E[\|Y + p\|] \geq S_0 + T.
\end{equation}

The two estimates \eqref{warmup:midpoint} and \eqref{warmup:means} are complementary: 
the former is stronger when $T$ is negative (the two directions tend to disagree), 
while the latter is stronger when $T$ is positive (the two directions tend to agree). 
Adding them eliminates the directional term $T$, yielding
$\frac{5\pi}{2}M \geq 2S_0$,
so that $S_0 \leq \frac{5 \pi}{4} M$. 
Substituting this into \eqref{warmup:bisector} gives
\[
    L \leq M + \frac{1}{2} S_0 \leq \left(1 + \frac{5\pi}{8}\right) M.
\]
Hence $M/L \geq 8/(8 + 5 \pi)$.
\end{proof}

\subsection{Upper bound}
\label{sec:improved-upper-bound}

Next, we turn to the planar upper bound.

\begin{proposition}\label{prp:upper-bound:improved-planar}
For the Euclidean distance,
\[
    \alpha(\cP^2) \leq \frac{5}{11}.
\]
\end{proposition}
\begin{proof}
To prove the proposition, it suffices to show that, for every $\eta > 0$, there exists a weighted affine general position configuration of red and blue points in $\R^2$ such that every line $H \in \Hh^2$ satisfies
\[
    C(H) \leq \prn{\frac{5}{11} + \eta} L.
\]
For $0 < \eps < 1/4$, consider the following unperturbed instance, which is displayed in Figure~\ref{fig:planar-improved-upper}:
\begin{enumerate}
    \item[(I)] There is a red point of mass $\eps$ at position $(-4/(3\eps), 0)$ and a blue point of mass $\eps$ at position $(4/(3 \eps), 0)$.
    \item[(II)] There is a red point of mass $1/2 - \eps / 3$ at position $(1, -\eps)$ and a blue point of mass $1/2 - \eps / 3$ at position $(-1, -\eps)$.
    \item[(III)] There is a red point of mass $1/3 - \eps / 3$ at position $(\eps, 0)$ and a blue point of mass $1/3 - \eps / 3$ at position $(-\eps, 0)$.
    \item[(IV)] There is a red point of mass $1/6 - \eps / 3$ at position $(-1, \sqrt{\eps})$ and a blue point of mass $1/6 - \eps / 3$ at position $(1, \sqrt{\eps})$.
\end{enumerate}

\begin{figure}
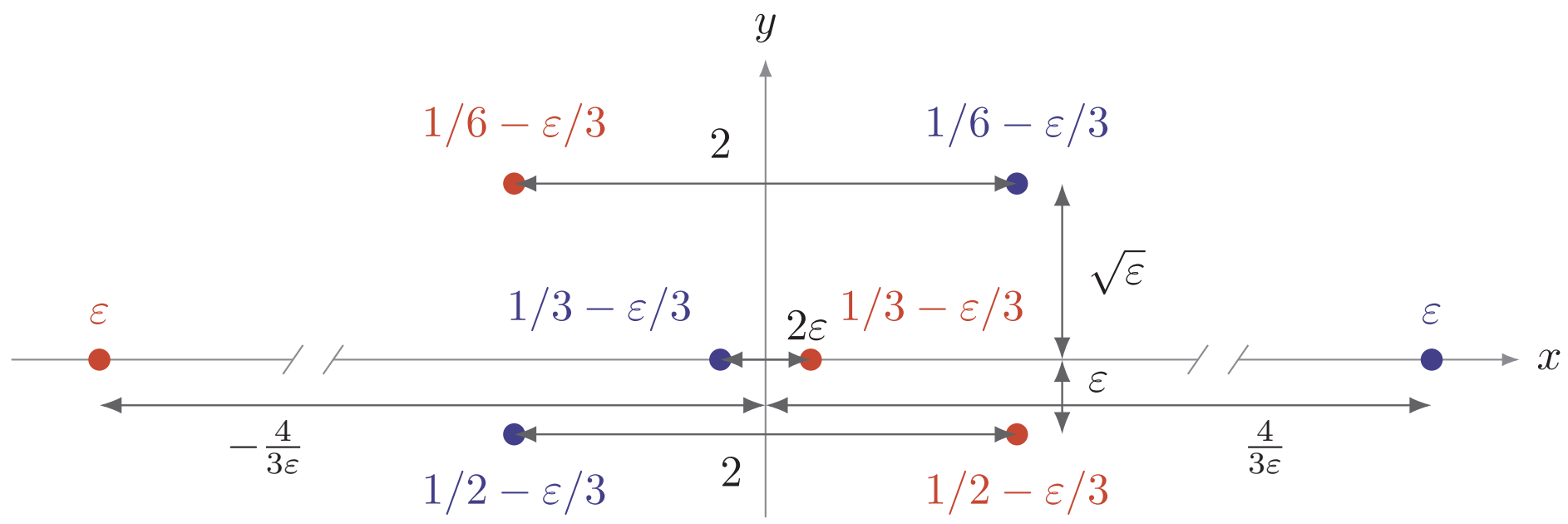

    \centering
    \planarinstance
    \caption{A planar instance in which the optimal cut ratio tends to $5/11$.}
    \label{fig:planar-improved-upper}
\end{figure}
We first compute $L$. Ignoring lower order terms, the contribution of the points in (I) is $2 \eps \cdot (4/(3 \eps)) \cdot 1 = 8/3$. The remaining contribution of the intervals with at least one point in (II) is approximately
\[
    \frac{1}{2} \cdot \frac{1}{2} \cdot 2 + 2 \cdot \frac{1}{2} \cdot \prn{\frac{1}{3} \cdot 1 + \frac{1}{6} \cdot \prn{\sqrt{\eps} + \eps}} \to \frac{5}{6}.
\]
The remaining contribution of the intervals with at least one point in (III) is approximately
\[
    \frac{1}{3} \cdot \frac{1}{3} \cdot (2 \eps) + 2 \cdot \frac{1}{3} \cdot \frac{1}{6} \cdot 1 \to \frac{1}{9}.
\]
Finally, the remaining contribution of the intervals formed by points in (IV) is approximately
\[
    \frac{1}{6} \cdot \frac{1}{6} \cdot 2 = \frac{1}{18}.
\]
Summing up all these contributions yields
\[
    L=\frac{8}{3}+\frac{5}{6}+\frac{1}{9}+\frac{1}{18}+o(1)=\frac{11}{3}+o(1),\qquad \eps\downarrow0.
\]

Next, label the red points $r_1, r_2, r_3, r_4$ and the blue points $b_1, b_2, b_3, b_4$, where $r_1, b_1$ correspond to the points in (I), 
$r_2, b_2$ correspond to the points in (II), and so on. The four points in (I) and (III) are collinear, so the displayed instance is not yet in affine general position. For fixed $\eps$, replace each of these four points $(x,0)$ by $(x,-\delta x^2)$, leaving the other four points unchanged. Choose $\delta>0$ sufficiently small and outside the finitely many values that create a collinear triple. Such a choice exists since triples involving at least one moved point have determinants that are non-zero polynomials in $\delta$, and triples involving only unmoved points are already non-collinear. Take $\delta$ small enough that all strict cut patterns of the unperturbed instance persist and the changes in $L$ and in the weight of each red-blue segment tend to zero. The limits below are understood with $\delta\downarrow0$ first and then $\eps\downarrow0$.

This perturbation does not introduce a larger limiting cut value since for fixed $\eps$, any new cut pattern has a sequence of realizing lines converging to a line through some original points. If that limiting line contains at most two points, any prescribed sides for those points can be achieved by a small perturbation of the line itself; hence the pattern was already a strict cut pattern of the unperturbed instance. The only remaining limiting line is the $x$-axis. Orient its positive side upwards. The points in (II) are then negative and those in (IV) positive. On the four moved points, a nearby line tests an expression
\[
    \alpha x-\beta\delta x^2-t,\qquad \beta>0.
\]
This is concave in $x$, so the moved points on the positive side form a consecutive block in the order $r_1,b_3,r_3,b_1$. For the empty block and the ten non-empty consecutive blocks, the limiting cut values are, respectively,
\[
\begin{array}{c|c@{\qquad}c|c}
\text{block}&\text{value}&\text{block}&\text{value}\\ \hline
\varnothing&5/9&\{r_1\}&13/9\\
\{b_3\}&10/9&\{r_3\}&10/9\\
\{b_1\}&13/9&\{r_1,b_3\}&10/9\\
\{b_3,r_3\}&5/3&\{r_3,b_1\}&10/9\\
\{r_1,b_3,r_3\}&5/3&\{b_3,r_3,b_1\}&5/3\\
\{r_1,b_3,r_3,b_1\}&5/3&&
\end{array}
\]
In particular, all of them are at most $5/3$. For sufficiently small $\delta$, there are 28 nontrivial strict cut patterns, identified up to exchanging sides. Representatives $H_i$, their positive-side sets $S(H_i)$, and their crossed segments are listed below. The last column gives the limiting value of $C(H_i)$.

\begingroup
\small
\setlength{\tabcolsep}{4pt}
\renewcommand{\arraystretch}{1.12}
\begin{center}
\begin{tabular}{@{}cclc@{}}
\hline
$H$ & $S(H)$ & Intervals crossed by $H$ & $\lim C(H)$\\
\hline
$H_1$ & $\{r_1\}$
& $(r_1,b_1),\ (r_1,b_2),\ (r_1,b_3),\ (r_1,b_4)$
& $4/3$\\[3pt]
$H_2$ & $\{r_2\}$
& $(r_2,b_1),\ (r_2,b_2),\ (r_2,b_3),\ (r_2,b_4)$
& $4/3$\\[3pt]
$H_3$ & $\{r_4\}$
& $(r_4,b_1),\ (r_4,b_2),\ (r_4,b_3),\ (r_4,b_4)$
& $1/3$\\[3pt]
$H_4$ & $\{b_1\}$
& $(r_1,b_1),\ (r_2,b_1),\ (r_3,b_1),\ (r_4,b_1)$
& $4/3$\\[3pt]
$H_5$ & $\{b_2\}$
& $(r_1,b_2),\ (r_2,b_2),\ (r_3,b_2),\ (r_4,b_2)$
& $4/3$\\[3pt]
$H_6$ & $\{b_4\}$
& $(r_1,b_4),\ (r_2,b_4),\ (r_3,b_4),\ (r_4,b_4)$
& $1/3$\\[3pt]
$H_7$ & $\{r_1,r_4\}$
& \begin{tabular}[c]{@{}l@{}}
  $(r_1,b_1),\ (r_1,b_2),\ (r_1,b_3),\ (r_1,b_4),$\\
  $(r_4,b_1),\ (r_4,b_2),\ (r_4,b_3),\ (r_4,b_4)$
  \end{tabular}
& $5/3$\\[3pt]
\hline
\end{tabular}
\end{center}

\begin{center}
\textit{All strict line cuts (continued).}\par\smallskip
\begin{tabular}{@{}cclc@{}}
\hline
$H$ & $S(H)$ & Intervals crossed by $H$ & $\lim C(H)$\\
\hline
$H_8$ & $\{r_1,b_2\}$
& \begin{tabular}[c]{@{}l@{}}
  $(r_1,b_1),\ (r_1,b_3),\ (r_1,b_4),$\\
  $(r_2,b_2),\ (r_3,b_2),\ (r_4,b_2)$
  \end{tabular}
& $4/3$\\[3pt]
$H_9$ & $\{r_2,b_1\}$
& \begin{tabular}[c]{@{}l@{}}
  $(r_1,b_1),\ (r_2,b_2),\ (r_2,b_3),$\\
  $(r_2,b_4),\ (r_3,b_1),\ (r_4,b_1)$
  \end{tabular}
& $4/3$\\[3pt]
$H_{10}$ & $\{r_2,b_2\}$
& \begin{tabular}[c]{@{}l@{}}
  $(r_1,b_2),\ (r_2,b_1),\ (r_2,b_3),$\\
  $(r_2,b_4),\ (r_3,b_2),\ (r_4,b_2)$
  \end{tabular}
& $5/3$\\[3pt]
$H_{11}$ & $\{r_4,b_4\}$
& \begin{tabular}[c]{@{}l@{}}
  $(r_1,b_4),\ (r_2,b_4),\ (r_3,b_4),$\\
  $(r_4,b_1),\ (r_4,b_2),\ (r_4,b_3)$
  \end{tabular}
& $5/9$\\[3pt]
$H_{12}$ & $\{b_1,b_4\}$
& \begin{tabular}[c]{@{}l@{}}
  $(r_1,b_1),\ (r_1,b_4),\ (r_2,b_1),\ (r_2,b_4),$\\
  $(r_3,b_1),\ (r_3,b_4),\ (r_4,b_1),\ (r_4,b_4)$
  \end{tabular}
& $5/3$\\[3pt]
$H_{13}$ & $\{r_1,r_2,b_2\}$
& \begin{tabular}[c]{@{}l@{}}
  $(r_1,b_1),\ (r_1,b_3),\ (r_1,b_4),\ (r_2,b_1),$\\
  $(r_2,b_3),\ (r_2,b_4),\ (r_3,b_2),\ (r_4,b_2)$
  \end{tabular}
& $5/3$\\[3pt]
$H_{14}$ & $\{r_1,r_4,b_2\}$
& \begin{tabular}[c]{@{}l@{}}
  $(r_1,b_1),\ (r_1,b_3),\ (r_1,b_4),\ (r_2,b_2),$\\
  $(r_3,b_2),\ (r_4,b_1),\ (r_4,b_3),\ (r_4,b_4)$
  \end{tabular}
& $5/3$\\[3pt]
$H_{15}$ & $\{r_1,r_4,b_4\}$
& \begin{tabular}[c]{@{}l@{}}
  $(r_1,b_1),\ (r_1,b_2),\ (r_1,b_3),\ (r_2,b_4),$\\
  $(r_3,b_4),\ (r_4,b_1),\ (r_4,b_2),\ (r_4,b_3)$
  \end{tabular}
& $13/9$\\[3pt]
$H_{16}$ & $\{r_1,b_2,b_3\}$
& \begin{tabular}[c]{@{}l@{}}
  $(r_1,b_1),\ (r_1,b_4),\ (r_2,b_2),\ (r_2,b_3),$\\
  $(r_3,b_2),\ (r_3,b_3),\ (r_4,b_2),\ (r_4,b_3)$
  \end{tabular}
& $10/9$\\[3pt]
$H_{17}$ & $\{r_2,r_3,b_1\}$
& \begin{tabular}[c]{@{}l@{}}
  $(r_1,b_1),\ (r_2,b_2),\ (r_2,b_3),\ (r_2,b_4),$\\
  $(r_3,b_2),\ (r_3,b_3),\ (r_3,b_4),\ (r_4,b_1)$
  \end{tabular}
& $10/9$\\[3pt]
$H_{18}$ & $\{r_2,b_1,b_2\}$
& \begin{tabular}[c]{@{}l@{}}
  $(r_1,b_1),\ (r_1,b_2),\ (r_2,b_3),\ (r_2,b_4),$\\
  $(r_3,b_1),\ (r_3,b_2),\ (r_4,b_1),\ (r_4,b_2)$
  \end{tabular}
& $5/3$\\[3pt]
$H_{19}$ & $\{r_2,b_1,b_4\}$
& \begin{tabular}[c]{@{}l@{}}
  $(r_1,b_1),\ (r_1,b_4),\ (r_2,b_2),\ (r_2,b_3),$\\
  $(r_3,b_1),\ (r_3,b_4),\ (r_4,b_1),\ (r_4,b_4)$
  \end{tabular}
& $5/3$\\[3pt]
$H_{20}$ & $\{r_3,r_4,b_4\}$
& \begin{tabular}[c]{@{}l@{}}
  $(r_1,b_4),\ (r_2,b_4),\ (r_3,b_1),\ (r_3,b_2),$\\
  $(r_3,b_3),\ (r_4,b_1),\ (r_4,b_2),\ (r_4,b_3)$
  \end{tabular}
& $10/9$\\[3pt]
$H_{21}$ & $\{r_4,b_1,b_4\}$
& \begin{tabular}[c]{@{}l@{}}
  $(r_1,b_1),\ (r_1,b_4),\ (r_2,b_1),\ (r_2,b_4),$\\
  $(r_3,b_1),\ (r_3,b_4),\ (r_4,b_2),\ (r_4,b_3)$
  \end{tabular}
& $13/9$\\[3pt]
$H_{22}$ & $\{r_4,b_3,b_4\}$
& \begin{tabular}[c]{@{}l@{}}
  $(r_1,b_3),\ (r_1,b_4),\ (r_2,b_3),\ (r_2,b_4),$\\
  $(r_3,b_3),\ (r_3,b_4),\ (r_4,b_1),\ (r_4,b_2)$
  \end{tabular}
& $10/9$\\[3pt]
$H_{23}$ & $\{r_1,r_3,b_2,b_3\}$
& \begin{tabular}[c]{@{}l@{}}
  $(r_1,b_1),\ (r_1,b_4),\ (r_2,b_2),\ (r_2,b_3),$\\
  $(r_3,b_1),\ (r_3,b_4),\ (r_4,b_2),\ (r_4,b_3)$
  \end{tabular}
& $13/9$\\[3pt]
$H_{24}$ & $\{r_1,r_4,b_2,b_3\}$
& \begin{tabular}[c]{@{}l@{}}
  $(r_1,b_1),\ (r_1,b_4),\ (r_2,b_2),\ (r_2,b_3),$\\
  $(r_3,b_2),\ (r_3,b_3),\ (r_4,b_1),\ (r_4,b_4)$
  \end{tabular}
& $4/3$\\[3pt]
$H_{25}$ & $\{r_1,r_4,b_2,b_4\}$
& \begin{tabular}[c]{@{}l@{}}
  $(r_1,b_1),\ (r_1,b_3),\ (r_2,b_2),\ (r_2,b_4),$\\
  $(r_3,b_2),\ (r_3,b_4),\ (r_4,b_1),\ (r_4,b_3)$
  \end{tabular}
& $13/9$\\[3pt]
$H_{26}$ & $\{r_1,r_4,b_3,b_4\}$
& \begin{tabular}[c]{@{}l@{}}
  $(r_1,b_1),\ (r_1,b_2),\ (r_2,b_3),\ (r_2,b_4),$\\
  $(r_3,b_3),\ (r_3,b_4),\ (r_4,b_1),\ (r_4,b_2)$
  \end{tabular}
& $10/9$\\[3pt]
$H_{27}$ & $\{r_3,r_4,b_1,b_4\}$
& \begin{tabular}[c]{@{}l@{}}
  $(r_1,b_1),\ (r_1,b_4),\ (r_2,b_1),\ (r_2,b_4),$\\
  $(r_3,b_2),\ (r_3,b_3),\ (r_4,b_2),\ (r_4,b_3)$
  \end{tabular}
& $10/9$\\[3pt]
$H_{28}$ & $\{r_3,r_4,b_3,b_4\}$
& \begin{tabular}[c]{@{}l@{}}
  $(r_1,b_3),\ (r_1,b_4),\ (r_2,b_3),\ (r_2,b_4),$\\
  $(r_3,b_1),\ (r_3,b_2),\ (r_4,b_1),\ (r_4,b_2)$
  \end{tabular}
& $5/3$\\[3pt]
\hline
\end{tabular}
\end{center}
\endgroup
The largest limiting value in the table is $5/3$. Since there are finitely many cut patterns, and a line passing through points can be shifted slightly to preserve all pairs it strictly cuts, the same upper estimate applies to every line. Thus
\[
    \max_H C(H)\leq\frac53+o(1),\qquad
    \max_H\frac{C(H)}{L}\leq\frac{5/3+o(1)}{11/3+o(1)}=\frac5{11}+o(1).
\]
Choosing $\eps$ sufficiently small and then $\delta$ sufficiently small proves the claim. \qedhere
\end{proof}
\section{Measure Equivalence}
\label{sec:measure-equivalence}

In this section, we prove Theorem~\ref{thm:measure-equivalence} which states that the families of measures introduced in Section~\ref{subsec:ratio-families} have the same optimal cut ratio. This allows us to transfer bounds between general measures, bounded absolutely continuous measures, and finite equal-weight distributions in affine general position.

Together with Theorem~\ref{thm:measure-equivalence}, Theorems~\ref{thm:lower-bound:dimension-one}, \ref{thm:lower-bound:dimension-three-or-above},~\ref{thm:lower-bound:dimension-two}, and~\ref{thm:intro:upper-bound} and Proposition~\ref{prp:upper-bound:dim-1} give $\alpha(\cP^1) = 1/3$ and $1/3 \leq \alpha(\cP^d) \leq 1/2$ for every $d \geq 2$. Also, Propositions~\ref{prp:lower-bound:improved-planar} and~\ref{prp:upper-bound:improved-planar} give $\alpha(\cP^d) \geq 8/(8+5\pi)$ and $\alpha(\cP^2) \leq 5/11$ for the Euclidean distance. The following result states that these bounds apply also to the families of distributions $\cP_b^d$, $\cP_{bc}^d$, and $\cP_\N^d$.

\begin{reptheorem}{thm:measure-equivalence}
For every $d \geq 1$ and every norm on $\R^d$,
\[
    \alpha(\cP^d) = \alpha(\cP_b^d) = \alpha(\cP_{bc}^d) = \alpha(\cP_\N^d).
\]
\end{reptheorem}

Fix a norm $\| \cdot \|$ on $\R^d$. Throughout this section, $\|\cdot\|_2$ and $\mathbb S^{d-1}$ denote the Euclidean norm and the Euclidean unit sphere, respectively.
Multiplying the norm by a positive constant leaves every cut ratio
unchanged, so we may assume $\|z\| \leq \|z\|_2$ for all $z$. In this section, the distances in balls $B(x, r)$, strips, and diameters are Euclidean. Since $\cP_{bc}^d \subseteq \cP_b^d \subseteq \cP^d$ and $\cP_\N^d \subseteq \cP^d$, we have
\begin{align*}
    \alpha(\cP^d) \leq \alpha(\cP_b^d) \leq \alpha(\cP_{bc}^d),
    \hbox{ and } 
    \alpha(\cP^d) \leq \alpha(\cP_\N^d).
\end{align*}

The rest of the section proves the reverse inequalities.
Lemma~\ref{lem:bounded} proves 
$\alpha(\cP^d) \geq \alpha(\cP_b^d)$,
Lemma~\ref{lem:continuous} proves 
$\alpha(\cP^d_b) \geq \alpha(\cP_{bc}^d)$,
and Lemma~\ref{lem:finite} proves 
$\alpha(\cP_{bc}^d) \geq \alpha(\cP_\N^d)$.

To prove that $\alpha(\cP^d) \geq \alpha(\cP_b^d)$, we use the fact that the distributions have finite first moments to truncate them to a sufficiently large bounded set, relocating the discarded mass inside that set. We show that, if the relocated mass is sufficiently small, 
neither $L$ nor $C(H)$ changes by much, uniformly over all hyperplanes $H$.

To prove that $\alpha(\cP_b^d) \geq \alpha(\cP_{bc}^d)$, we approximate distributions $(\mu,\nu)$ with bounded support by absolutely continuous distributions $(\mu',\nu')$, by adding to each random variable an independent random variable that is uniformly distributed on a small ball around $\vec 0$. 
This operation changes $L$ by only a small amount. Unfortunately, $C(H)$ might change significantly, because a hyperplane that does not pass through an atom of $\mu$ or $\nu$ may intersect one of the small balls created around that atom. We use the affine general position assumption to show that such a hyperplane can be slightly perturbed so as to eliminate these intersections one at a time, without creating new ones and while changing $C$ by only a small amount. After finitely many such perturbations we obtain a hyperplane that avoids all those small atomic balls. This allows us to compare the optimal crossing ratios for $(\mu,\nu)$ and $(\mu',\nu')$.

To prove that $\alpha(\cP_{bc}^d)\ge\alpha(\cP_\N^d)$, we approximate any absolutely continuous distribution with bounded support by an atomic distribution all of whose atoms have the same weight. We partition the support into sufficiently small sets and place in each set a number of atoms approximately proportional to the probability assigned to that set. By choosing the partition sufficiently fine and the number of atoms sufficiently large, we show that this operation changes neither $L$ nor $C(H)$ significantly, uniformly over all hyperplanes $H$.

\subsection{Reduction to bounded support}

\begin{lemma}\label{lem:bounded}
$\alpha(\cP^d) \geq \alpha(\cP_b^d)$.
\end{lemma}

\begin{proof}
Fix $\varepsilon\in(0,1/2)$.  Choose $(\mu,\nu) \in \cP^d$ such that
\begin{equation}\label{eq:bounded-choice}
\alpha(\cP^d) \geq \sup_{H\in\Hh^d}\frac{C_{\mu,\nu}(H)}{L_{\mu,\nu}}-\varepsilon,
\end{equation}
and
\[
0<\delta<\frac{\varepsilon L_{\mu,\nu}}{14}.
\]
The idea is to move the small amount of mass far away from the origin
into a large bounded ball, while changing both $L_{\mu,\nu}$ and
$C_{\mu,\nu}(H)$ only by a small amount.
Let $X\sim \mu$ and $Y\sim \nu$ be independent.  Since $\mu$ and $\nu$ have finite
first moments, there exists $M>0$ such that
\[
\E\bigl[\|X\|\1_{\{\|X\|>M\}}\bigr]<\delta,
\qquad
\E\bigl[\|Y\|\1_{\{\|Y\|>M\}}\bigr]<\delta,
\]
and 
\[
M>\E\bigl[\|X\|\bigr],
\qquad
M>\E\bigl[\|Y\|\bigr].
\]
Consequently,
\begin{equation}
\label{equ:bound:5}
\begin{split}
\E\bigl[\|Y\|\bigr]\,\Pr[\|X\|>M]
&\leq
M \Pr[\|X\|>M]
\le\E\bigl[\|X\|\1_{\{\|X\|>M\}}\bigr]
<\delta,\\
\E\bigl[\|X\|\bigr]\,\Pr[\|Y\|>M]
&
\leq M \Pr[\|Y\|>M]
\le\E\bigl[\|Y\|\1_{\{\|Y\|>M\}}\bigr]
<\delta.
\end{split}
\end{equation}

Let
\[
\cB_M\coloneqq \{x:\|x\|\le M\},
\]
and $U_M$ be the uniform probability measure on $\cB_M$. 
Let $\mu_{|\cB_M}$ 
(resp., $\nu_{|\cB_M}$) be the restricted (not necessarily probability) measure of $\mu$ (resp., $\nu$) to $\cB_M$.
Define
\[
\mu'\coloneqq \mu_{|\cB_M}+\mu(\cB_M^c)U_M,
\qquad
\nu'\coloneqq \nu_{|\cB_M}+\nu(\cB_M^c)U_M,
\]
where $\cB_M^c$ is the complement of $\cB_M$.
The atoms of $\mu'$ and $\nu'$ are the atoms of $\mu$ and $\nu$ that lie
in $\cB_M$.
Hence, the points in $\atom(\mu') \cup \atom(\nu')$ remain in affine general position.  
The non-atomic
part of $\mu'$ (resp., $\nu'$) is the restriction of the original non-atomic part of $\mu$ (resp., $\nu$)
plus a multiple of
$U_M$.
Since $U_M$ assigns zero mass to every hyperplane, we have
$\mu'_{\na}(H) = 0$ (resp., $\nu'_{\na}(H) = 0$) for every hyperplane $H \in \mathbb H^d$.
Thus $(\mu',\nu')\in\cP_b^d$.

Choose two couplings for $(X,X')$ and $(Y,Y')$ independently.
For $(X,X')$, leave $X$ unchanged on $\{\|X\|\le M\}$ and replace it by an independent point with law $U_M$ on $\{\|X\|>M\}$.
Then
\[
\E[\|X-X'\|]
\le
\E[\|X\|\1_{\{\|X\|>M\}}]+M \Pr[\|X\|>M]
<2\delta.
\]
Hence $W_1(\mu,\mu')<2\delta$.
Construct $(Y,Y')$ analogously. Then, $W_1(\nu,\nu')<2\delta$.
In particular, $X$ and $Y$ are independent with laws $\mu$ and $\nu$,
and $X'$ and $Y'$ are independent with laws $\mu'$ and $\nu'$.
By Lemma~\ref{lem:L-cont},
\begin{equation}\label{eq:bounded-L}
|L_{\mu,\nu}-L_{\mu',\nu'}|<4\delta.
\end{equation}

For any
$H\in\Hh^d$, the two pairs $(X,Y)$ and $(X',Y')$ coincide on
$\{\|X\|\le M,\|Y\|\le M\}$.
Hence a difference in their contributions
to $C(H)$ can occur only if at least one of $X$ and $Y$ lies outside
$\cB_M$.
Therefore,
\begin{align}
\nonumber
|C_{\mu,\nu}(H)-&C_{\mu',\nu'}(H)|
\le
\E\!\left[
(\|X-Y\|+\|X'-Y'\|)
\1_{\{\|X\|>M\}\cup\{\|Y\|>M\}}
\right] \\
&\le
\E\!\left[(\|X\|+\|Y\|+2M)\1_{\{\|X\|>M\}}\right]+
\E\!\left[(\|X\|+\|Y\|+2M)\1_{\{\|Y\|>M\}}\right]. 
\label{equ:diff:c}
\end{align}
By independence and \eqref{equ:bound:5},
\begin{align*}
    \E\!\left[(\|X\|+\|Y\|+2M) \1_{\{\|X\|>M\}}\right]
    &= \E\!\left[\|X\|\1_{\{\|X\|>M\}}\right] + \E[\|Y\|]\Pr[\|X\|>M] \\
    &\qquad + 2M \Pr[\|X\|>M] 
    < 4 \delta.
\end{align*}
The analogous bound 
for the second term in \eqref{equ:diff:c} holds with $X$ and $Y$ interchanged.  Hence,
\begin{equation}
\label{equ:bound:6}
|C_{\mu,\nu}(H)-C_{\mu',\nu'}(H)|<8\delta.
\end{equation}

Since $L_{\mu,\nu} > 0$ and $L_{\mu',\nu'} + 4\delta > 0$, \eqref{eq:bounded-L} and~\eqref{equ:bound:6} imply that, for every $H$,
\[
\frac{C_{\mu,\nu}(H)}{L_{\mu,\nu}}
\ge
\frac{C_{\mu',\nu'}(H)-8\delta}{L_{\mu',\nu'}+4\delta}.
\]

Since $C_{\mu',\nu'}(H)\le L_{\mu',\nu'}$,
\begin{align*}
\frac{C_{\mu',\nu'}(H)}{L_{\mu',\nu'}}
-
\frac{C_{\mu',\nu'}(H)-8\delta}{L_{\mu',\nu'}+4\delta}
&=
\frac{4\delta C_{\mu',\nu'}(H)+8\delta L_{\mu',\nu'}}{L_{\mu',\nu'}(L_{\mu',\nu'}+4\delta)}\le
\frac{12\delta}{L_{\mu',\nu'}+4\delta}
\le
\frac{12\delta}{L_{\mu',\nu'}}.
\end{align*}
Therefore
\[
\frac{C_{\mu,\nu}(H)}{L_{\mu,\nu}}
\ge
\frac{C_{\mu',\nu'}(H)}{L_{\mu',\nu'}}
-
\frac{12\delta}{L_{\mu',\nu'}}.
\]

Since $\varepsilon<1/2$ and $\delta<\frac{\varepsilon L_{\mu,\nu}}{14}$, we have $4\delta<\frac17L_{\mu,\nu}$.
Hence, by \eqref{eq:bounded-L},
\[
L_{\mu',\nu'}\ge L_{\mu,\nu}-4\delta>\frac67L_{\mu,\nu},
\]
and the choice of $\delta$ gives
\[
\frac{12\delta}{L_{\mu',\nu'}}<
\frac{12(\varepsilon L_{\mu,\nu}/14)}{6L_{\mu,\nu}/7}
=\varepsilon.
\]
Thus, we obtain
\[
\sup_{H \in \mathbb H^d}\frac{C_{\mu,\nu}(H)}{L_{\mu,\nu}}
\ge
\sup_{H \in \mathbb H^d}\frac{C_{\mu',\nu'}(H)}{L_{\mu',\nu'}}-\varepsilon.
\]
Since $(\mu',\nu')\in\cP_b^d$, the definition of $\alpha(\cP_b^d)$ gives
\[
\sup_{H \in \mathbb H^d}
\frac{C_{\mu',\nu'}(H)}{L_{\mu',\nu'}}
\ge
\alpha(\cP_b^d).
\]
Together with \eqref{eq:bounded-choice},
\[
\alpha(\cP^d)\ge \alpha(\cP_b^d)-2\varepsilon.
\]
Taking $\varepsilon \to 0$ proves the lemma.
\end{proof}

\subsection{Reduction to absolutely continuous distributions}
\label{section:abs:cont}

Translation and dilation of $\R^d$
multiply $L_{\mu,\nu}$ and every $C_{\mu,\nu}(H)$ by the same positive constant.  
Thus, whenever $\mu$ and $\nu$ have bounded support, we may assume
without loss of generality that both supports lie in $[-1,1]^d$.

\begin{lemma}\label{lem:continuous}
$\alpha(\cP_b^d)\ge \alpha(\cP_{bc}^d)$.
\end{lemma}

To prove Lemma~\ref{lem:continuous}, we use the strip notation $B(H,\eta)$ from Section~\ref{subsec:measures-cuts} and two technical lemmas.

The first technical lemma states that given a probability distribution $\mu$ with bounded support whose non-atomic part assigns zero mass on every hyperplane, then for every $\varepsilon > 0$ there is $\eta > 0$ such that the non-atomic part of $\mu$ assigns probability at most $\varepsilon$ to all strips of radius $\eta$ around any hyperplane.

\begin{lemma}\label{lem:na-strip}
Let $\mu$ be supported on $[-1,1]^d$ and suppose
$\mu_{\na}(H)=0$ for every hyperplane $H \in \mathbb H^d$.  Then, for every
$\varepsilon>0$, there exists $\eta>0$ such that
\[
    \mu_{\na}(B(H,\eta))<\varepsilon, \qquad H \in \mathbb H^d.
\]
\end{lemma}
\begin{proof}
It is enough to consider hyperplanes $H$ with
$B(H,1)\cap[-1,1]^d\neq\emptyset$, since otherwise
$\mu_{\na}(B(H,\eta))=0$ for every $\eta\le1$.
If
\[
H=\{x \in \mathbb R^d:u\cdot x=c\},
\qquad \|u\|_2 = 1,
\]
satisfies this condition, then $|c|\le \sqrt d+1$.  Hence these
hyperplanes can be represented by parameters
\[
(u,c)\in \mathbb S^{d-1}\times[-\sqrt d-1,\sqrt d+1],
\]
which form a compact set.

If the lemma were wrong, there would be 
$\varepsilon > 0$ and hyperplanes $(H_n)_{n \in \mathbb N}$ such that
$\mu_{\na}(B(H_n,1/n))\ge\varepsilon$. 
Thus, even though the strips become arbitrarily thin, each of them
contains at least $\varepsilon$ nonatomic mass.
By compactness, there is a subsequence of
$(H_n)_{n \in \mathbb N}$ that converges to a hyperplane $H$.

For every
$\rho>0$, 
convergence of the parameters defining $(H_n)_{n \in \mathbb N}$ to
those defining $H$, together with $1/n\to0$, implies that
for all sufficiently large $n$,
\[
B(H_n,1/n)\cap[-1,1]^d
\subseteq
B(H,\rho).
\]
That is, on the bounded support, a sufficiently thin strip around a
hyperplane close to $H$ is contained in any prescribed wider strip
around $H$.
Hence,
$\mu_{\na}(B(H,\rho))\ge\varepsilon$,
for every $\rho>0$.  By continuity from above,
$\mu_{\na}(H)
=
\lim_{\rho\downarrow0}\mu_{\na}(B(H,\rho))
\ge\varepsilon$,
contradicting the assumption that $\mu_{\na}(H)=0$.
\end{proof}

The second technical lemma considers $n$ small balls whose centers lie in affine general position, and a hyperplane that may cut some of those balls. The lemma states that, provided the radius of the balls is sufficiently small (as a function of their centers), one can tilt the hyperplane in such a way that (i) the tilted hyperplane intersects (at least) one fewer ball than the original hyperplane, and (ii) for each ball that the tilted hyperplane intersects, the volume of the ball that lies in each of the two half spaces defined by the original and tilted hyperplane is roughly the same.

\begin{lemma}\label{lem:perturb}
Let $x_1,\dots,x_n\in\R^d$ be in affine general position, and let
$K\subset\R^d$ be bounded.  For every $\varepsilon>0$ there are
$r_\varepsilon>0$ and $C_\varepsilon<\infty$ such that, whenever
$0<r<r_\varepsilon$, the following holds.
For each $i = 1,\dots,n$ let
\[
    B_i\coloneqq B(x_i,r),
\]
let $H$ be a hyperplane intersecting at least one of  $(B_i)_{i=1}^n$, and choose
$i_0$ with $H\cap B_{i_0}\ne\varnothing$.  For each
$\sigma\in\{+,-\}$ there exists a hyperplane $\widetilde H$ such that:
\begin{enumerate}[label=(\alph*)]
\item $\widetilde H$ intersects no ball that $H$ did not intersect;
\item $\widetilde H\cap B_{i_0}=\varnothing$ and
$B_{i_0}$ lies in $\widetilde H^\sigma$;
\item for every $i\ne i_0$,
\[
 \frac{\Leb\bigl(B_i\cap(H^+\mathbin{\triangle}\widetilde H^+)\bigr)}
      {\Leb(B_i)}<\varepsilon;
\]
\item[(d)] Every $x\in K$ that lies in different open half-spaces
determined by $H$ and $\widetilde H$ satisfies
$\dist(x,H)\leq C_\varepsilon r$.
\end{enumerate}
\end{lemma}
\begin{proof}
For $d=1$ a hyperplane is a point,
and the lemma holds trivially.
We therefore assume that $d \geq 2$.

Fix $\varepsilon>0$.
The idea is to perturb the equation of $H$ by an affine function, chosen so as to move the distinguished ball completely to
the half-space prescribed by $\sigma$, while leaving the other nearby balls essentially
unchanged.
Write
\[
H=\{x \in \mathbb R^d:u\cdot x=c\},
\qquad \|u\|_2 = 1.
\]
Let
\[
\tau \coloneqq
\begin{cases}
1, & \text{if } \sigma=+,\\
-1, & \text{if } \sigma=-.
\end{cases}
\]
Thus, $\tau$ indicates on which side of the perturbed hyperplane
$B_{i_0}$ should lie.

We first record two consequences of affine general position.
Every subset of at most $d$ of the points $x_1,\dots,x_n$ is affinely
independent. Hence,
for every set $J$ of at most $d$ indices and every collection
$(a_i)_{i\in J}$ with $|a_i|\leq 3$, there is an affine function
$h(x)=w\cdot x-b$,
where $w\in\R^d$, and $b\in\R$,
such that
$h(x_i)=a_i$, 
for every $i\in J$.
Since there are only finitely many choices of $J$, 
$w$ and $b$ can be chosen with a uniform bound:
$\|w\|_2 + |b|\leq C_0$,
for some constant $C_0<\infty$. 
The bound $3$ on $(a_i)_{i\in J}$ is only a convenient choice: below we move the normalized
signed distance of the distinguished center to $\pm2$, and its original
normalized signed distance lies in $[-1,1]$.
Moreover, for every fixed $A<\infty$, if $r$ is sufficiently small,
every hyperplane intersects at most $d$ of the balls
$B(x_i,Ar)$.
Indeed, otherwise, as $r\to0$, 
there is a subsequence of such
hyperplanes that converges to a hyperplane that contains $d+1$ of the points $x_i$ in the limit,
contradicting affine general position.
Let
$R_0\coloneqq \max_{1\leq i\leq n}\|x_i\|_2$.
Since $\|w\|_2 + |b|\leq C_0$, we have
$|h(x_i)|\leq C_0(R_0+1)$
for every $1 \leq i \leq n$.
Choose a fixed constant
\[
A>C_0(R_0+1)+2
\]
and set
\[
J\coloneqq \{1 \leq i \leq n:\dist(x_i,H)\leq Ar\}.
\]
The set $J$ contains all centers that are close enough to $H$ to be
potentially affected by a perturbation.
For sufficiently small $r$, 
we have $|J|\leq d$.

For $i \in [n]$, set
\[
s_i\coloneqq \frac{u\cdot x_i-c}{r}.
\]
Thus
\[
I\coloneqq \{1 \leq i \leq n:H\cap B_i\neq\varnothing\}
=
\{1 \leq i \leq n:|s_i|\leq1\}.
\]
Fix $i_0\in I$. Define numbers $(a_i)_{i\in J}$ by
\[
a_{i_0}\coloneqq 2\tau-s_{i_0},
\]
\[
a_i\coloneqq 0,\qquad \hbox{for every } i\in I\setminus\{i_0\},
\]
and
\[
a_i\coloneqq \operatorname{sgn}(s_i),
\qquad \hbox{for every } i\in J\setminus I.
\]
Choose an affine function
\[
h(x)=w\cdot x-b
\]
as above, and define
\[
\widetilde H
\coloneqq
\{x \in \R^d:(u+rw)\cdot x=c+rb\}.
\]
Set
\[
\widetilde u\coloneqq \frac{u+rw}{\|u+rw\|_2},
\qquad
\widetilde c\coloneqq \frac{c+rb}{\|u+rw\|_2}.
\]
Since $H$ intersects at least one of the balls $B_i$, we have
$|c|\leq R_0+r$, and hence $c$ is uniformly bounded for $r<1$.
Together with the uniform bound on $\|w\|_2 + |b|$, this implies
\[
\|\widetilde u-u\|_2+|\widetilde c-c|=O(r).
\]
Consequently, since $K$ is bounded, there is a
constant $C<\infty$ such that
\begin{equation}
\label{eq:perturb-distance}
\sup_{x\in K}
\left|
(u\cdot x-c)
-
(\widetilde u\cdot x-\widetilde c)
\right|
\leq Cr.
\end{equation}

For $i\in J$, we moreover have the more explicit identity
\begin{align}
\label{eq:perturb-normalized}
\frac{\widetilde u\cdot x_i-\widetilde c}{r}
=
\frac{s_i+h(x_i)}{\|u+rw\|_2}=\frac{s_i+a_i}{\|u+rw\|_2}
=
s_i+a_i+O(r).
\end{align}
For $i=i_0$, the right-hand side tends to $2\tau$ as $r$ goes to 0. Hence, for sufficiently
small $r$,
\[
\tau(\widetilde u\cdot x_{i_0}-\widetilde c)>r,
\]
and therefore
$\widetilde H$ does not intersect $B_{i_0}$, and the entire
ball lies in $\widetilde H^\sigma$.
This proves property {\rm (b)}.

If $i\in J\setminus I$, then $|s_i|>1$, 
and our choice
$a_i=\operatorname{sgn}(s_i)$ implies that $s_i+a_i$ has the same sign
as $s_i$ and absolute value larger than $2$.  By
\eqref{eq:perturb-normalized}, for sufficiently small $r$, the ball
$B_i$ therefore remains entirely on the same side of the perturbed
hyperplane and is not intersected by $\widetilde H$.
If $i\notin J$, then
\[
\dist(x_i,H)>Ar.
\]
Moreover,
\[
|h(x_i)|\leq C_0(R_0+1).
\]
Since $A>C_0(R_0+1)+2$, the perturbation term $rh(x_i)$ is too small
to change the sign of $u\cdot x_i-c$ or to bring the center within
distance $r$ of the perturbed hyperplane, provided $r$ is sufficiently
small.  Thus $B_i$ again remains entirely on the same side and is not
intersected by $\widetilde H$.
Hence, $\widetilde H$ intersects no ball
that $H$ did not intersect. This proves property {\rm (a)}.

It remains to prove property {\rm (c)}. Let
\[
 f(x)=u\cdot x-c,\qquad
 g(x)=(u+rw)\cdot x-(c+rb)=f(x)+rh(x).
\]
For $i\in I\setminus\{i_0\}$, we have $h(x_i)=0$. If $x\in B_i$ changes side, then
\[
 |f(x)|\leq|g(x)-f(x)|
 =r|w\cdot(x-x_i)|\leq r\|w\|_2\|x-x_i\|_2 \leq C_0r^2.
\]
The changing points therefore lie in a strip of width $2C_0r^2$.
Its volume inside a radius-$r$ ball is at most
$2C_0v_{d-1}r^{d+1}$, where $v_k$ is the volume of the Euclidean unit ball
in $\R^k$. Its fraction of the ball is at most
$2C_0(v_{d-1}/v_d)r$, uniformly in $H$ and $i$.
For $i\notin I$, the existing proof of (a) shows that the whole ball
stays on the same side. Thus
\[
 \frac{\Leb\bigl(B_i\cap(H^+\mathbin{\triangle}\widetilde H^+)\bigr)}
      {\Leb(B_i)}<\varepsilon;
\]
holds for sufficiently small $r$.

Finally, \eqref{eq:perturb-distance} proves property {\rm (d)}.
Indeed, if 
$x\in\widetilde H\cap K$, then $\widetilde u\cdot x-\widetilde c=0$.
If $x \in K$ lies in different open half-spaces of $H$ and$\widetilde H$, then either $u \cdot x - c$ and $\widetilde u \cdot x - \widetilde c$ have opposite signs or one vanishes, so
\[
    \dist(x,H) = |u\cdot x-c| \leq |(u\cdot x-c) - (\widetilde u \cdot x - \widetilde c)| \leq C r,
\]
where the inequality follows by \eqref{eq:perturb-distance}.
Hence
\[
\widetilde H\cap K
\subseteq
B(H,Cr)\cap K.
\]
Taking $C_\varepsilon\coloneqq C$ proves property {\rm (d)}.
Thus $\widetilde H$ satisfies all the required properties.
\end{proof}

We are now ready to prove Lemma~\ref{lem:continuous}.
\begin{proof}[Proof of Lemma~\ref{lem:continuous}]
Fix $\varepsilon>0$.  Choose $(\mu,\nu)\in\cP_b^d$ such that
$\supp(\mu)\cup\supp(\nu)\subseteq[-1,1]^d$ and
\begin{equation}\label{eq:cont-choice}
\alpha(\cP_b^d)
\ge
\sup_{H\in\Hh^d}\frac{C_{\mu,\nu}(H)}{L_{\mu,\nu}}-\varepsilon.
\end{equation}
Choose
$0<\tau<\frac{\varepsilon L_{\mu,\nu}}{d+2}$,
$0<\delta<\frac{\tau}{64d^{3/2}}$,
and two finite sets
$A(\mu)\subseteq\atom(\mu)$ and
$A(\nu)\subseteq\atom(\nu)$
such that
\begin{align*}
\sum_{x\in\atom(\mu)\setminus A(\mu)}\mu(\{x\})<\delta,
\qquad
\sum_{x\in\atom(\nu)\setminus A(\nu)}\nu(\{x\})<\delta.
\end{align*}
Let $A(\mu)\cup A(\nu)=\{x_1,\dots,x_n\}$. By assumption, these points are in affine general position.

By Lemma~\ref{lem:na-strip}, there exists $s>0$ such that, for every
hyperplane $H$,
\begin{equation}\label{eq:cont-strip}
\mu_{\na}(B(H,s))<\delta,
\qquad
\nu_{\na}(B(H,s))<\delta.
\end{equation}

Set $\kappa\coloneqq \tau/(16\sqrt d)$, and let $r_\kappa$ and $C_\kappa$ be the
constants from Lemma~\ref{lem:perturb}, applied with
$K=[-2,2]^d$.
Choose
\[
0<r<
\min\left\{
\delta,\,
1,\,
r_\kappa,\,
\frac{s}{C_\kappa+1}
\right\}
\]
sufficiently small such that every hyperplane intersects at most $d$ of the balls $B_i\coloneqq B(x_i,r)$ for $i=1,\dots,n$.

Let $U$ and $V$ be independent random variables, each uniformly distributed on $B(0,r)$, and
independent of $X\sim \mu$ and $Y\sim \nu$. 
Define
\[
X'\coloneqq X+U,\qquad Y'\coloneqq Y+V,
\]
and let $\mu',\nu'$ denote their laws.  
Since $r<1$, both $\mu'$ and $\nu'$ are supported on $[-2,2]^d$, whose
diameter is $4\sqrt d$.  
Both $\mu'$ and $\nu'$ are absolutely continuous.
Moreover, $W_1(\mu,\mu')\le r$ and $W_1(\nu,\nu')\le r$, and thus
\begin{equation}\label{eq:cont-L}
|L_{\mu,\nu}-L_{\mu',\nu'}|\le 2r.
\end{equation}

For a hyperplane $H$, denote
$m(H)\coloneqq \#\{i:H\cap B_i\ne\varnothing\}$.
By the choice of $r$,
$m(H)\le d$ for every $H \in \mathbb H^d$.
We first establish a one-step rounding statement.  

\medskip
\begin{claim}\label{cl:rounding}
    If $\delta$ and $r$ are sufficiently small, then whenever $m(H)>0$ there
    exists a hyperplane $\widetilde H$ such that $m(\widetilde H)\le m(H)-1$ and
    \[
    C_{\mu',\nu'}(\widetilde H)\ge C_{\mu',\nu'}(H)-\tau.
    \]
\end{claim}
\begin{proof}[Proof of Claim~\ref{cl:rounding}]
    To prove the claim, choose $i_0$ with $H\cap B_{i_0}\ne\varnothing$.
Suppose, without loss of generality, that $x_{i_0}\in A(\mu)$. Orient $H$ and define
\[
t_i(H)\coloneqq
\frac{\Leb(B_i\cap H^+)}{\Leb(B_i)}.
\]
When changing $\mu$ to $\mu'$, 
the mass $\mu(\{x_{i_0}\})$ becomes $\mu(\{x_{i_0}\})U_{B_{i_0}}$.
If, in the contribution of this component to
$C_{\mu',\nu'}(H)$ we replace $\|x'-y'\|$ by $\|x_{i_0}-y'\|$, the total error is at most $r$.  
After this replacement,
the approximate contribution of this component is an affine function of
$t_{i_0}(H)$:
\[
a\,t_{i_0}(H)+b(1-t_{i_0}(H))
\]
for some $a,b\in [0,4\sqrt{d}]$.
Hence one of the two endpoint choices
$t_{i_0}=0$ or $t_{i_0}=1$ does not decrease this approximate
contribution.

Apply Lemma~\ref{lem:perturb} 
with accuracy
$\kappa=\tau/(16\sqrt d)$,
so that $B_{i_0}$ is moved entirely to the
preferred side and is no longer intersected.
Then, for every $i\ne i_0$

\begin{equation}
\label{equ:t:1}
 \frac{\Leb\bigl(B_i\cap(H^+\mathbin{\triangle}\widetilde H^+)\bigr)}
      {\Leb(B_i)}
 < \frac{\tau}{16\sqrt d}.
\end{equation}

We now discuss the difference between 
$C_{\mu',\nu'}(H)$ and 
$C_{\mu',\nu'}(\widetilde H)$.

\begin{itemize}
\item
As discussed above, rounding the ball $B_{i_0}$ to the preferred side
does not decrease its approximate contribution. 
The total error caused by the approximation is at most $2r$.
\item
For every $i \neq i_0$ such that $B_i \cap H \neq \varnothing$,
\eqref{equ:t:1} controls the change in the fraction of each
other selected component that changes side:
the total error from all these atoms is at most
\[
2(4\sqrt d)\frac{\tau}{16\sqrt d}
=
\frac{\tau}{2}.
\]
\item 
The total mass of the atoms in $(\atom(\mu) \setminus A(\mu)) \cup (\atom(\nu) \setminus A(\nu))$
is less than $2\delta$.
Hence their total contribution can change by at most
$8\sqrt d\,\delta$.
\item
If a point from one of the non-atomic parts changes side when passing
from $H$ to $\widetilde H$, then its original location
lies in the $s$-neighborhood of $H$, since
$(C_\kappa+1)r<s$.
By \eqref{eq:cont-strip}, the total such mass for $\mu$ and $\nu$ is less
than $2\delta$.  Hence, the resulting error is at most
$8\sqrt d\,\delta$.
\end{itemize}
Altogether,
\[
C_{\mu',\nu'}(\widetilde H)
\ge
C_{\mu',\nu'}(H)
-
\left(
2r+\frac{\tau}{2}+16\sqrt d\,\delta
\right).
\]
By the choices of $\delta$ and $r$,
\[
16\sqrt d\,\delta<\frac{\tau}{4}
\qquad\text{and}\qquad
2r<2\delta<\frac{\tau}{4}.
\]
Hence
\[
C_{\mu',\nu'}(\widetilde H)
\ge
C_{\mu',\nu'}(H)-\tau.
\]
This proves the claim.
\end{proof}

For $k=0,\dots,d$, define
\[
S_k\coloneqq \sup\{
C_{\mu',\nu'}
(H) \colon m(H)\le k\}.
\]
Claim~\ref{cl:rounding} implies
$S_{k-1}\ge S_k-\tau$ for every $k=1,\dots,d$.
Since
$S_d=\sup_{H\in\mathbb H^d}C_{\mu',\nu'}(H)$,
we obtain
$S_0\ge
\sup_{H\in\mathbb H^d}C_{\mu',\nu'}(H)-d\tau$.
By the definition of $S_0$, there exists a hyperplane $H_0$ with
$m(H_0)=0$ such that
\begin{equation}\label{eq:cont-H0}
C_{\mu',\nu'}(H_0)
\ge
\sup_{H\in\mathbb H^d}C_{\mu',\nu'}(H)-(d+1)\tau.
\end{equation}

We argue that
\begin{equation}\label{eq:cont-Cfinal}
C_{\mu,\nu}(H_0)
\ge
C_{\mu',\nu'}(H_0)-16\sqrt d\,\delta-2r.
\end{equation}
Indeed, because $H_0$ intersects none of the balls $(B_i)_{i=1}^n$, every selected atom
$x_i$ and every point in $B_i$ lie on the same side of $H_0$.  
The total
mass of atoms in $(\atom(\mu)\setminus A(\mu)) \cup (\atom(\nu)\setminus A(\nu))$ is less than $2\delta$.
If $X$ and $X'$ lie on different
sides of $H_0$, then $X\in B(H_0,r)$; by
\eqref{eq:cont-strip}, the total non-atomic mass for which this can happen
is less than $2\delta$.  Finally,
\[
\bigl|\|X'-Y'\|-\|X-Y\|\bigr|\le 2r.
\]
Combining \eqref{eq:cont-H0} and \eqref{eq:cont-Cfinal},
\[
\sup_{H\in\mathbb H^d}C_{\mu,\nu}(H)
\ge
\sup_{H\in\mathbb H^d}C_{\mu',\nu'}(H)
-
\left(
(d+1)\tau+16\sqrt d\,\delta+2r
\right).
\]
Together with \eqref{eq:cont-L} and
$C_{\mu',\nu'}(H)\le L_{\mu',\nu'}$, this gives
\[
\sup_{H\in\mathbb H^d}
\frac{C_{\mu,\nu}(H)}{L_{\mu,\nu}}
\ge \sup_{H\in\mathbb H^d} \frac{C_{\mu',\nu'}(H)}{L_{\mu',\nu'}}
- \frac{(d+1)\tau+16\sqrt d\,\delta+4r}{L_{\mu,\nu}}.
\]
By the choices of $\delta$ and $r$,
\[
16\sqrt d\,\delta+4r
<\frac{\tau}{4}+4\delta
<\tau.
\]
Hence
\[
(d+1)\tau+16\sqrt d\,\delta+4r
<(d+2)\tau
<\varepsilon L_{\mu,\nu}.
\]
Consequently,
\[
\sup_{H\in\mathbb H^d}
\frac{C_{\mu,\nu}(H)}{L_{\mu,\nu}}
\ge \sup_{H\in\mathbb H^d}
\frac{C_{\mu',\nu'}(H)}{L_{\mu',\nu'}}-\varepsilon
\ge\alpha(\cP_{bc}^d)-\varepsilon.
\]
The desired result follows from \eqref{eq:cont-choice} and by taking $\varepsilon \to 0$.
\end{proof}

\subsection{Reduction to equal-weight finite distributions}

In this section we prove the inequality 
$\alpha(\cP_{bc}^d)\ge\alpha(\cP_\N^d)$.
We start with an auxiliary result.

\begin{lemma}
\label{lemma:empirical}
Let $\mu$ and $\nu$ be absolutely continuous probability distributions
whose supports are contained in $[-1,1]^d$. Then, for every
$\varepsilon>0$, there exists an $n \in \N$ and $(\mu_n,\nu_n)\in\cP_n^d$ such that
\begin{equation}
\label{equ:l1}
|L_{\mu,\nu}-L_{\mu_n,\nu_n}|<\varepsilon
\end{equation}
and
\begin{equation}
\label{equ:c1}
\sup_{H\in\mathbb H^d}
|C_{\mu,\nu}(H)-C_{\mu_n,\nu_n}(H)|<\varepsilon.
\end{equation}
\end{lemma}

\begin{proof}
Since $\mu$ and $\nu$ are absolutely continuous,
\[
\mu(H)=\nu(H)=0,
\qquad\text{for every }H\in\mathbb H^d.
\]
Fix $\varepsilon>0$.
By Lemma~\ref{lem:na-strip}, we may choose
$\delta>0$ sufficiently small such that

\begin{equation}\label{eq:empirical-strip}
2\delta
+
2\sqrt d\,
\sup_{H\in\mathbb H^d}
\bigl(\mu(B(H,\delta))+\nu(B(H,\delta))\bigr)
<
\frac{\varepsilon}{2}.
\end{equation}
Intuitively, $\delta$ controls both the displacement within each subset and the amount of mass that can change sides of a hyperplane under such a displacement.
Partition the set $[-1,1]^d$ into finitely many Borel subsets 
$(F_i)_{i=1}^I$
with non-empty interior
and diameter smaller than $\delta$.

Choose $n \geq 16I\sqrt d/\varepsilon$
and
non-negative integers $(n_i^\mu)_{i=1}^I$ and $(n_i^\nu)_{i=1}^I$ such that
\[
\sum_{i=1}^I n_i^\mu=\sum_{i=1}^I n_i^\nu=n,
\qquad
\left|\mu(F_i)-\frac{n_i^\mu}{n}\right|<\frac1n,
\qquad
\left|\nu(F_i)-\frac{n_i^\nu}{n}\right|<\frac1n.
\]
Such integers can be obtained by rounding down 
$(\mu(F_i),\nu(F_i))_{i=1}^I$ and distributing the
remaining units among subsets $(F_i)_{i=1}^I$ with the largest positive fractional parts, namely, 
those that maximize $n\mu(F_i) - \lfloor n \mu(F_i)\rfloor$
(for $(n_i^\mu)_{i=1}^I$)
and
$n \nu(F_i) - \lfloor n \nu(F_i)\rfloor$
(for $(n_i^\nu)_{i=1}^I$).
Let
\[
e_\mu\coloneqq
\sum_{i=1}^I
\left|\mu(F_i)-\frac{n_i^\mu}{n}\right| \leq \frac{2I}{n},
\qquad \text{and} \qquad
e_\nu\coloneqq
\sum_{i=1}^I
\left|\nu(F_i)-\frac{n_i^\nu}{n}\right| \leq \frac{2I}{n}.
\]
Thus, $e_\mu$ and $e_\nu$ measure the total error introduced by rounding the masses of the subsets to multiples of $1/n$.

For each $1 \leq i \leq I$, choose $n_i^\mu$ distinct points in $F_i$
which will be atoms of $\mu_n$,
and
$n_i^\nu$ distinct points in $F_i$ which will be atoms of $\nu_n$.
The points can be chosen so that all $2n$ points are in affine general position. Indeed, choose them sequentially. When choosing a new point, the previously chosen points determine only finitely many affine subspaces of dimension at most $d-1$ whose union must be avoided in order to preserve affine general position. Since each $F_i$ has non-empty interior, it is not contained in this finite union, so a valid new point can always be chosen. Let $\mu_n$ (resp., $\nu_n$) be the atomic probability measure on $\R^d$ whose atoms are the $n = \sum_{i=1}^I n_i^\mu$ (resp., $n = \sum_{i=1}^I n_i^\nu$) points selected in the first sentence of this paragraph, and where all atoms have the same mass $\frac{1}{n}$.

In particular, all $2n$ points are distinct, and hence
$\atom(\mu_n)\cap\atom(\nu_n)=\varnothing$.
Thus $(\mu_n,\nu_n)\in\cP_n^d$.

It remains to show that these distributions satisfy the desired property.
Couple $\mu$ and $\mu_n$ by matching, in each subset $F_i$, mass
\[
\min\left\{\mu(F_i),\frac{n_i^\mu}{n}\right\}
\]
within that subset, and coupling the remaining mass arbitrarily.
Since both $\mu$ and $\mu_n$ have total mass $1$, the total unmatched mass is
$e_\mu/2$.  
The factor $1/2$ appears because each unit of mass moved between subsets is counted once as excess mass and once as missing mass in the definition of $e_\mu$.
Since the diameter of $F_i$ is smaller than $\delta$ for every $i$,
and the
diameter of $[-1,1]^d$ is $2\sqrt d$,
\[
W_1(\mu,\mu_n)\le \delta+\sqrt d\,e_\mu,
\qquad \text{and} \qquad
W_1(\nu,\nu_n)\le \delta+\sqrt d\,e_\nu.
\]
Therefore, by Lemma~\ref{lem:L-cont},
\[
|L_{\mu,\nu}-L_{\mu_n,\nu_n}|
\le
2\delta+\sqrt d\,(e_\mu+e_\nu)
<\varepsilon,
\]
where the last inequality follows from
\eqref{eq:empirical-strip} and the choice of $n$.
Thus, \eqref{equ:l1} holds.

We turn to prove \eqref{equ:c1}.
Let
$(X,X_n)$ 
be the random variable that couples $\mu$ and $\mu_n$,
and let 
$(Y,Y_n)$ be the random variable that couples $\nu$ and $\nu_n$,
and assume that these two random variables are independent.
Then,
\[
    \E\left[\|X-X_n\|\right] \leq \frac{e_\mu}{2} \cdot 2\sqrt d  + \left(1-\frac{e_\mu}{2}\right) \delta \leq \delta + \sqrt{d} \: e_\mu,
\]
and an analogous inequality holds for $\E[\|Y-Y_n\|]$.

Fix $H\in\mathbb H^d$.  Let $A_X(H)$ be the event that $X$ and $X_n$
do not lie in the same open half-space determined by $H$, and define
$A_Y(H)$ analogously.  If $X$ and $X_n$ lie in the same
subset $F_i$ and $A_X(H)$ occurs, then $X\in B(H,\delta)$, because the diameter of $F_i$ is smaller than $\delta$. In other words, a point that is moved by less than $\delta$ can cross $H$ only if it was initially within distance $\delta$ of $H$.
Hence
\[
\Pr[A_X(H)]
\le
\frac{e_\mu}{2}+\mu(B(H,\delta)),
\]
and similarly
\[
\Pr[A_Y(H)]
\le
\frac{e_\nu}{2}+\nu(B(H,\delta)).
\]
If neither $A_X(H)$ nor $A_Y(H)$ occurs, then
\[
\1_{\text{Sep}(H)}(X,Y)=\1_{\text{Sep}(H)}(X_n,Y_n).
\]
Moreover,
\[
\bigl|\|X-Y\|-\|X_n-Y_n\|\bigr|
\le
\|X-X_n\|+\|Y-Y_n\|,
\]
and all relevant distances are at most $2\sqrt d$.  Consequently,
\begin{align*}
|C_{\mu,\nu}(H)-C_{\mu_n,\nu_n}(H)|
&\le
\E[\|X-X_n\|]+\E[\|Y-Y_n\|]+
2\sqrt d\,
\bigl(\Pr[A_X(H)]+\Pr[A_Y(H)]\bigr)\\
&\le
2\delta+2 \sqrt d\,(e_\mu+e_\nu)+
2\sqrt d\,
\bigl(\mu(B(H,\delta))+\nu(B(H,\delta))\bigr).
\end{align*}
The three terms above correspond respectively to the displacement within subsets, the rounding error of the subset masses, and the mass lying close enough to $H$ to change sides.
By \eqref{eq:empirical-strip} and
the choice of $n$,
the right-hand side is smaller than
$\varepsilon$, uniformly over $H$.  Therefore
\eqref{equ:c1} is proved.
\end{proof}

\begin{lemma}\label{lem:finite}
\[
\alpha(\cP_{bc}^d)\ge \alpha(\cP_\N^d).
\]
\end{lemma}

\begin{proof}
Fix $\varepsilon \in (0,1)$, and choose $(\mu,\nu)\in\cP_{bc}^d$ such that
\begin{equation}
\label{equ:red:d}
\alpha(\cP_{bc}^d)
\geq
\sup_{H\in\mathbb H^d}
\frac{C_{\mu,\nu}(H)}{L_{\mu,\nu}}
-\varepsilon.
\end{equation}
By translation and dilation, we may assume without loss of generality that $\supp(\mu)\cup\supp(\nu)\subseteq[-1,1]^d$. Choose $\delta>0$ such that
\begin{equation}
\label{equ:red:c}
\delta<\frac{\varepsilon L_{\mu,\nu}}{4}.
\end{equation}
By Lemma~\ref{lemma:empirical}, there exists an $n \in \N$ and
$(\mu_n,\nu_n)\in\cP_n^d$ such that
\[
|L_{\mu,\nu}-L_{\mu_n,\nu_n}|<\delta
\]
and
\[
\sup_{H\in\mathbb H^d}
|C_{\mu,\nu}(H)-C_{\mu_n,\nu_n}(H)|<\delta.
\]
Since $\varepsilon<1$ and
$|L_{\mu,\nu}-L_{\mu_n,\nu_n}|<\delta$, we have
\begin{equation}
\label{equ:red:a}
L_{\mu_n,\nu_n}
\geq L_{\mu,\nu}-\delta
>
\frac34L_{\mu,\nu}>0.
\end{equation}
For every hyperplane $H \in \mathbb H^d$,
\begin{equation}
\label{equ:ddd}
C_{\mu,\nu}(H)
\ge
C_{\mu_n,\nu_n}(H)-\delta.
\end{equation}
Since both $C_{\mu,\nu}(H)$ and $L_{\mu,\nu}$ are approximated additively, and $L_{\mu,\nu}$ is bounded away from zero relative to $\delta$, the corresponding ratios are close as well.
By \eqref{equ:ddd},
\begin{align*}
\frac{C_{\mu,\nu}(H)}{L_{\mu,\nu}}
&\ge
\frac{C_{\mu_n,\nu_n}(H)}{L_{\mu,\nu}}
-
\frac{\delta}{L_{\mu,\nu}}
=
\frac{C_{\mu_n,\nu_n}(H)}{L_{\mu_n,\nu_n}}
\frac{L_{\mu_n,\nu_n}}{L_{\mu,\nu}}
-
\frac{\delta}{L_{\mu,\nu}}.
\end{align*}
Eq.~\eqref{equ:red:a} implies that
$\frac{L_{\mu_n,\nu_n}}{L_{\mu,\nu}}
\ge
1-\frac{\delta}{L_{\mu,\nu}}$.
Since
$C_{\mu_n,\nu_n}(H)\le L_{\mu_n,\nu_n}$,
and by \eqref{equ:red:c},
we then obtain
\[
\frac{C_{\mu,\nu}(H)}{L_{\mu,\nu}}
\ge
\frac{C_{\mu_n,\nu_n}(H)}{L_{\mu_n,\nu_n}}
-
\frac{2\delta}{L_{\mu,\nu}}
>
\frac{C_{\mu_n,\nu_n}(H)}{L_{\mu_n,\nu_n}}
-\varepsilon.
\]
Therefore,
\[
\sup_{H\in\mathbb H^d}
\frac{C_{\mu,\nu}(H)}{L_{\mu,\nu}}
\geq
\sup_{H\in\mathbb H^d}
\frac{C_{\mu_n,\nu_n}(H)}{L_{\mu_n,\nu_n}}
-\varepsilon.
\]
Since $(\mu_n,\nu_n)\in\cP_n^d$,
\begin{equation}
\label{equ:red:e}
\sup_{H\in\mathbb H^d}
\frac{C_{\mu_n,\nu_n}(H)}{L_{\mu_n,\nu_n}}
\geq \alpha(\cP_n^d).
\end{equation}
By \eqref{equ:red:d} and \eqref{equ:red:e},
\[
\alpha(\cP_{bc}^d)
\geq
\alpha(\cP_n^d)-2\varepsilon
\geq
\alpha(\cP_{\mathbb N}^d)-2\varepsilon.
\]
Taking $\varepsilon \to 0$ proves $\alpha(\cP_{bc}^d) \geq \alpha(\cP_\N^d)$.
\end{proof}

\section*{AI Disclosure Statement}

The authors developed the problem formulation in January of 2024 and have been continuously working on the problem since. The proofs of Theorems~\ref{thm:measure-equivalence} and~\ref{thm:intro:upper-bound} were developed solely by the authors. The proof of Theorem~\ref{thm:dim-1:exact} was originally obtained by Claude Code (Opus 4.6). Subsequently, the authors came up with a completely different and much simpler proof, which is the one presented here. For Theorems~\ref{thm:lower-bound:dimension-three-or-above} and~\ref{thm:lower-bound:dimension-two}, a weaker guarantee was originally obtained solely by the authors. Subsequently, ChatGPT (GPT 5.6 Sol and GPT 6 Astra) was used to arrive at a simpler proof, which is the one presented here. The proofs in Sections~\ref{subsec:large-dimension} and~\ref{sec:improved-bounds} were developed by ChatGPT (GPT 6 Astra) after conversations with the authors, who identified that the lower bound of $1/3$ and upper bound of $1/2$ were not tight on the plane. The paper was written by the authors, who take full responsibility for its correctness and logical integrity.

\bibliographystyle{alpha}
\bibliography{refs}

\appendix

\section{Projections}
\label{app:projections}

In this section, we show that simple coordinate projection does not guarantee a constant in higher dimensions and in fact its guarantee can be polynomially worse, for any norm on $\R^d$. The next example shows that using only the coordinate directions can lose a factor of $d$.

\begin{proposition}\label{prop:coordinate-projection-hardness}
For every $d \geq 2$ and every norm on $\R^d$, there are red and blue point sets $R,B \subset \R^d$, with $|R| = |B| = d$, such that some hyperplane separates every red--blue pair, while every coordinate-projection optimum lifts to a hyperplane $H$ satisfying
\[
    \frac{C(H)}{L} \leq \frac{3}{d}.
\]
\end{proposition}
\begin{proof}
Let $e_1, \dots, e_d$ be the standard coordinate vectors. For each $i \in [d]$, let
\[
    a_i = \frac{1}{\|e_i\|}, \qquad r_i = a_i e_i, \qquad b_i = - a_i e_i,
\]
$R = \set{r_1, \dots, r_d}$ and $B = \set{b_1, \dots, b_d}$. Note that every atom in $R \cup B$ has norm $1$ and that the hyperplane $H = \set{x \in \R^d \midd \sum_{\ell = 1}^d x_\ell = 0}$ strictly separates all red points from all blue points, since the sum over all coordinates of each red point $r_i$ is $a_i > 0$, while that of each blue point $b_i$ is $- a_i < 0$, so the two colors lie strictly on opposite sides of $H$. Therefore, the optimal cut ratio is $1$.

Fix a coordinate $k \in [d]$. The projected points take one of three values: $b_k$ projects to $- a_k$, $r_k$ projects to $a_k$, and all other points project to $0$. A threshold in either gap $(-a_k, 0)$ or $(0, a_k)$ isolates
$b_k$ or $r_k$, respectively, and has projected value
\[
    2 a_k + (d - 1) a_k = (d + 1) a_k,
\]
and every other threshold has smaller value. Therefore, every projected optimum isolates a single point. For either isolating cut, the lifted hyperplane $H$ has value
\[
    C(H) = \sum_{j = 1}^d \|r_k - b_j\| = \sum_{j = 1}^d \|r_k + r_j\| \leq \sum_{j = 1}^d \prn{\|r_k\| + \|r_j\|} = 2 d.
\]
To bound $L$ from below, we apply the triangle inequality to any $i,j,\ell \in [d]$:
\[
    2 = 2\|r_i\| = \|(r_i + r_j) + (r_i + r_\ell) - (r_j + r_\ell)\| \leq \|r_i + r_j\| + \|r_i + r_\ell\| + \|r_j + r_\ell\|.
\]
Summing the above inequality over all $d^3$ triples $i,j,\ell \in [d]$ yields
\[
    2 d^3 \leq d \sum_{i,j = 1}^d \|r_i + r_j\| + d \sum_{i,\ell = 1}^d \|r_i + r_\ell\| + d \sum_{j,\ell = 1}^d \|r_j + r_\ell\| = 3 d \sum_{i,j = 1}^d \|r_i + r_j\| = 3 d L.
\]
Thus $L \geq (2/3) d^2$, and every coordinate-projection optimum lifts to a hyperplane $H$ satisfying
\[
    \frac{C(H)}{L} \leq \frac{2d}{\frac{2}{3} d^2} = \frac{3}{d}. \qedhere
\]
\end{proof}

\end{document}